\documentclass[a4paper,11pt]{article}
\pdfoutput=1 

\usepackage{jheppub} 

\usepackage[T1]{fontenc} 
\usepackage{placeins}
\usepackage{rotating}
\usepackage{dblfloatfix}
\usepackage{graphicx, caption, lscape, threeparttable}

\newcommand{\comment}[1]{}

\usepackage{xspace} 
\usepackage{upgreek}

\usepackage{subfloat}
\usepackage{subfig}
\usepackage{graphicx}
\usepackage{amsmath}

\def\PD      {\ensuremath{D}\xspace}            
\def\Dbar    {{\kern 0.2em\overline{\kern -0.2em \PD}{}}\xspace}
\def\Dzb     {{\ensuremath{\Dbar{}^0}}\xspace}

\def\CP {{\ensuremath{C\!P}}\xspace}

\def\PB      {\ensuremath{B}\xspace}            
\def\Bbar    {{\kern 0.2em\overline{\kern -0.2em \PB}{}}\xspace}

\title{\boldmath Model-independent method for measuring the angular coefficients of $B^0 \to D^{*-} \tau^+ \nu_{\tau}$ decays}

\author[a,1]{Donal Hill,\note{Corresponding author.}}
\author[a]{Malcolm John,}
\author[b]{Wenqi Ke,}
\author[c]{Anton Poluektov}

\affiliation[a]{University of Oxford, United Kingdom}
\affiliation[b]{D\'epartement de Physique, \'Ecole Normale Sup\'erieure, Paris, France}
\affiliation[c]{Aix Marseille Univ, CNRS/IN2P3, CPPM, Marseille, France}

\emailAdd{donal.hill@cern.ch}
\emailAdd{malcolm.john@physics.ox.ac.uk}
\emailAdd{wke@clipper.ens.fr}
\emailAdd{poluektov@cppm.in2p3.fr}

\abstract{
Reconstruction of the $B^0 \to D^{*-} \tau^+ \nu_{\tau}$ angular distribution is complicated by the strongly biasing effect of losing the neutrino information from both the $B$ and $\tau$ decays. In this work, a novel method for making unbiased measurements of the angular coefficients while preserving the model independence of the angular technique is demonstrated. The twelve angular functions that describe the signal decay, in addition to background terms, are modelled in a multidimensional fit, using template probability density functions that encapsulate all resolution and acceptance effects. Sensitivities at the LHCb and Belle II experiments are estimated, and sources of systematic uncertainty are discussed, notably in the extrapolation to a measurement of $R(D^{*})$.}

\begin{document} 
\maketitle
\flushbottom

\section{Introduction}
\label{sec:intro}

An indication of new physics (NP) is emerging though the evidence that $b \to c\tau\nu_\tau$ transitions proceed at a higher rate than would be expected from a comparison to the same process with the lighter leptons. The ratio of branching fractions
\begin{equation*}
R(D^{(*)}) = \frac{\mathcal{B}(B \to D^{(*)} \tau \nu)}{\mathcal{B}(B \to D^{(*)} l \nu)} \hspace{0.8cm} l \in \{e,\mu\}
\end{equation*}
is measured above the Standard
Model (SM) expectation in results from the $B$-factories~\cite{Lees:2012xj,Lees:2013uzd,Huschle:2015rga,Hirose:2016wfn,Hirose:2017dxl,Abdesselam:2019dgh} and LHCb~\cite{Aaij:2015yra,PhysRevD.97.072013}. The current global averages of experimental results agree with the SM predictions \cite{HFLAV16} at only $\sim3$ standard deviations:
\begin{eqnarray*}
R(D^{*})^{\rm exp} = 0.295 \pm 0.014\,, &\hspace{1.0cm}& R(D)^{\rm exp} = 0.340 \pm 0.030\,,\\
R(D^{*})^{\rm SM} = 0.258 \pm 0.005\,, &\hspace{1.0cm}& R(D)^{\rm SM} = 0.299 \pm 0.003\,.
\end{eqnarray*}

In order to further characterise the underlying physics in $b \to c\tau\nu_\tau$ transitions, it is necessary to study the kinematics of semitauonic $B$ decays in addition to their rates. Many polarisation and asymmetry observables have been shown to discriminate between the SM and NP scenarios~\cite{Biancofiore:2013ki,Duraisamy2013,PhysRevD.85.094025,Tanaka1995,PhysRevD.82.034027,PhysRevD.87.054002,PhysRevD.86.034027,PhysRevD.92.114022,Becirevic:2016hea,PhysRevD.94.094021,PhysRevD.95.115038,Bardhan2017,PhysRevD.95.036021,PhysRevD.95.093006,PhysRevD.99.035015,Colangelo:2018cnj}.  
One such example is the $D^{*}$ longitudinal polarisation fraction, which has recently been measured to be $F_{L}^{D^*} = 0.60 \pm 0.09$~\cite{Abdesselam:2019wbt}. Several calculations of the SM expectation exist, which centre around 0.45~\cite{Tanaka:2012nw,PhysRevD.92.114022,PhysRevD.95.115038,PhysRevD.98.095018,Bhattacharya:2018kig,Becirevic:2019tpx}; this tension constitutes another potential indication of deviation from the SM.

Complete information on the \mbox{$B^0 \to D^{*-}\tau^{+}\nu_\tau$} decay kinematics is ultimately obtained from the full angular decay rate~\cite{Becirevic:2019tpx}
\begin{equation}
\begin{split}
\frac{d^{4}\Gamma}{d q^2\, d(\cos{\theta_D})\, d(\cos{\theta_L})\, d\chi } &\propto I_{1c}\cos^2{\theta_D} + I_{1s}\sin^2{\theta_D} \\ 
&+ [I_{2c}\cos^2{\theta_D} + I_{2s}\sin^2{\theta_D}]\cos2\theta_L \\
&+ [I_{6c}\cos^2{\theta_D} + I_{6s}\sin^2{\theta_D}]\cos\theta_L \\
&+ [I_{3}\cos2\chi + I_{9}\sin2\chi]\sin^2{\theta_L}\sin^2{\theta_D} \\
&+ [I_{4}\cos\chi + I_{8}\sin\chi]\sin2\theta_L\sin2\theta_D \\
&+ [I_{5}\cos\chi + I_{7}\sin\chi]\sin\theta_L\sin2\theta_D
\,,\end{split}
\label{eq:decay_rate}
\end{equation}
where the angles (${\theta_D}$, ${\theta_L}$, $\chi$) parameterise the spin-0 $B^0$ meson decay topology, and are defined in App.~\ref{sec:angles}. This expression involves a sum of twelve independent angular functions, each of which is multiplied by a coefficient $I_X$ ($X \in \{1c,1s,2c,2s,3,4,5,6c,6s,7,8,9\}$) that encapsulates the dependence on the square of the dilepton invariant mass, $q^2$, form factors, and the fundamental couplings. The angular distribution can reveal the influence of NP even if $R(D^*)$ becomes fully compatible with the SM.

\begin{sloppypar}
Angular analysis is well established in the study of rare dimuon decays such as \mbox{$B^0 \to K^{*0} \mu^+\mu^-$~\cite{Alok:2011gv,Aaij2013}}.  The principal advantage of the technique is that the coefficients contain all form factor dependence, so there is no experimental uncertainty due to a choice of form factor scheme. Combinations of the angular coefficients can also reduce dependence on the form factors in subsequent phenomenological interpretations. The difficulty that arises in applying angular analysis methods to semitauonic decays is the missing information due to the lost neutrinos in both the $B$ and $\tau$ decays, which strongly sculpts the angular distribution and makes a parametric fit to data impossible. 

In this paper, a novel approach is presented that uses a multidimensional template fit in the angular basis to measure the $I_X$ coefficients in a model-independent manner without statistical biases. 
The technique assumes and requires excellent agreement between data and simulated samples for the construction of the templates, which must describe all reconstruction, resolution and migration effects. In this case the fit reduces to a linear sum of twelve independent templates, preserving the inherent model independence of the angular method. Eleven $I_X$ parameters are measured (one is fixed by $\sum I_X=1$), and the model dependence is confined to one overall signal yield. The expected resolutions and covariance matrix of the $I_X$ parameters are determined under realistic experimental conditions emulating both hadron collider and $B$-factory scenarios; assessment of the $I_X$ sensitivity to NP is beyond the scope of the work.

For this demonstration, the angular coefficients are measured integrated over all values of $q^2$. This is not required and parallel angular fits in several $q^2$ bins could follow a similar procedure. NP may also be sought by measuring $I_X$ and $\bar{I}_X$ (the $\CP$ conjugate) values separately with the same template fit by tagging according to the $\tau$ lepton charge.  

It is also noted that the methods developed in this work are applicable to the light lepton $B^0 \to D^{*-}l^+\nu_l$ modes. These decays suffer from lower background levels, and superior angular resolution due to the stable charged lepton coming directly from a well-defined $B$ decay vertex. A measurement of the $I_X$ coefficients of these modes is well motivated to validate form factor schemes and provide a null test of the SM.

\end{sloppypar}

\subsection{Monte Carlo simulation}
\label{sec:samples}

Monte Carlo (MC) signal samples of $B^0 \to D^{*-}\tau^+\nu_{\tau}$ decays are generated using the \mbox{RapidSim} package~\cite{Cowan:2016tnm}. RapidSim is a fast MC generator for simulating heavy-quark hadron decays. It uses TGenPhaseSpace~\cite{Brun:1997pa} to generate $b$-quark hadron decays and FONLL~\cite{Cacciari:2001td} to give the $b$ quark the correct production kinematics for the Large Hadron Collider (LHC). Exclusive decays of the $D^{*-}$ meson decay to $\Dzb\pi^{-}$ are generated, with the $\Dzb$ meson decaying to $K^+\pi^-$. The three-prong $\tau^+ \to \pi^+ \pi^+ \pi^- \bar{\nu}_{\tau}$ decay, rather than the more abundant muonic $\tau$ decay, is the focus of this study. This is because the presence of a $\tau$ decay vertex results in lower backgrounds, and with only two neutrinos in the final state, this mode has the best decay 
angle resolution.

Signal $B^0$ mesons are decayed using EvtGen~\cite{Lange:2001uf}. The ISGW2~\cite{PhysRevD.39.799} model distributes $B^0$ decays according to Eq.~\eqref{eq:decay_rate}, with SM values for each $I_X$ coefficient. The VSS model is used to model the vector $D^{*-}$ decay, and TAUOLA~\cite{Chrzaszcz:2016fte} (model number 5) produces the correct kinematic and invariant mass structure for the three-prong $\tau$ decay. To emulate the effects of detector resolution, the track momenta and decay vertex coordinates are smeared according to RapidSim LHCb resolution presets~\cite{Cowan:2016tnm}. The missing momentum due to the presence of two final state neutrinos is modelled by ignoring both particles in any calculations of reconstructed quantities. Detector acceptance effects are modelled by restricting generated $B^0$ mesons to the momentum range $[0,100]$ GeV$/c$ and the pseudorapidity range $[1,6]$, which is similar to the geometrical acceptance of the LHCb detector.

\FloatBarrier

\section{Kinematic reconstruction}
\label{sec:reco}

Due to the lost neutrinos in the final state and the absence of a constraint from the initial state, neither the $\tau$ nor $B^0$ momentum can be fully reconstructed at a hadron collider. The best calculation of the decay angles uses estimates of the $\tau$ and $B^0$ momentum that are determined from the topology of the decay.  
As the $\tau$ lepton mass is well known~\cite{PhysRevD.98.030001}, its momentum can be estimated up to a two-fold ambiguity from its line of flight between the reconstructed $B^0$ and $\tau$ vertices. The $\tau$ momentum magnitude in the laboratory frame is
\begin{equation}
|\vec{p}_{\tau}| = \frac{(m_{3\pi}^2 + m_{\tau}^2)|\vec{p}_{3\pi}|\cos\theta_{\tau,3\pi} \pm E_{3\pi}\sqrt{(m_{\tau}^2-m_{3\pi}^2)^2 - 4m_{\tau}^2|\vec{p}_{3\pi}|^2\sin^2\theta_{\tau,3\pi}}}{2(E_{3\pi}^2 - |\vec{p}_{3\pi}|^2\cos^2\theta_{\tau,3\pi})}\,,
\label{eq:tau_mom}
\end{equation}
where $m_{3\pi}$, $|\vec{p}_{3\pi}|$, and $E_{3\pi}$ are the reconstructed invariant mass, momentum, and energy of the three-prong system, $m_{\tau}$ is the known $\tau$ mass, and $\theta_{\tau,3\pi}$ is the angle between the three-prong momentum vector and the flight vector of the $\tau$. Eq.~\eqref{eq:tau_mom} has a single solution when $\theta_{\tau,3\pi}$ takes the maximum allowed value such that the square-root term is zero, i.e.
\begin{equation}
\theta_{\tau,3\pi}^{\text{max}} = \arcsin \bigg(\frac{m_{\tau}^2-m_{3\pi}^2}{2m_{\tau}|\vec{p}_{3\pi}|}\bigg)\,.
\end{equation}
Combined with the $\tau$ line of flight, this provides an estimate of the $\tau$ momentum components with minimal bias. In a similar fashion, the $B^0$ momentum is estimated using
\begin{equation}
|\vec{p}_{B^0}| = \frac{(m_{Y}^2 + m_{B^0}^2)|\vec{p}_{Y}|\cos\theta_{B^0,Y} \pm E_{Y}\sqrt{(m_{B^0}^2-m_{Y}^2)^2 - 4m_{B^0}^2|\vec{p}_{Y}|^2\sin^2\theta_{B^0,Y}}}{2(E_{Y}^2 - |\vec{p}_{Y}|^2\cos^2\theta_{B^0,Y})}\,,
\label{eq:b_mom}
\end{equation}
where $Y$ represents the $D^{*-}\tau^+$ system as reconstructed using Eq.~\eqref{eq:tau_mom}. The maximum opening angle between the $B^0$ flight vector and the momentum vector of $Y$ is
\begin{equation}
\theta_{B^0,Y}^{\text{max}} = \arcsin \bigg(\frac{m_{B^0}^2-m_{Y}^2}{2m_{B^0}|\vec{p}_{Y}|}\bigg)\,,
\label{eq:b_max}
\end{equation}
which is used in Eq.~\eqref{eq:b_mom} to provide a single estimate for the $B^0$ momentum magnitude. The $\tau$ and $B^0$ momentum bias and resolution are tabulated in Tab.~\ref{tab:tau_b_mom_res_vals}.

\begin{table}
\centering
\begin{tabular}{c | c | c | c}
Particle & Mean $p_{\text{True}}$ [GeV/$c$] & Res. $\mu$ & Res. $\sigma$ \\ \hline
$B^0$ & 150.6 & 0.09 & 0.19 \\
$\tau$ & 61.9 & 0.01 & 0.25 \\
\end{tabular}
\caption{Summary of $\tau$ and $B^0$ momentum bias and resolution, defined as the mean and width of $(p_{\text{Reco}} - p_{\text{True}})/p_{\text{True}}$, respectively.}
\label{tab:tau_b_mom_res_vals}
\end{table}

\begin{figure}
\centering
\includegraphics[width = 0.32\textwidth]{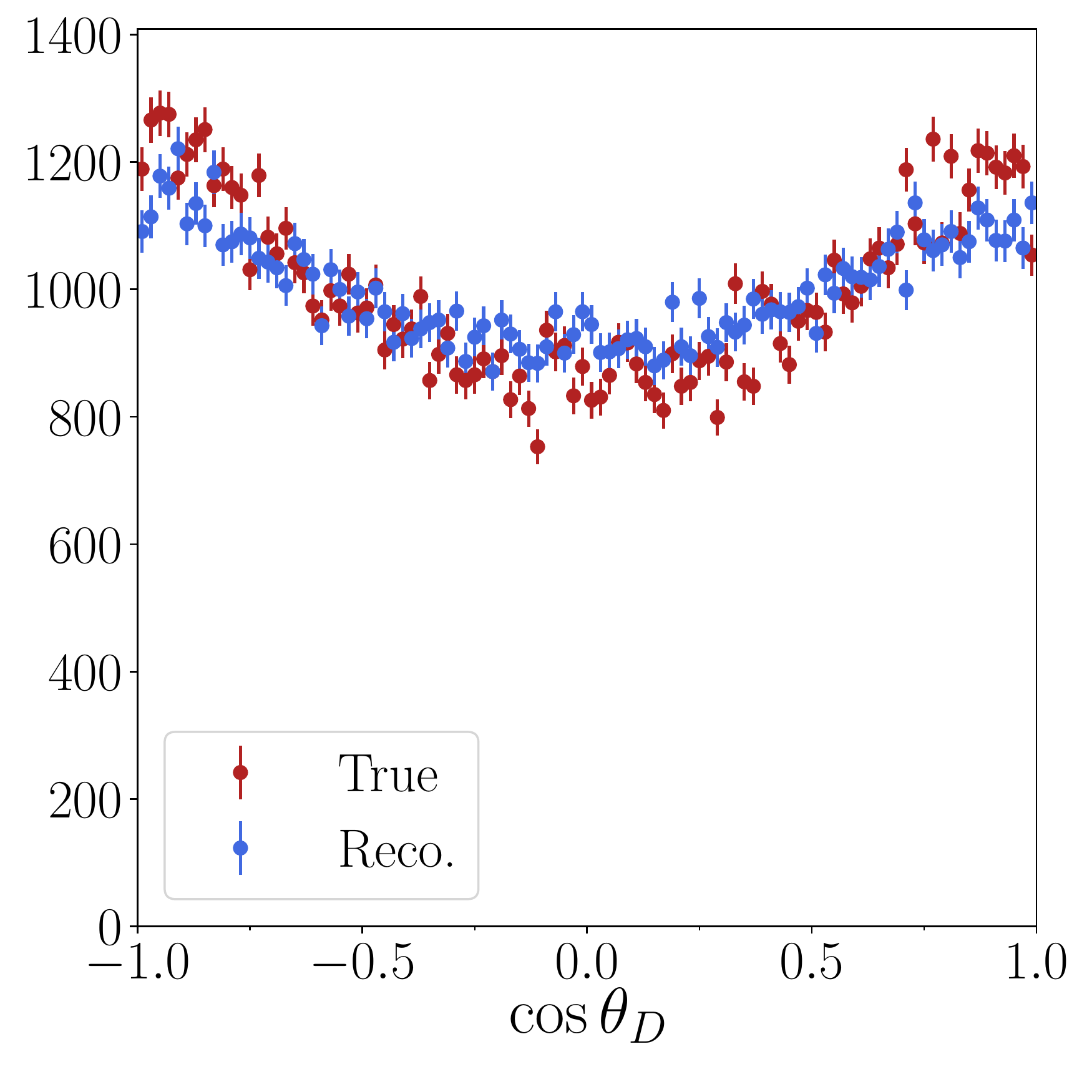}
\includegraphics[width = 0.32\textwidth]{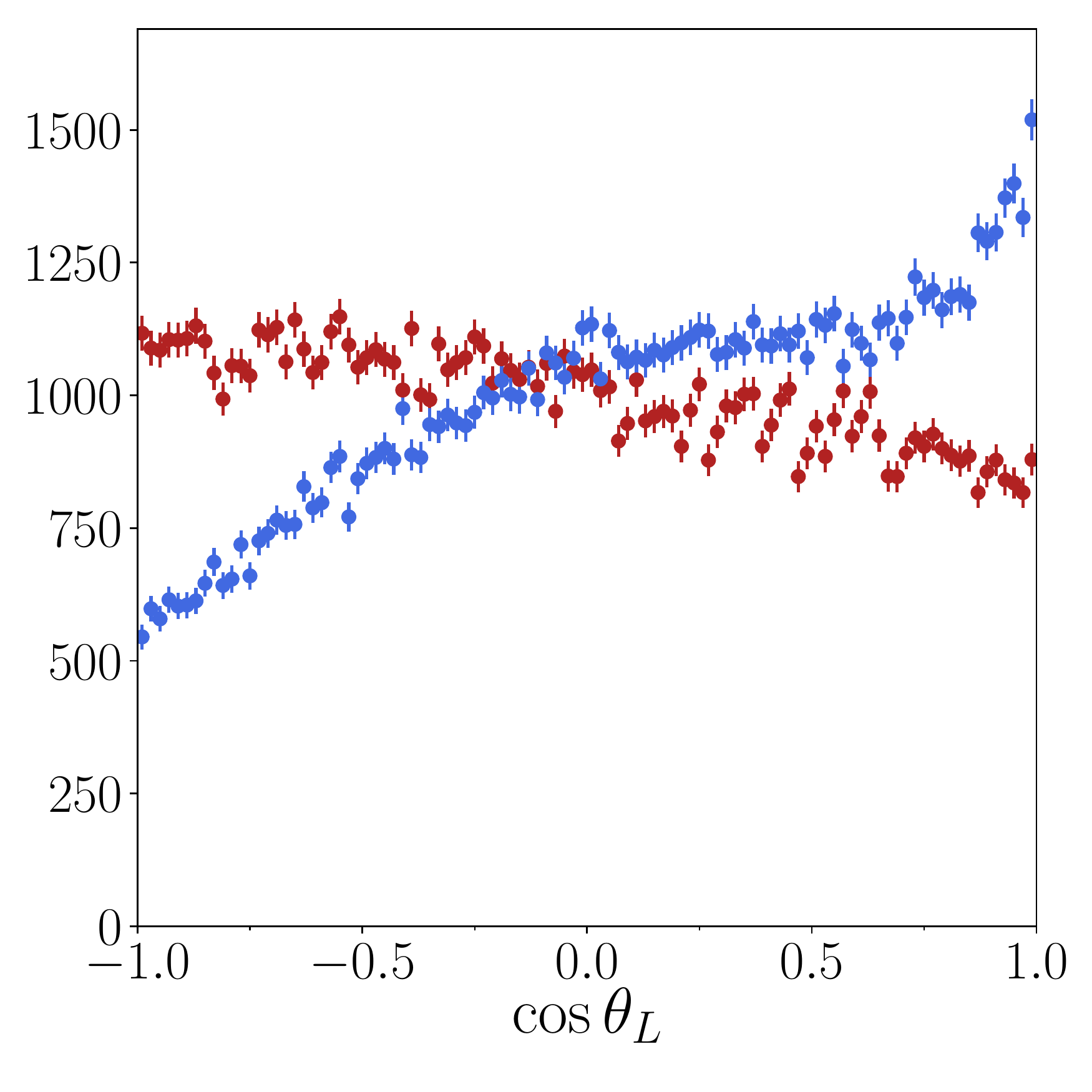}
\includegraphics[width = 0.32\textwidth]{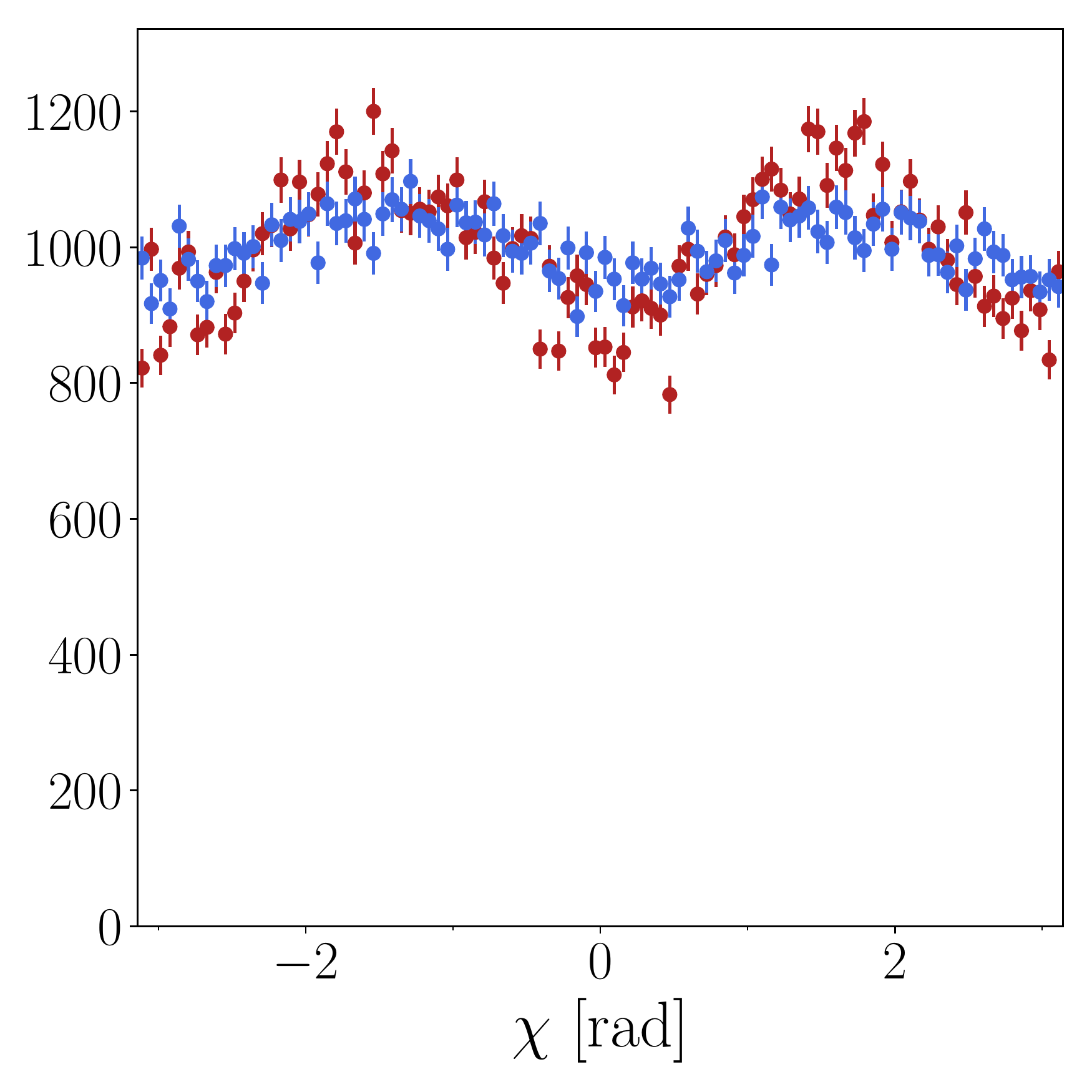} \\
\includegraphics[width = 0.32\textwidth]{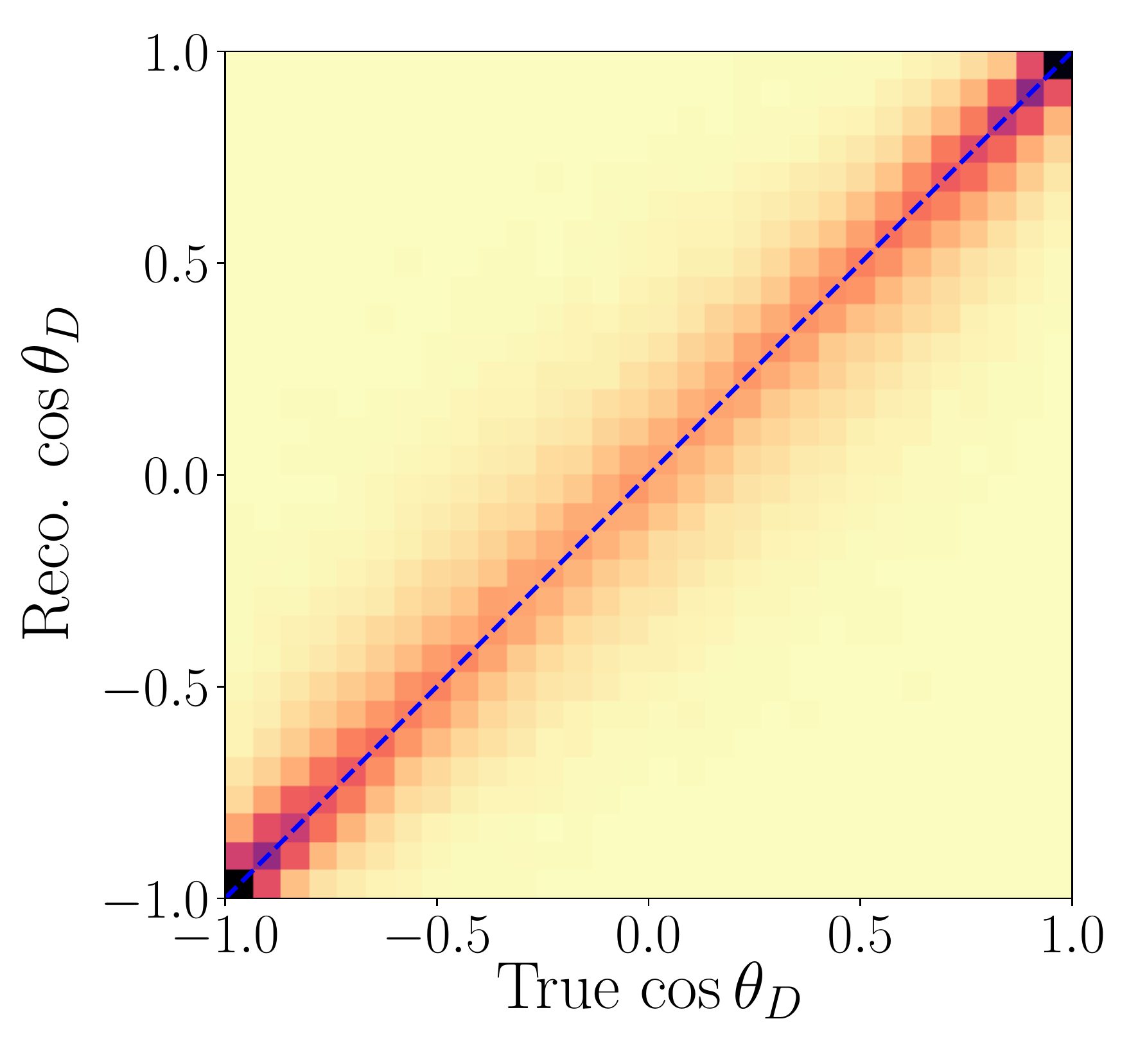}
\includegraphics[width = 0.32\textwidth]{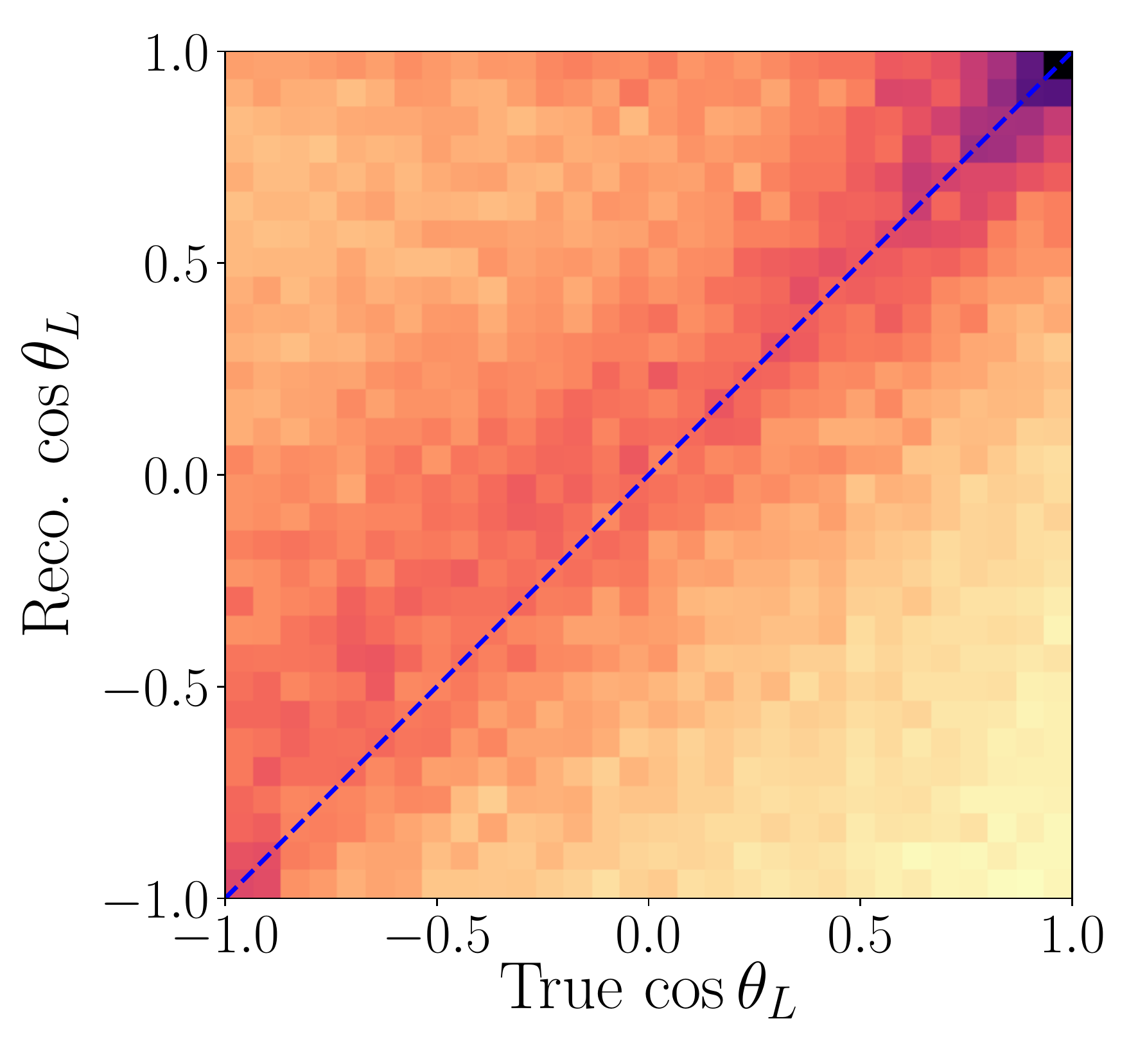}
\includegraphics[width = 0.3\textwidth]{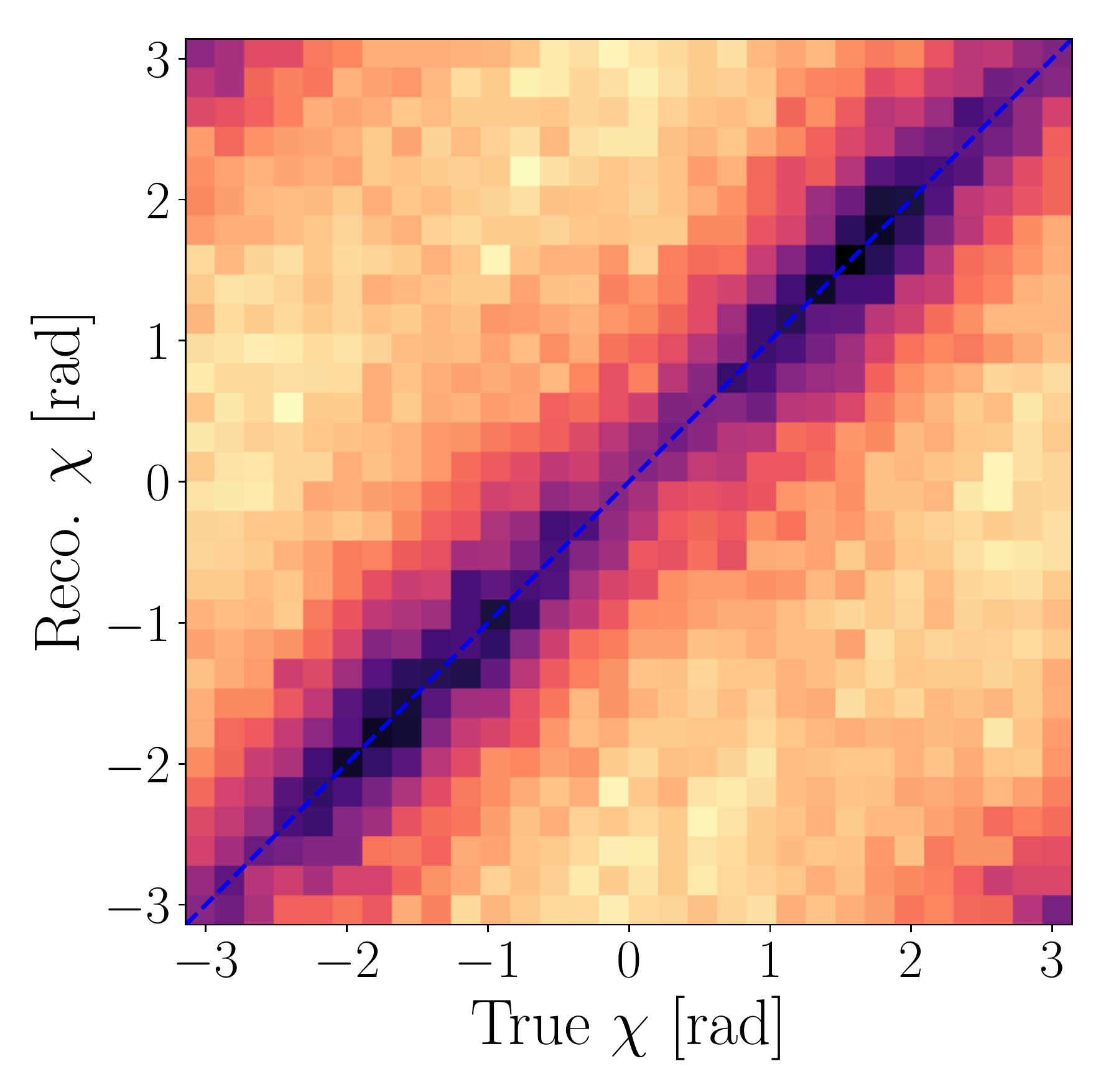}

\caption{(Top) True (red) and reconstructed (blue) angular distributions from 100,000 generated \mbox{$B^0 \to D^{*-}\tau^+\nu_\tau$} events. (Bottom) Distributions of reconstructed angular variables versus true, where darker colours indicate a higher density of events.}
\label{fig:angle_res}
\end{figure}

\begin{figure}
\centering
\includegraphics[width = 0.32\textwidth]{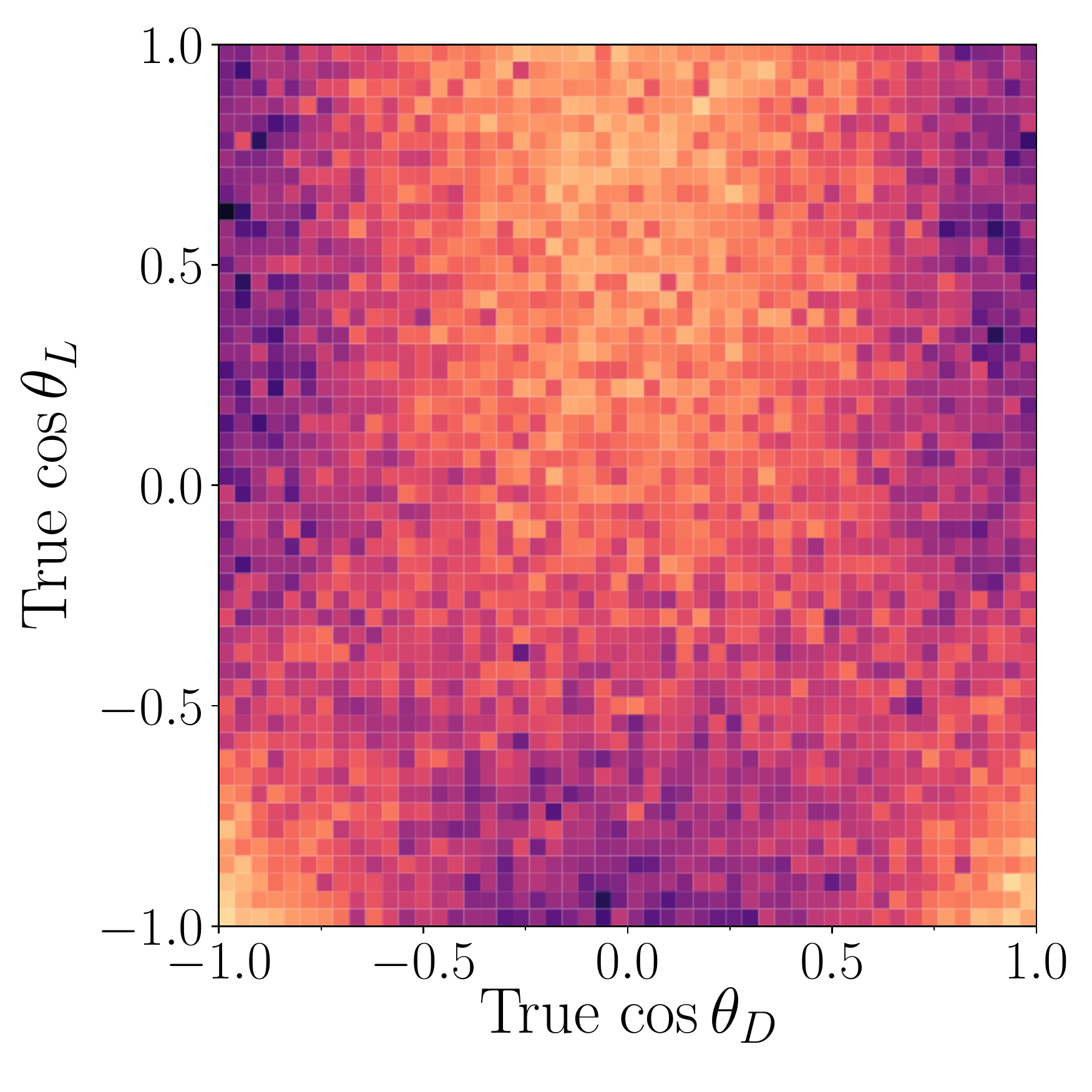}
\includegraphics[width = 0.32\textwidth]{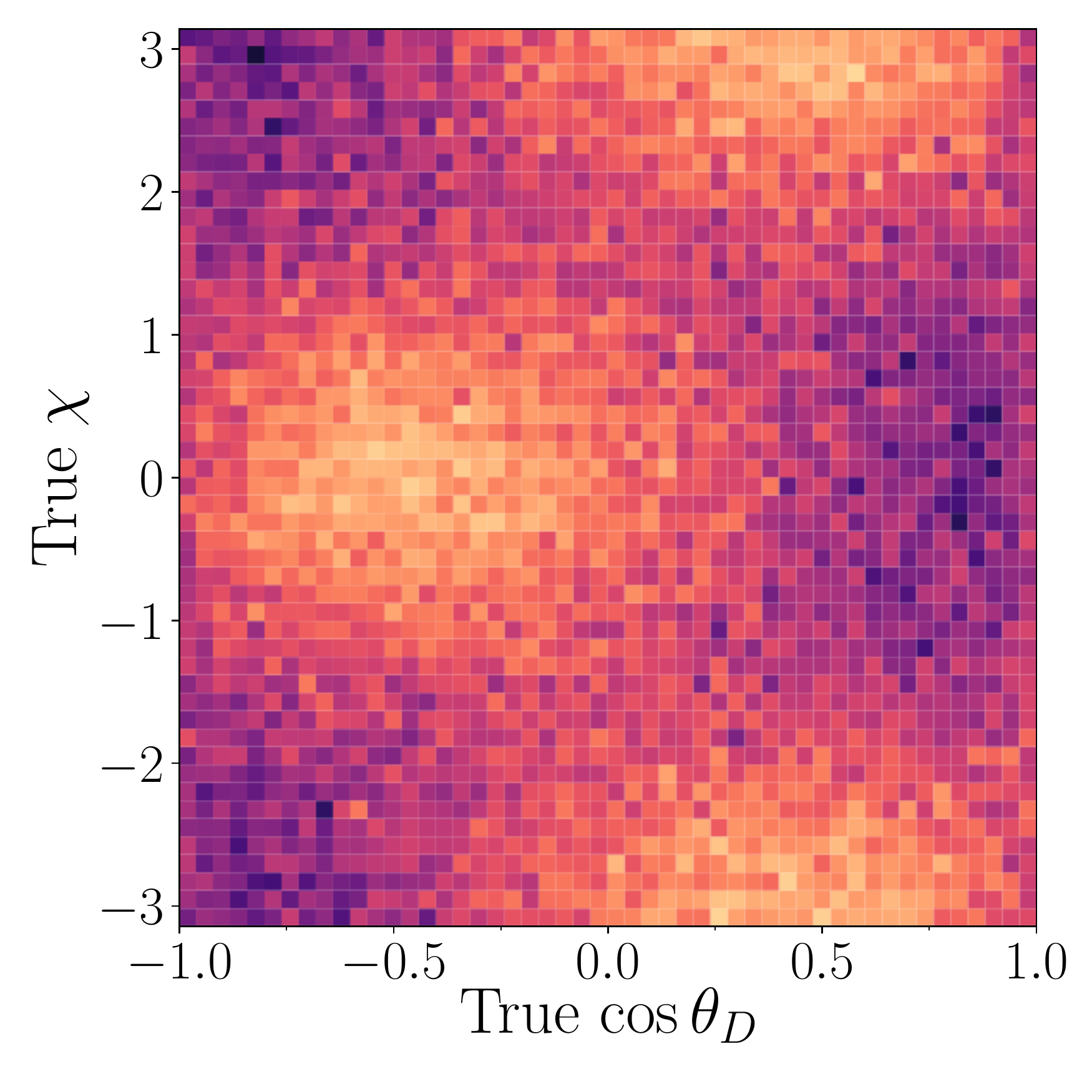}
\includegraphics[width = 0.32\textwidth]{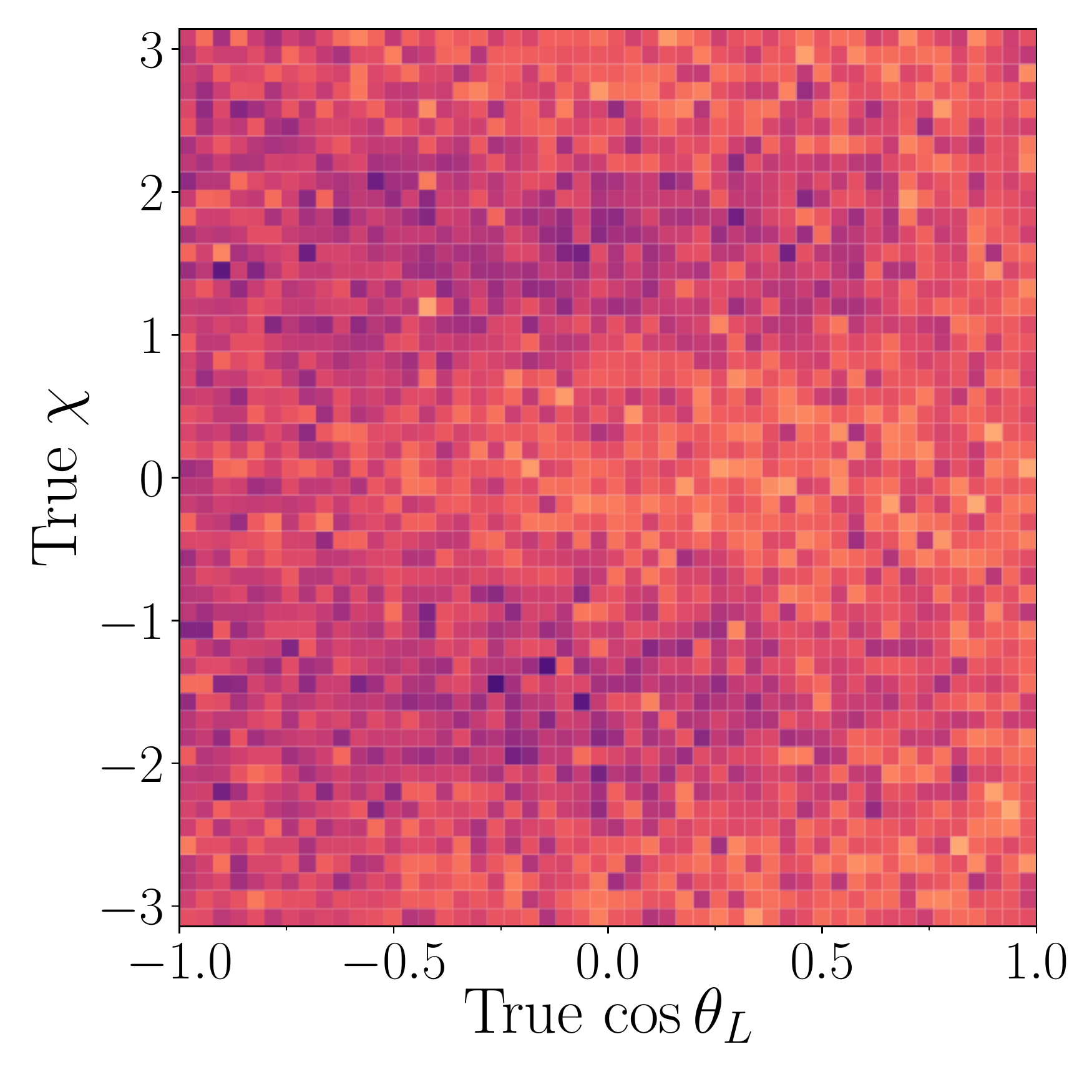} \\
\includegraphics[width = 0.32\textwidth]{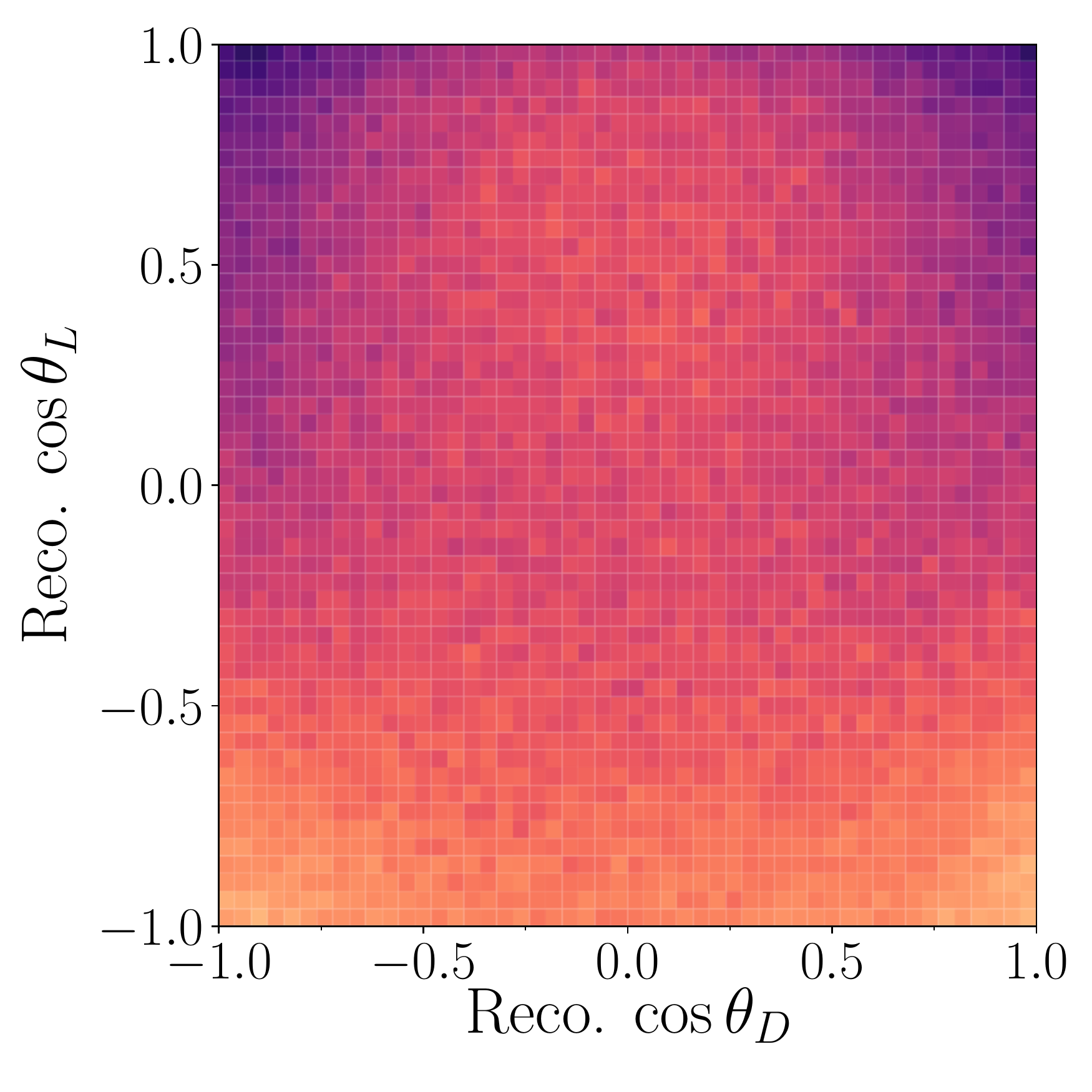}
\includegraphics[width = 0.32\textwidth]{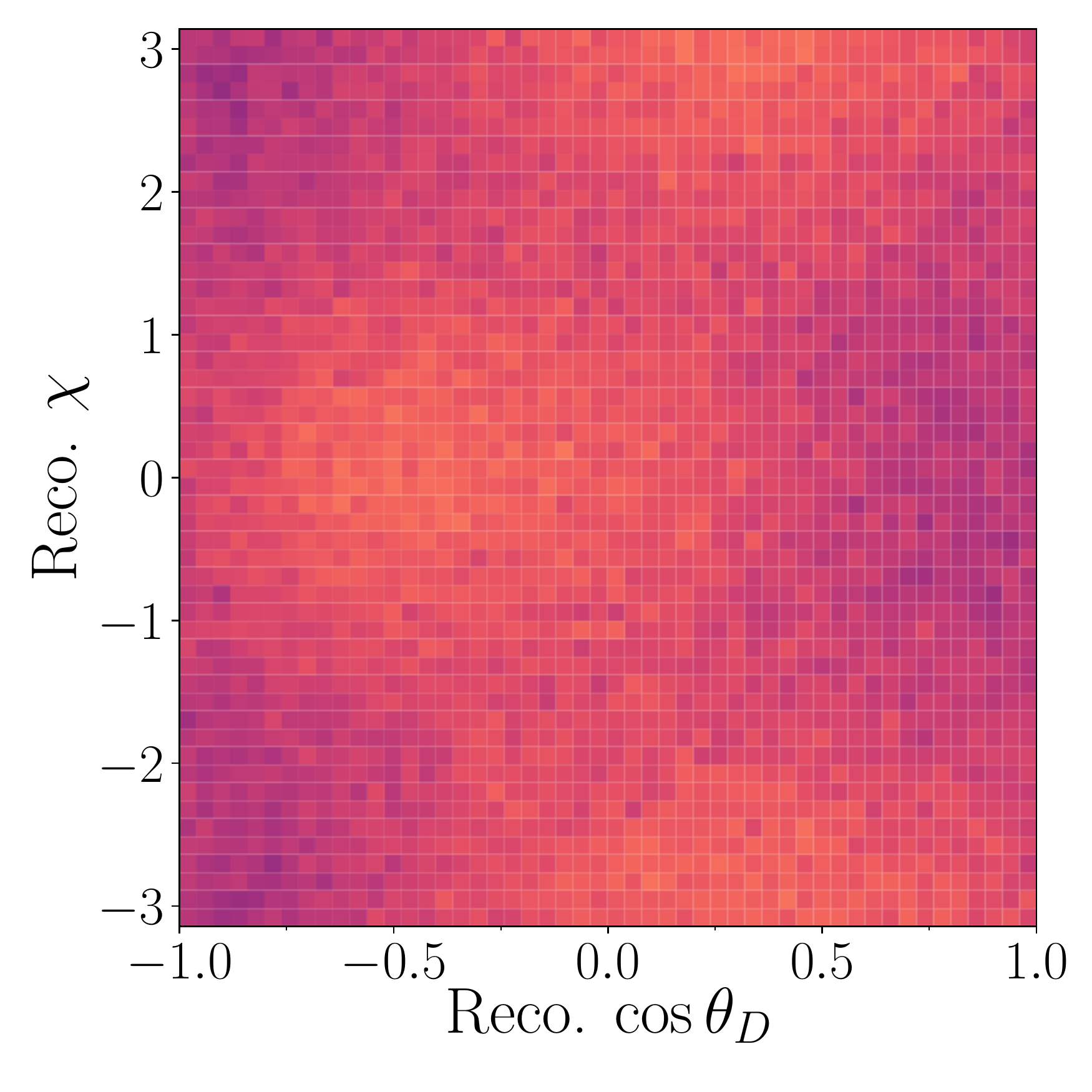}
\includegraphics[width = 0.32\textwidth]{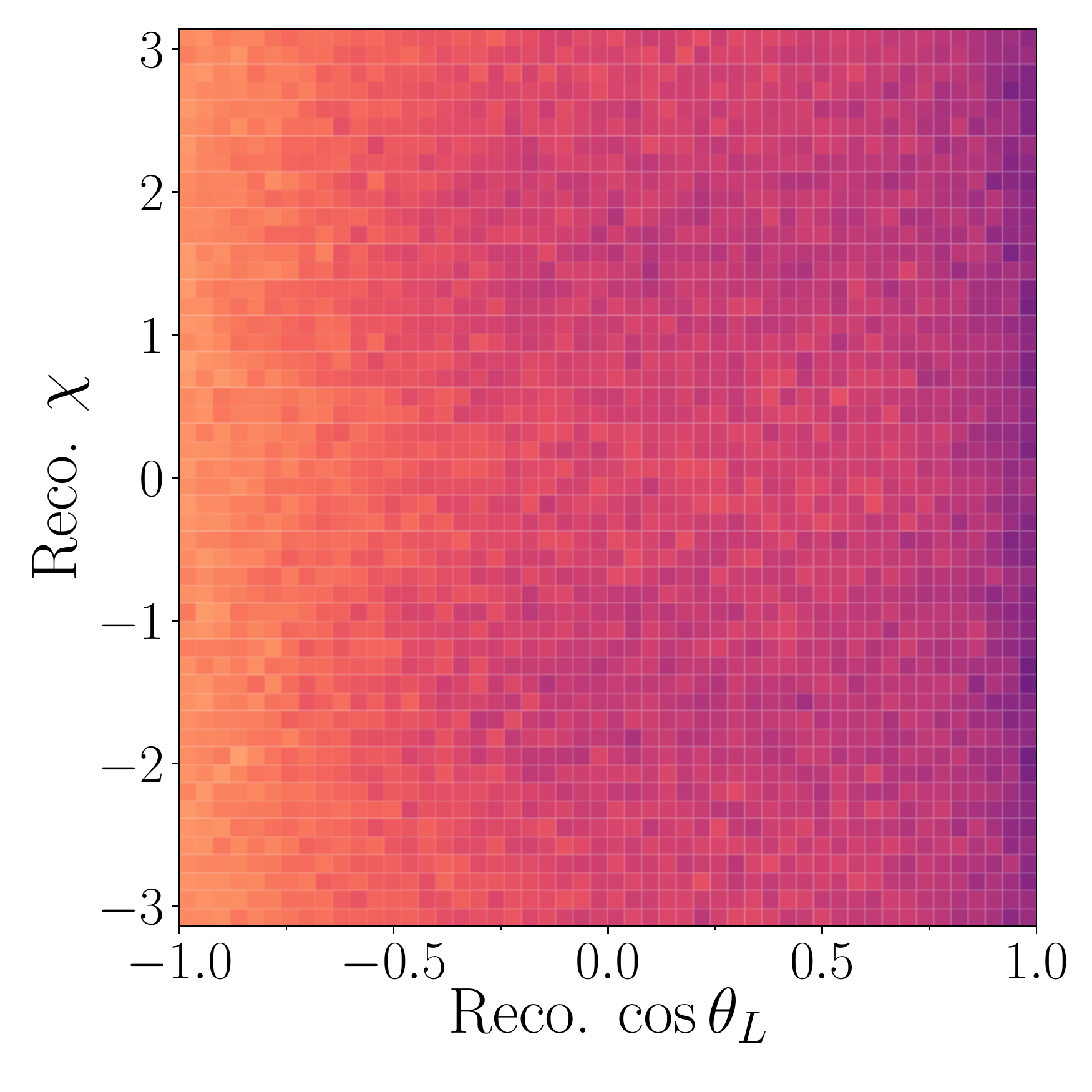} \\
\includegraphics[width = 0.32\textwidth]{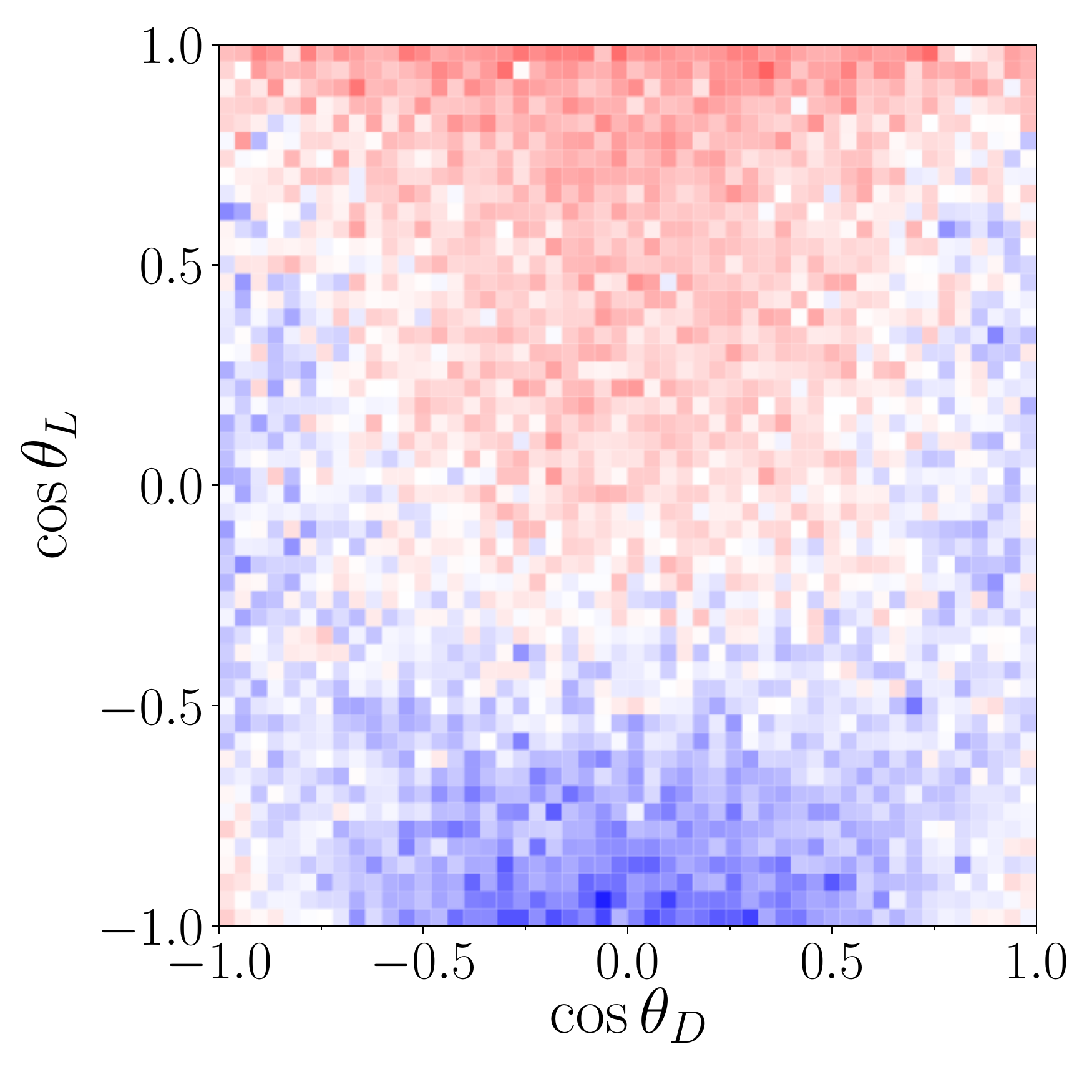}
\includegraphics[width = 0.32\textwidth]{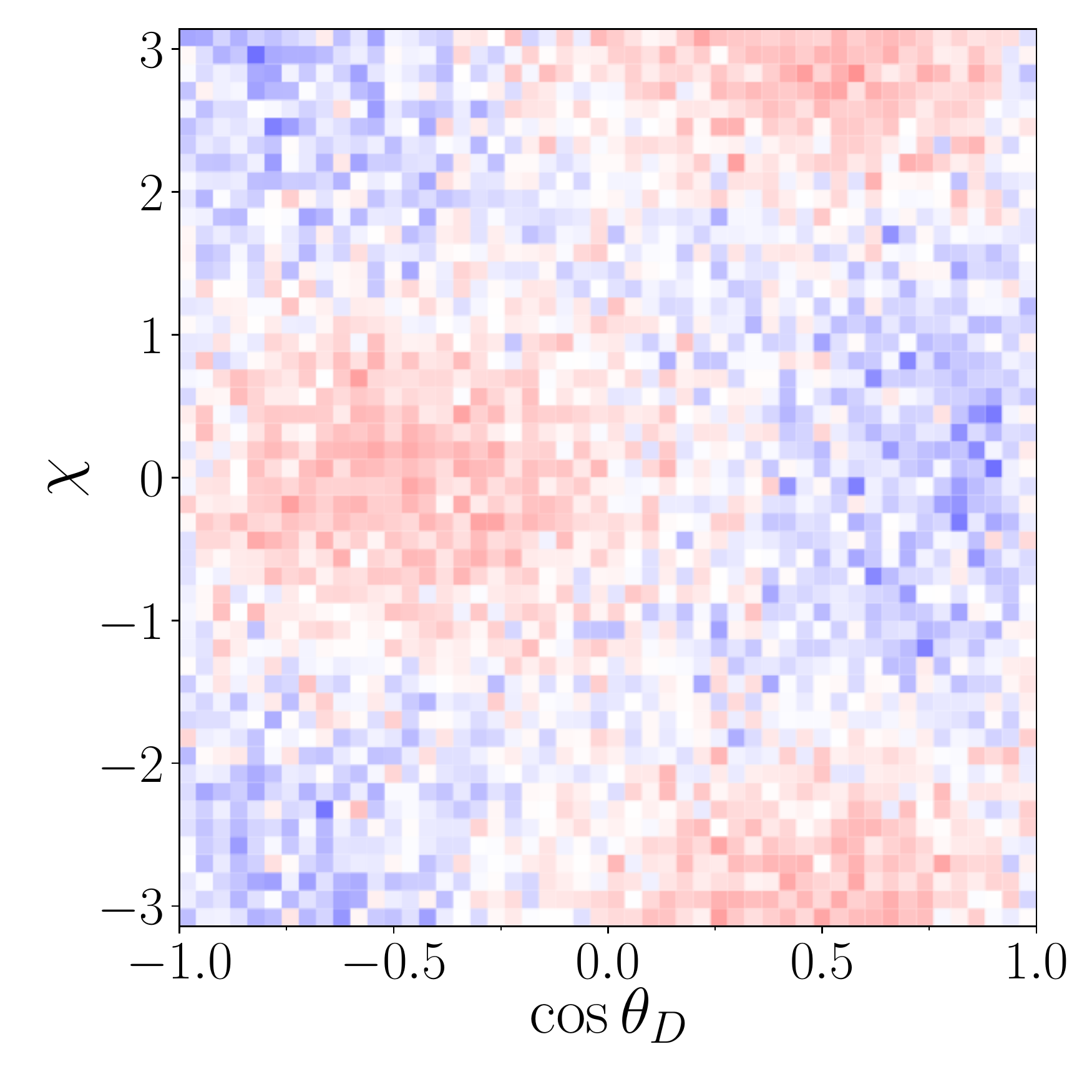}
\includegraphics[width = 0.32\textwidth]{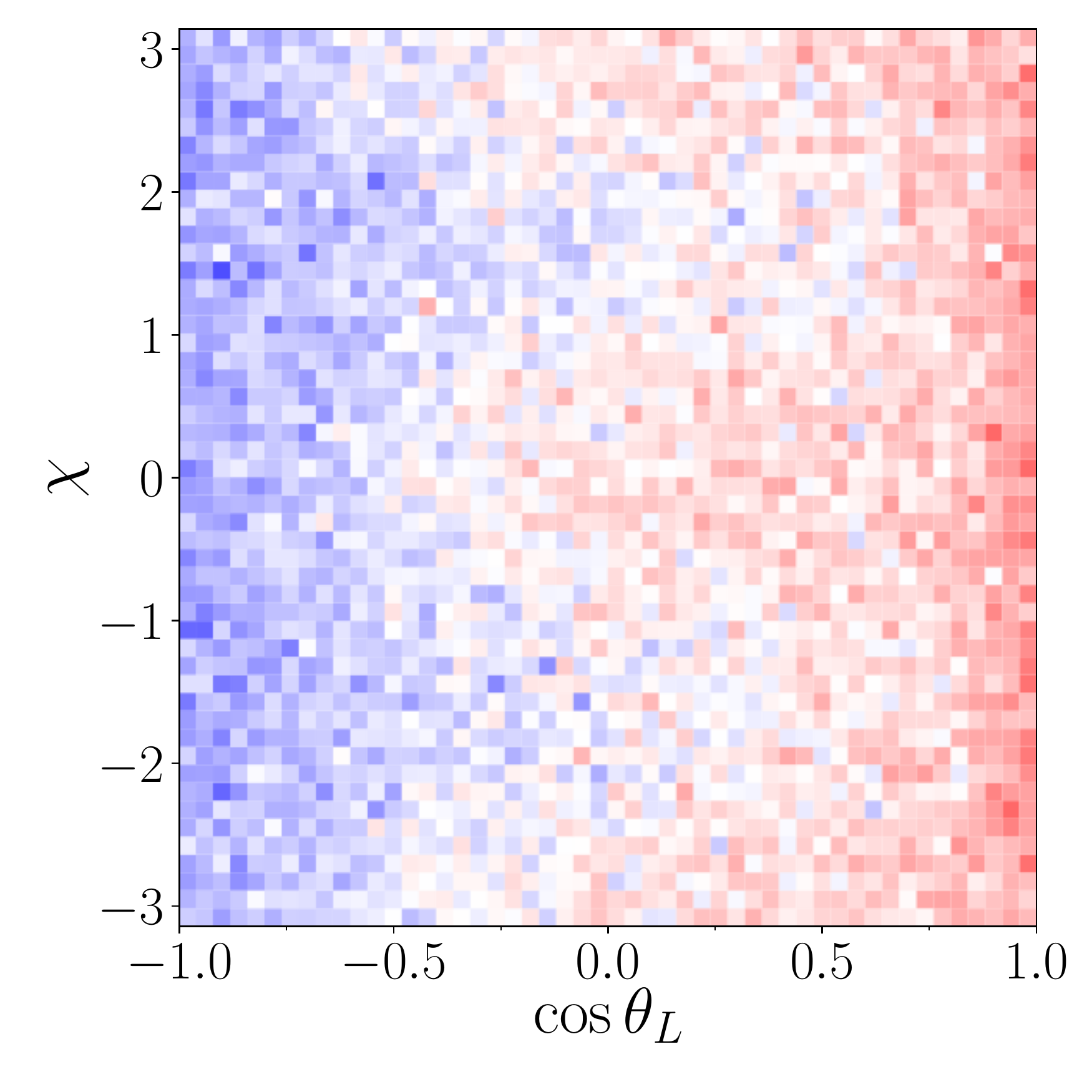} \\
\caption{Two-dimensional projections of the truth-level (top) and reconstructed (middle) angular distributions; darker colours indicate regions of higher density. The density difference (bottom) in each bin indicates where the density increases (red) and decreases (blue) as a result of the reconstruction.}
\label{fig:event_migration}
\end{figure}

Since the four-vectors of the $\tau$ and $B^0$ are necessary inputs for the calculation of the decay angles, they too suffer substantial resolution effects. This is shown in Fig.~\ref{fig:angle_res}, where the true and reconstructed angular distributions of $B^0 \to D^{*-}\tau^+\nu_\tau$ are compared. The angular resolutions are quantified in Tab.~\ref{tab:angle_res_vals}, which shows that $\cos\theta_D$ is the most well-reconstructed quantity. As a result of the large resolution effects on $\cos\theta_L$ and $\chi$, considerable migration of events occurs within the angular phase space. This is illustrated in Fig.~\ref{fig:event_migration}, where the two-dimensional projections of the truth-level and reconstructed angular distributions are shown. The density difference within each bin is also shown, where red (blue) regions indicate increases (decreases) in density caused by the reconstruction. The overall effect of the event migration is to reduce the density variation across the phase space, but a bias in $\cos\theta_L$ towards more positive values is also evident.

Due to the reconstruction-induced event migration, a parametric fit to the reconstructed decay angles using Eq.~\eqref{eq:decay_rate} cannot be used to measure the $I_X$ coefficients. Any attempt to correct the reconstruction biases leads to a dependence on the model used in the Monte Carlo from which the correction is derived. Instead, it is demonstrated that the $I_X$ coefficients can be measured with a binned fit using multidimensional histogram templates, where the angular degradation and other detector effects are included directly in each of the twelve templates that describe the signal probability density function (PDF). 

\begin{table}[h]
\centering
\begin{tabular}{c | c | c}
Angle & Res. $\mu$ & Res. $\sigma$ \\ \hline
$\cos\theta_D$ & 0.00 & 0.23 \\
$\cos\theta_L$ & 0.15 & 0.65 \\
$\chi$ & -0.01 rad & 2.24 rad \\
\end{tabular}
\caption{Angular variable resolution mean ($\mu$) and width ($\sigma$) determined using generated \mbox{$B^0 \to D^{*-}\tau^+\nu_\tau$} events. The resolution is defined as $a_{Reco} - a_{True}$, where \mbox{$a \in \{\cos\theta_D,\cos\theta_L,\chi\}$}.}
\label{tab:angle_res_vals}
\end{table}

\section{Building and using templates}
\label{sec:templates}

The decay rate defined in Eq.~\eqref{eq:decay_rate} involves a sum of twelve independent angular functions in the $(\cos\theta_D,\cos\theta_L,\chi)_{\text{True}}$ variable space. 
Because of the bias and resolution effects that arise in the reconstruction of semitauonic decays, a template fit in the $(\cos\theta_D,\cos\theta_L,\chi)_{\text{Reco}}$ variable space must be used. The twelve angular functions become twelve multidimensional template histograms each scaled by an $I_X$ coefficient. In this way, the fit is performed with the reconstruction effects included directly within the PDF.

The template histograms are created by first filling twelve density histograms $D_{I_{X}}$. Each density histogram contains a large number of bins across $(\cos\theta_D,\cos\theta_L,\chi)_{\text{True}}$ space; $30 \times 30 \times 30$ uniform bins are used here. The $D_{I_{X}}$ histograms are filled according to the angular function associated with each $I_X$ in Eq.~\eqref{eq:decay_rate}. 
The twelve $D_{I_{X}}$ histograms are divided by a density histogram, $M$, of the total signal model given by Eq.~\eqref{eq:decay_rate}. By taking the ratio $R_{I_{X}} = {D_{I_{X}}}/{M}$, the model used in the simulation cancels and ensures that the $R_{I_{X}}$ histograms are model independent. 
The $R_{I_{X}}$ histograms are then used to assign weights to simulated signals events based on their $(\cos\theta_D,\cos\theta_L,\chi)_{\text{True}}$ value. A simulated event falling within the true angular bin $i$ will be assigned twelve weights, $w_{I_{X}} = R_{I_{X}}(i)$.

\begin{figure}[t]
\centering
\includegraphics[width = 0.32\textwidth]{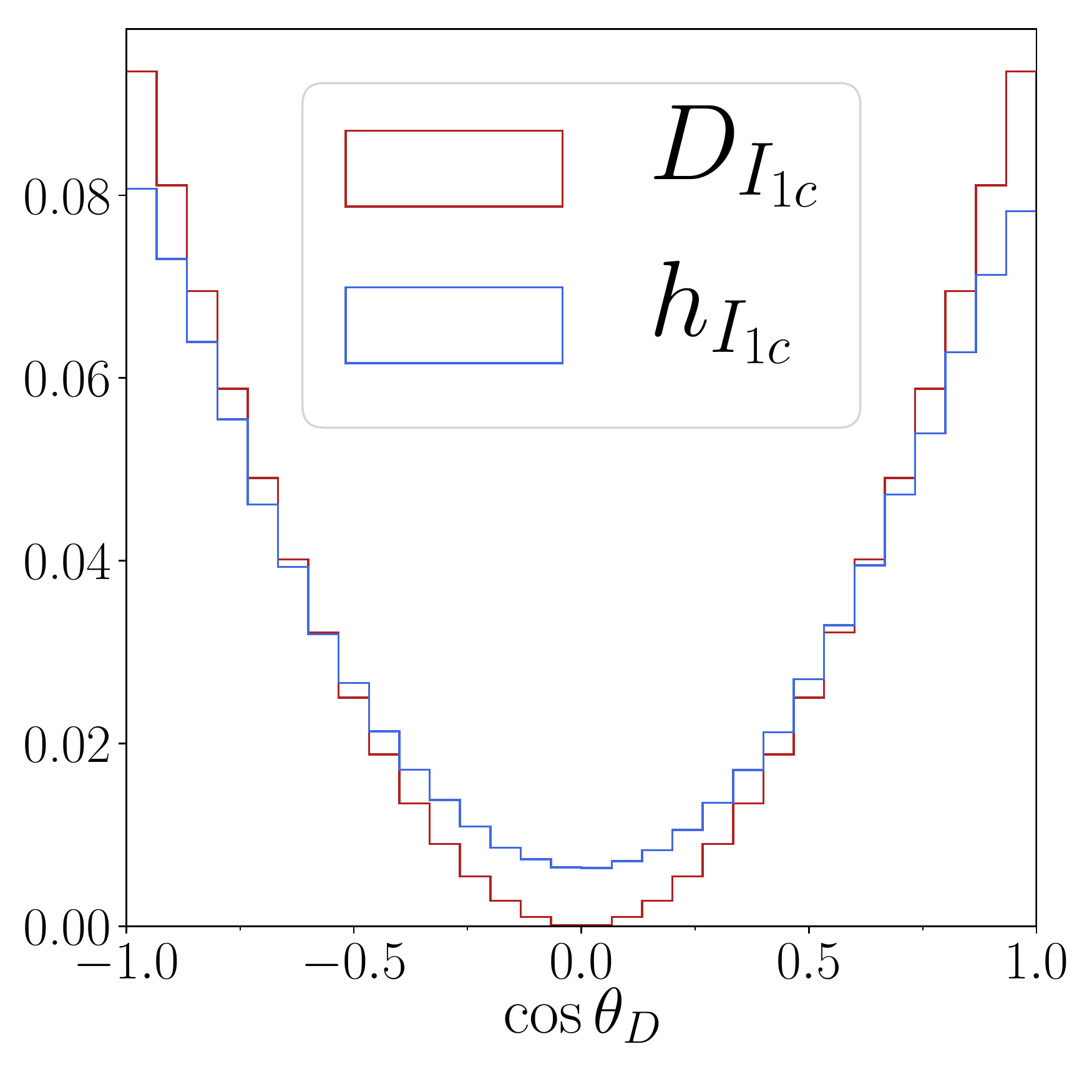}
\includegraphics[width = 0.32\textwidth]{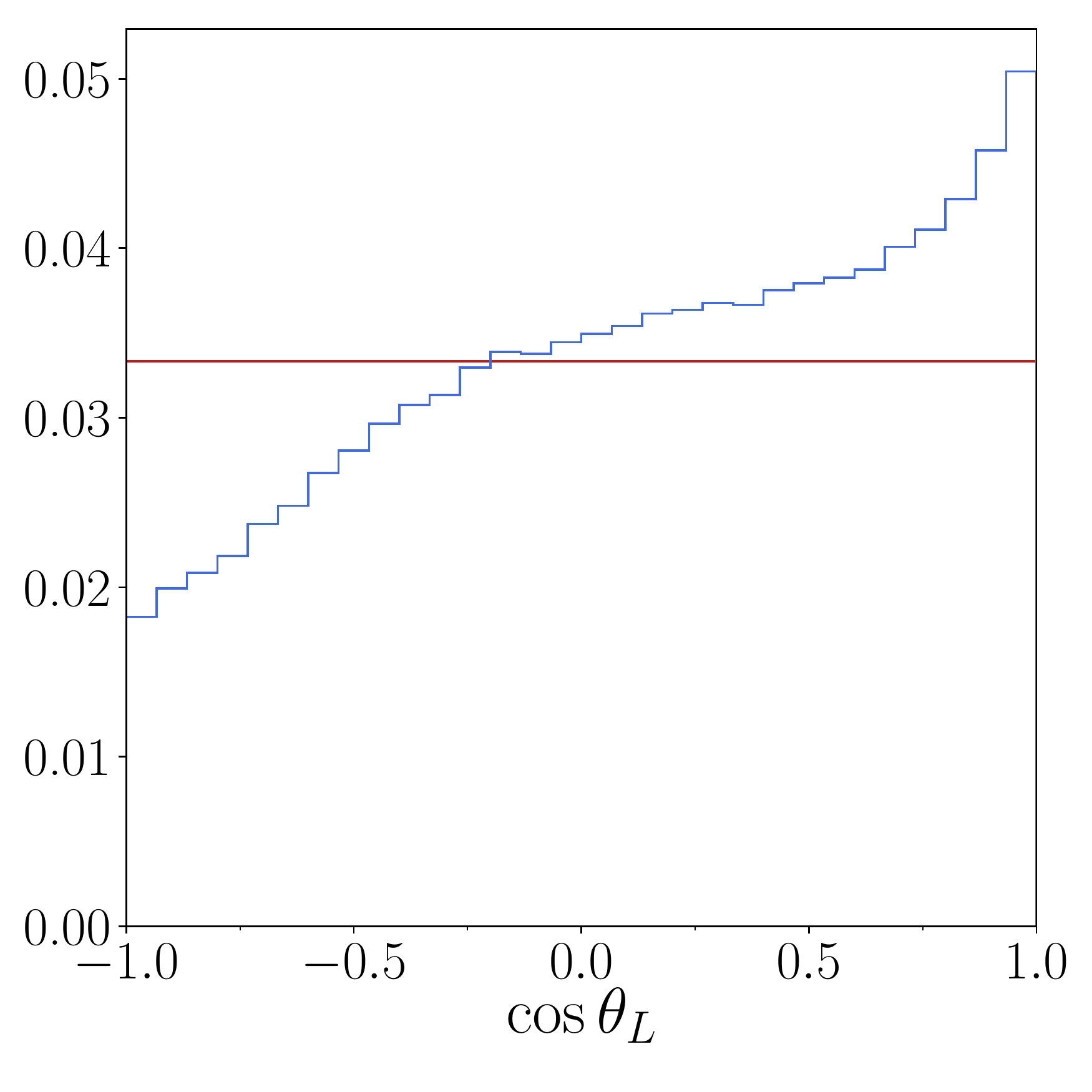}
\includegraphics[width = 0.32\textwidth]{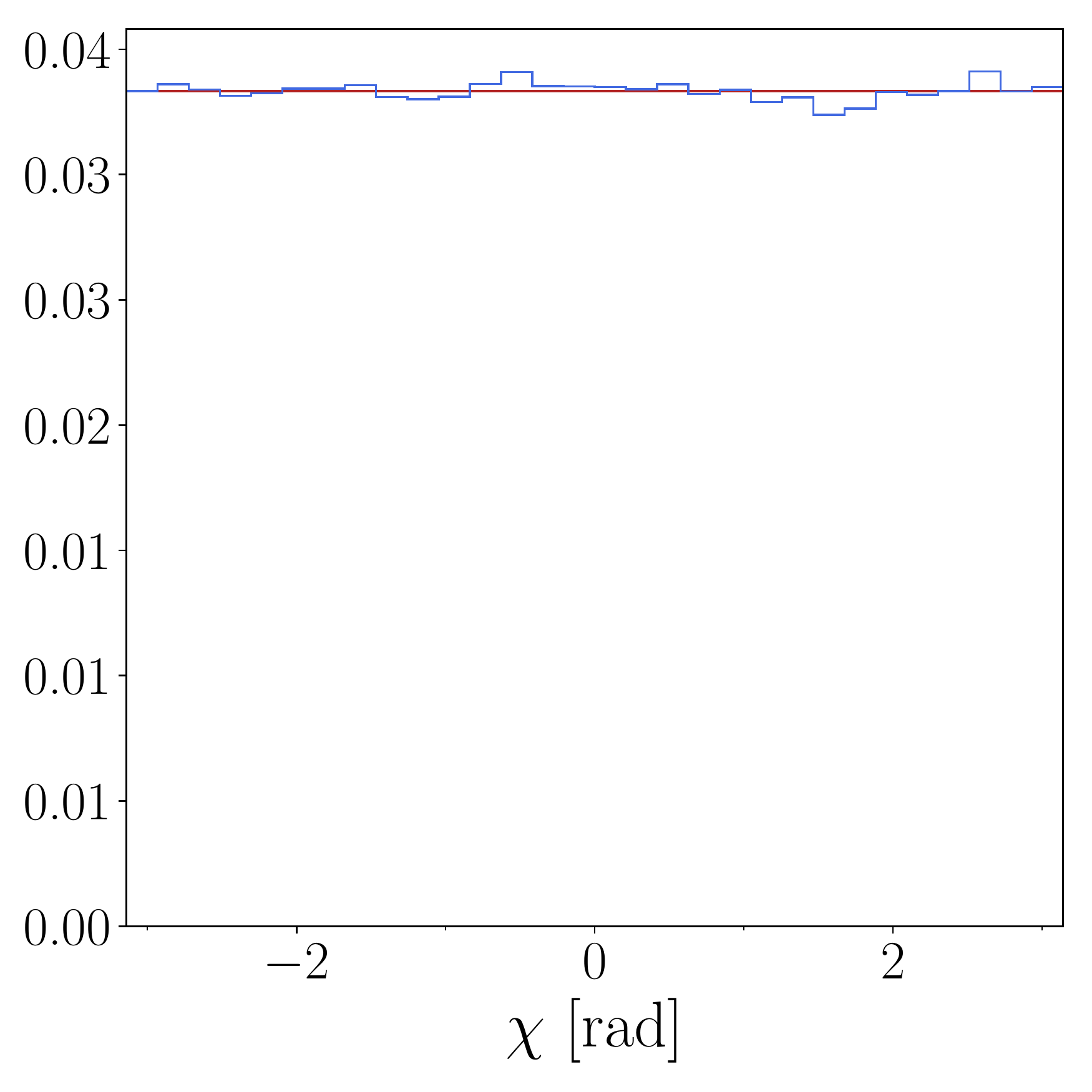} 
\includegraphics[width = 0.32\textwidth]{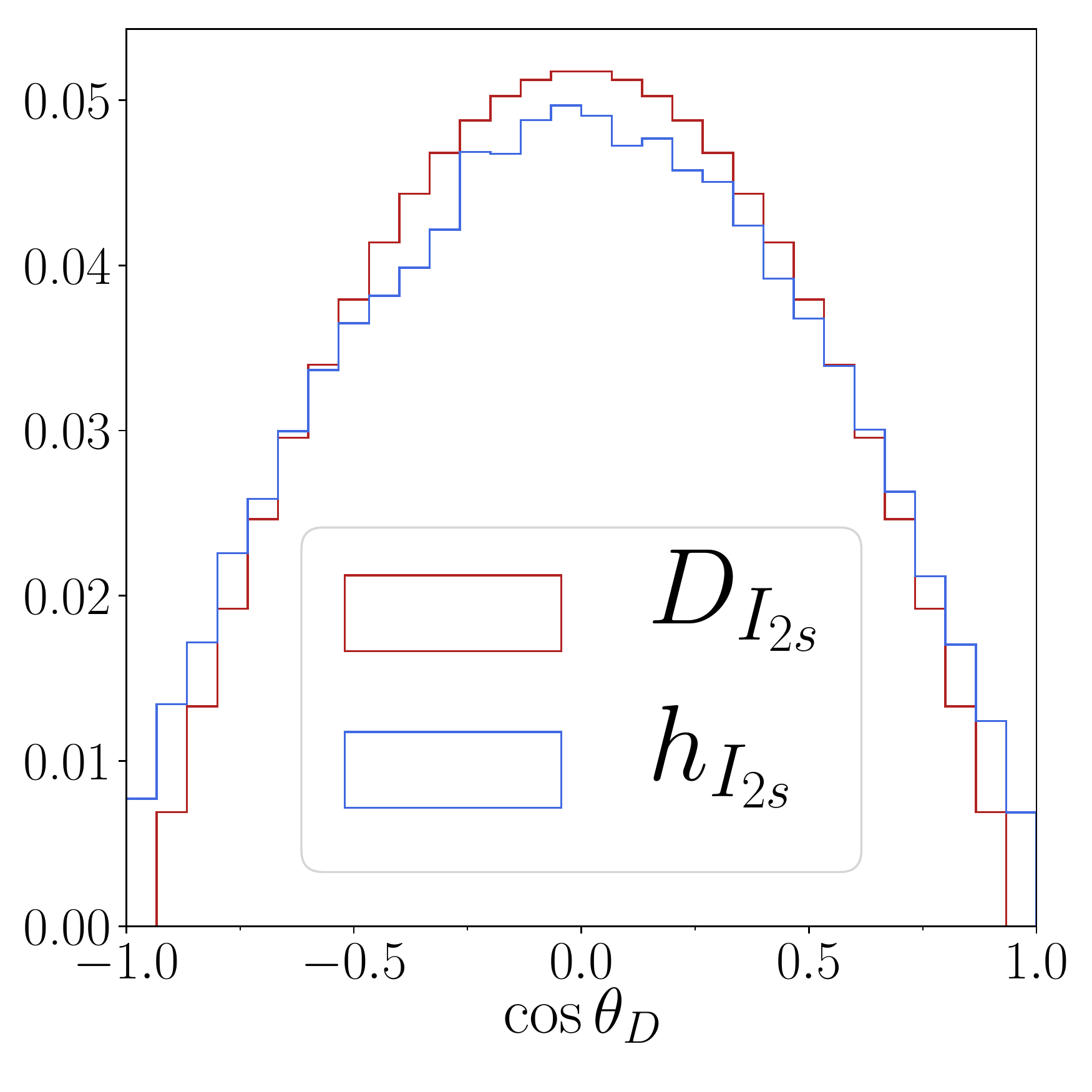}
\includegraphics[width = 0.32\textwidth]{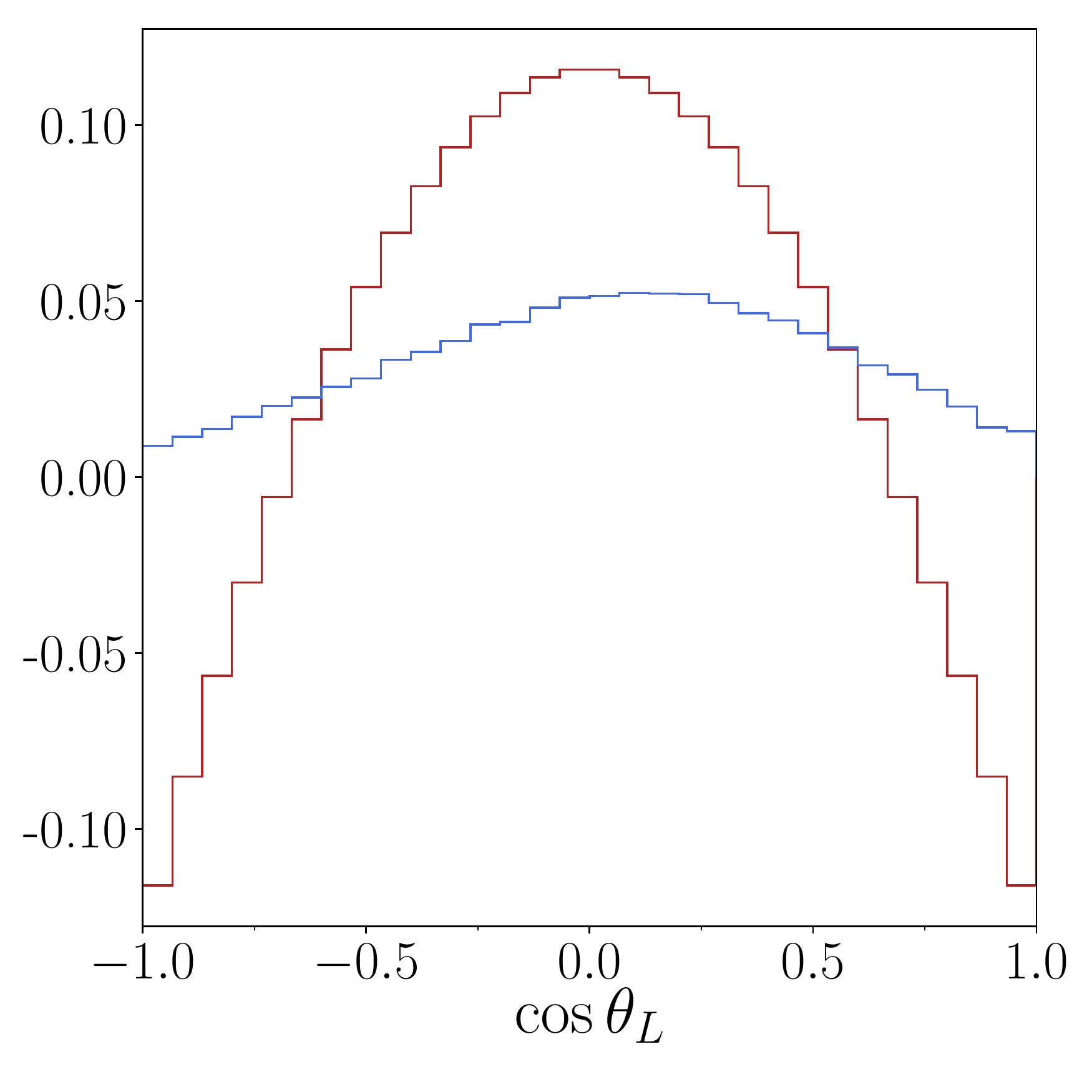}
\includegraphics[width = 0.32\textwidth]{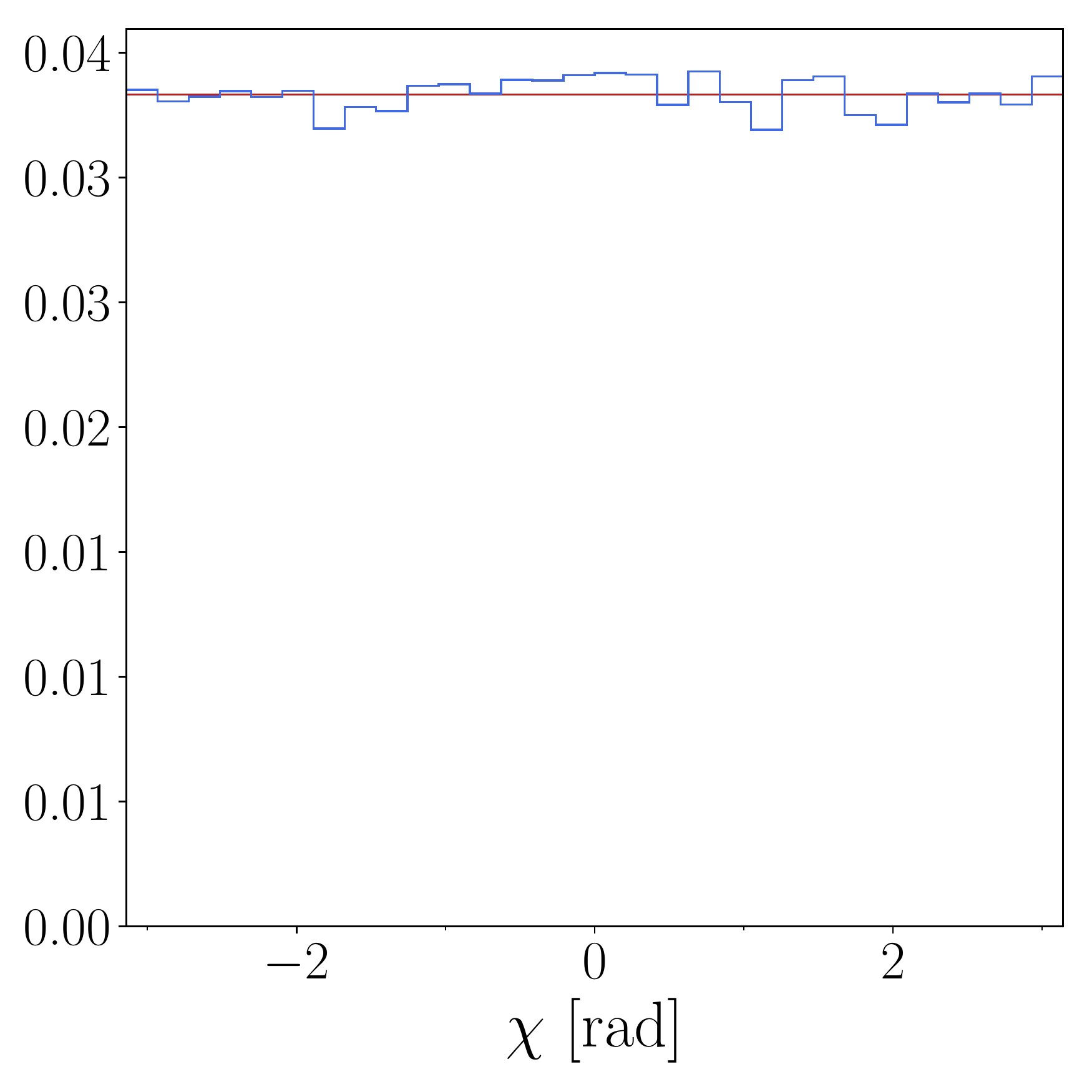}
\caption{Projections of the $D_{I_X}$ density histograms (red) and the $h_{I_X}$ templates (blue) for $X = 1c$ (top) and $X = 2s$ (bottom). The differences between $D_{I_X}$ and $h_{I_X}$ are caused by the bias and resolution effects introduced by the reconstruction.}
\label{fig:template_example}
\end{figure}

Subsequently, the per-event weights $w_{I_{X}}$ are applied when constructing histogram templates, $h_{I_{X}}$, in the reconstructed angular variables $(\cos\theta_D,\cos\theta_L,\chi)_{\text{Reco}}$. The result of this procedure is illustrated in Fig.~\ref{fig:template_example}, where the $h_{I_{1c}}$ and $h_{I_{2s}}$ templates created using a sample of one million generated events are shown. The large size of the sample ensures that the template statistical uncertainty in each bin is negligible. 
The corresponding density histograms $D_{I_{1c}}$ and $D_{I_{2s}}$ are also shown, to illustrate the sculpting effect of the reconstruction. Note that the $h_{I_{X}}$ templates may not be positive in all bins, but the sum of all twelve templates in any given bin is always positive, and is proportional to the total decay rate in that bin. 

\subsection{Signal-only template fit}
\label{sec:binned_fit}

The twelve $h_{I_{X}}$ templates are normalised to have unit integral, then multiplied by their corresponding $I_X$ coefficients to build the total signal PDF
\begin{equation}
\begin{split}
P((\cos\theta_D,\cos\theta_L,\chi)_{\text{Reco}}) &= \frac{1}{3}(4 - 6I_{1s} + I_{2c} + 2I_{2s})h_{1_{1c}} \\ 
& + I_{1s}h_{I_{1s}} + I_{2c}h_{I_{2c}} + I_{2s}h_{I_{2s}} + I_{6c}h_{I_{6c}} + I_{6s}h_{I_{6s}} \\
&+ I_{3}h_{I_{3}} + I_{9}h_{I_{9}} + I_{4}h_{I_{4}}  + I_{8}h_{I_{8}} + I_{7}h_{I_{7}} + I_{5}h_{I_{5}}\,.
\end{split}
\label{eq:decay_rate_binned}
\end{equation}
This expression is analogous to Eq.~\eqref{eq:decay_rate}, but where the angular functions are replaced with the $h_{I_{X}}$ templates. The normalisation condition from Ref.~\cite{Becirevic:2019tpx} is imposed,
\begin{equation}
\Gamma = \frac{1}{4}(3I_{1c} + 6I_{1s} - I_{2c} - 2I_{2s}) = 1\,,
\label{eq:PDFnorm}
\end{equation}
which constrains the value of $I_{1c}$ from the other $I_X$ coefficients. Importantly, the form of each $h_{I_X}$ template remains the same regardless of the underlying physics model; only the values of the $I_X$ coefficients are modified by the presence of NP.

Using this PDF, a binned maximum likelihood fit to the reconstructed decay angle distribution is demonstrated using the TensorFlowAnalysis package~\cite{TFA}, which provides an interface between TensorFlow~\cite{tensorflow2015-whitepaper} and the MINUIT~\cite{James:1975dr} minimisation package. For this demonstration, \mbox{$N_{\text{sig}} = 100,000$} signal events are generated across $n_{\text{bins}}$ bins in each of the three angular variables, where
$n_{\text{bins}} = 16$ is chosen to ensure that there are approximately 25 signal events in each bin. An alternative binning scheme based on placing approximately 50 events into each bin has been tested and shows consistent behaviour.  
The $I_X$ values found by the binned fit are summarised in Tab.~\ref{tab:Signal_only_I} (a).

\begin{table}[h]
\centering
\scriptsize
\subfloat{(a)}{%
\begin{tabular}{c | c}
Coefficient & Value \\ \hline
$I_{1c}$ & $\phantom{-}0.53 \pm 0.01$ \\
$I_{1s}$ & $\phantom{-}0.40 \pm 0.00$ \\
$I_{2c}$ & $-0.17 \pm 0.02$ \\
$I_{2s}$ & $\phantom{-}0.08 \pm 0.01$ \\
$I_{3}$ & $-0.10 \pm 0.01$ \\
$I_{4}$ & $-0.14 \pm 0.02$ \\
$I_{5}$ & $\phantom{-}0.28 \pm 0.01$ \\
$I_{6c}$ & $\phantom{-}0.28 \pm 0.02$ \\
$I_{6s}$ & $-0.24 \pm 0.01$ \\
$I_{7}$ & $-0.01 \pm 0.01$ \\
$I_{8}$ & $\phantom{-}0.01 \pm 0.02$ \\
$I_{9}$ & $-0.01 \pm 0.02$ \\
\end{tabular}
}%
\qquad\qquad
\subfloat{(b)}{%
\begin{tabular}{c | c}
Coefficient & Value \\ \hline
$I_{1c}$ & $\phantom{-}0.52$ \\
$I_{1s}$ & $\phantom{-}0.40$ \\
$I_{2c}$ & $-0.16$ \\
$I_{2s}$ & $\phantom{-}0.06$ \\
$I_{3}$ & $-0.12$ \\
$I_{4}$ & $-0.14$ \\
$I_{5}$ & $\phantom{-}0.28$ \\
$I_{6c}$ & $\phantom{-}0.32$ \\
$I_{6s}$ & $-0.25$ \\
$I_{7}$ & $-0.01$ \\
$I_{8}$ & $\phantom{-}0.00$ \\
$I_{9}$ & $-0.00$ \\
\end{tabular}
}
\caption{Results of the (a) binned template fit to the reconstructed angles, and (b) unbinned parametric fit to the truth-level angles. In both cases only the $B^0 \to D^{*-}\tau^+ \nu_\tau$ signal sample is used.  The uncertainties quoted in (a) are statistical, and arise from the use of a finite number of bins. The uncertainties in (b) are due only to finite sample size, and are negligibly small compared to those of (a); they are thus omitted for this comparison.}
\label{tab:Signal_only_I}
\end{table}

\subsection{Validating with a truth-level fit}

\begin{figure}[h]
\centering
\includegraphics[width = 0.32\textwidth]{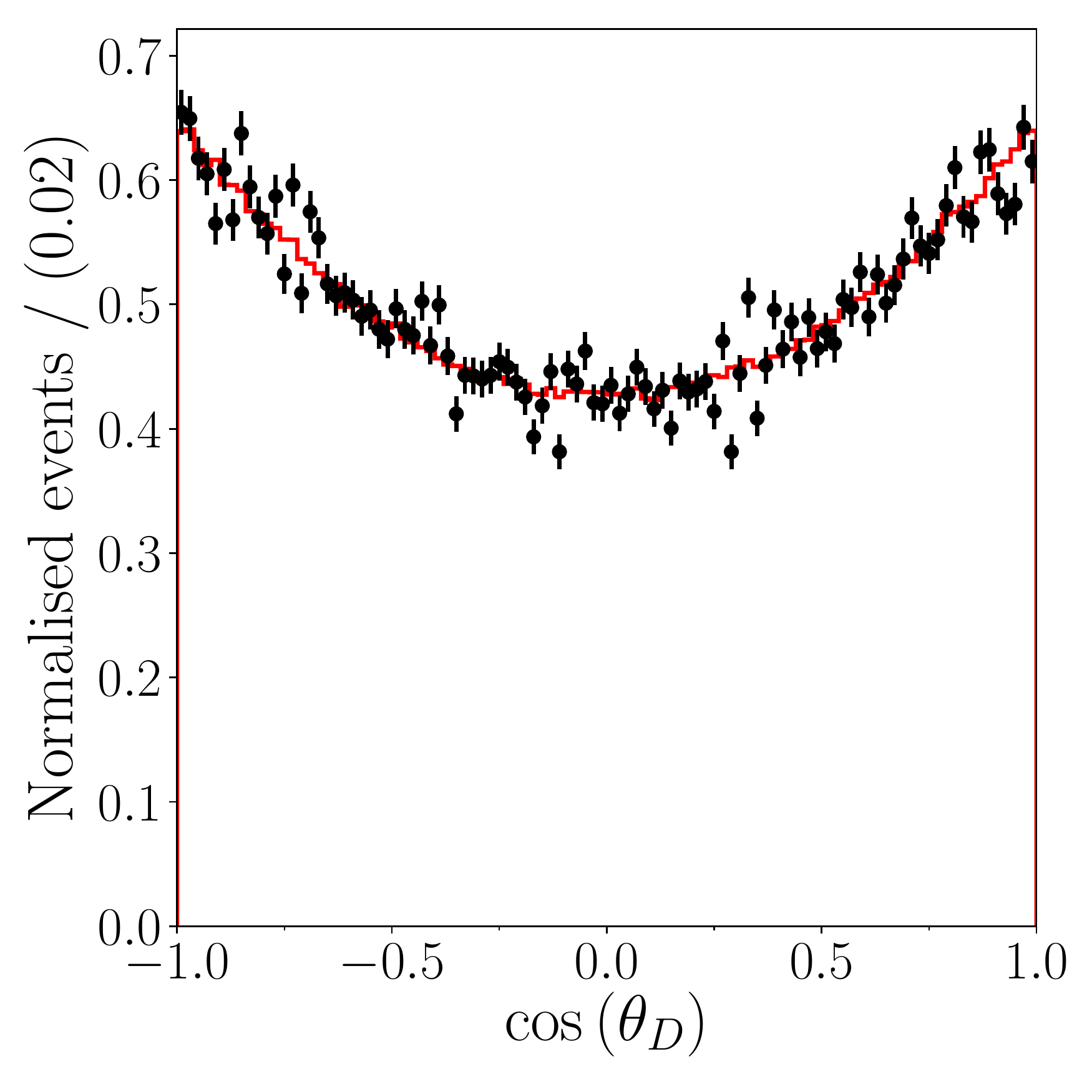}
\includegraphics[width = 0.32\textwidth]{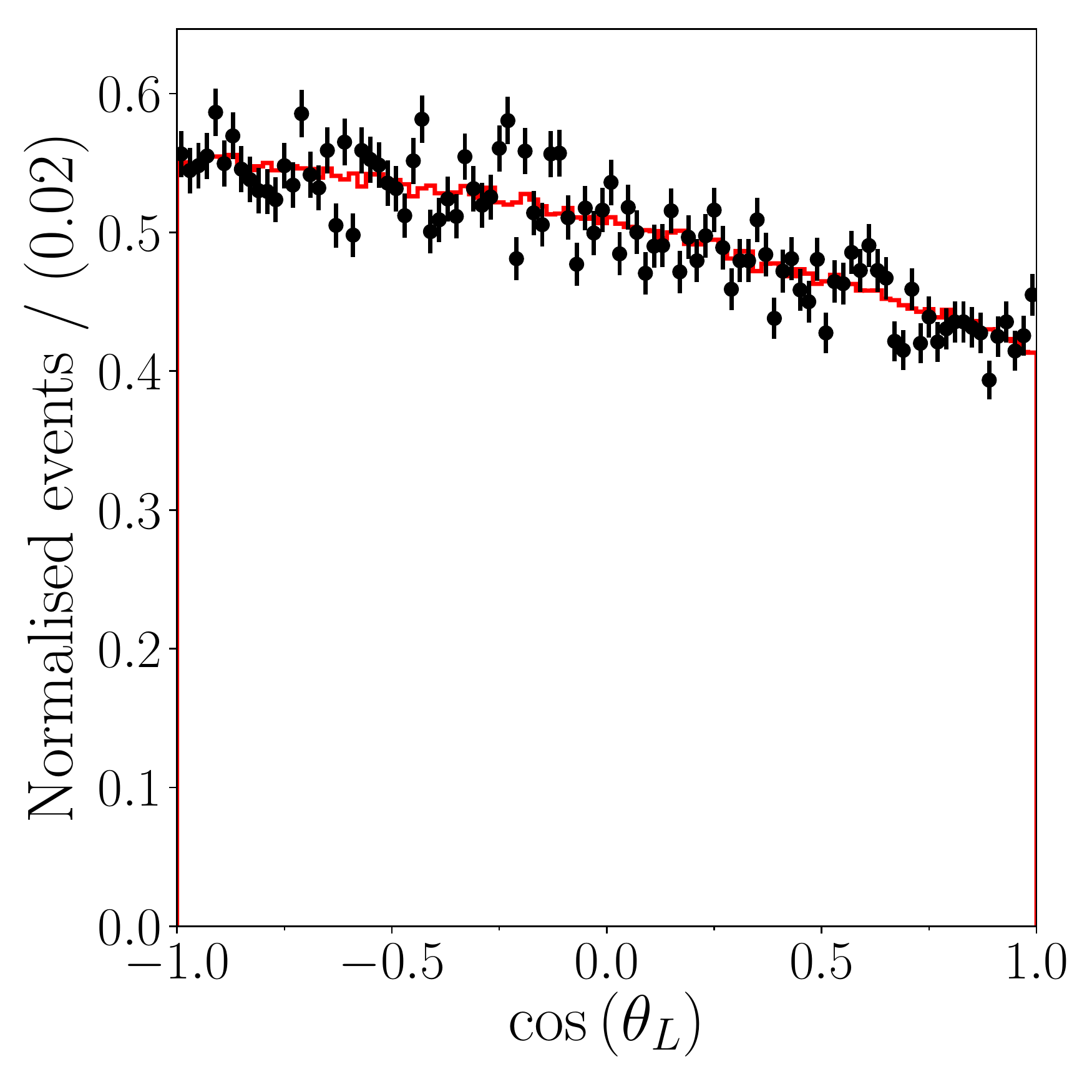}
\includegraphics[width = 0.32\textwidth]{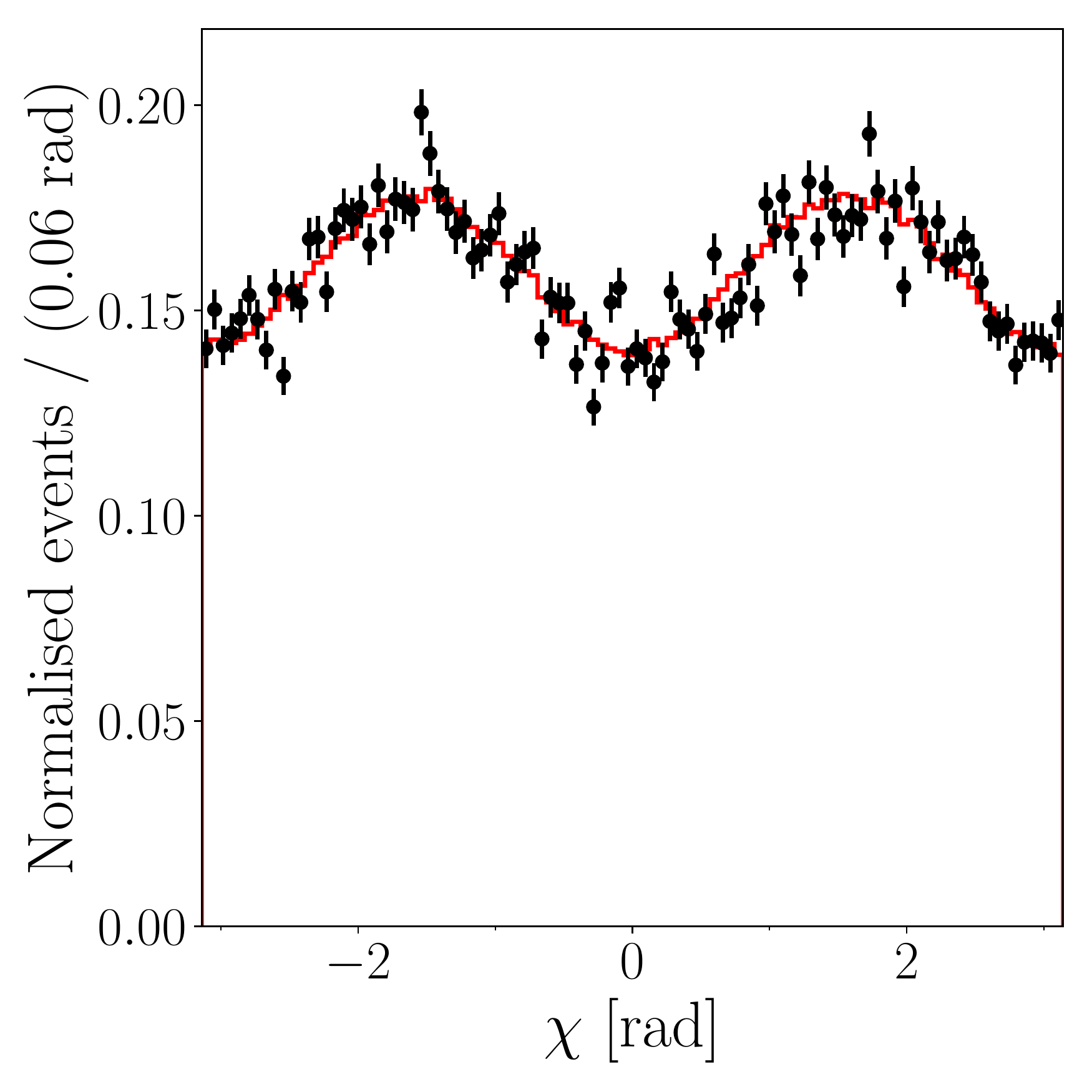}
\caption{Normalised one-dimensional projections of the unbinned parametric fit to the truth-level angular distribution in a sample of $100,000$ $B^0 \to D^{*-}\tau^+\nu_{\tau}$ decays. The black points are the signal sample, and the red solid line represents the best fit using Eq.~\eqref{eq:decay_rate}, where the $I_X$ coefficient values listed in Tab.~\ref{tab:Signal_only_I} (b) are found.
\label{fig:true_angles_fit}}
\end{figure}

To validate the binned template fit results, a second sample of $100,000$ signal decays is generated without acceptance cuts, resolution smearing, or the effect of the missing neutrino in the four-vector calculation. Using the true angles and Eq.~\eqref{eq:decay_rate} as the PDF, an unbinned parametric fit is performed with the normalisation condition of Eq.~\eqref{eq:PDFnorm} again imposed.  
The total fit projections are shown in Figure~\ref{fig:true_angles_fit} and the fitted $I_X$ values are recorded in \mbox{Table~\ref{tab:Signal_only_I} (b)}. The level of agreement between the truth-level parametric fit and the binned fit to the reconstructed angular variables is excellent. By correctly describing reconstruction biases and resolution effects using templates, the binned fit correctly measures the angular coefficients. 

\begin{figure}[h!]
\centering
\includegraphics[width = 0.7\textwidth]{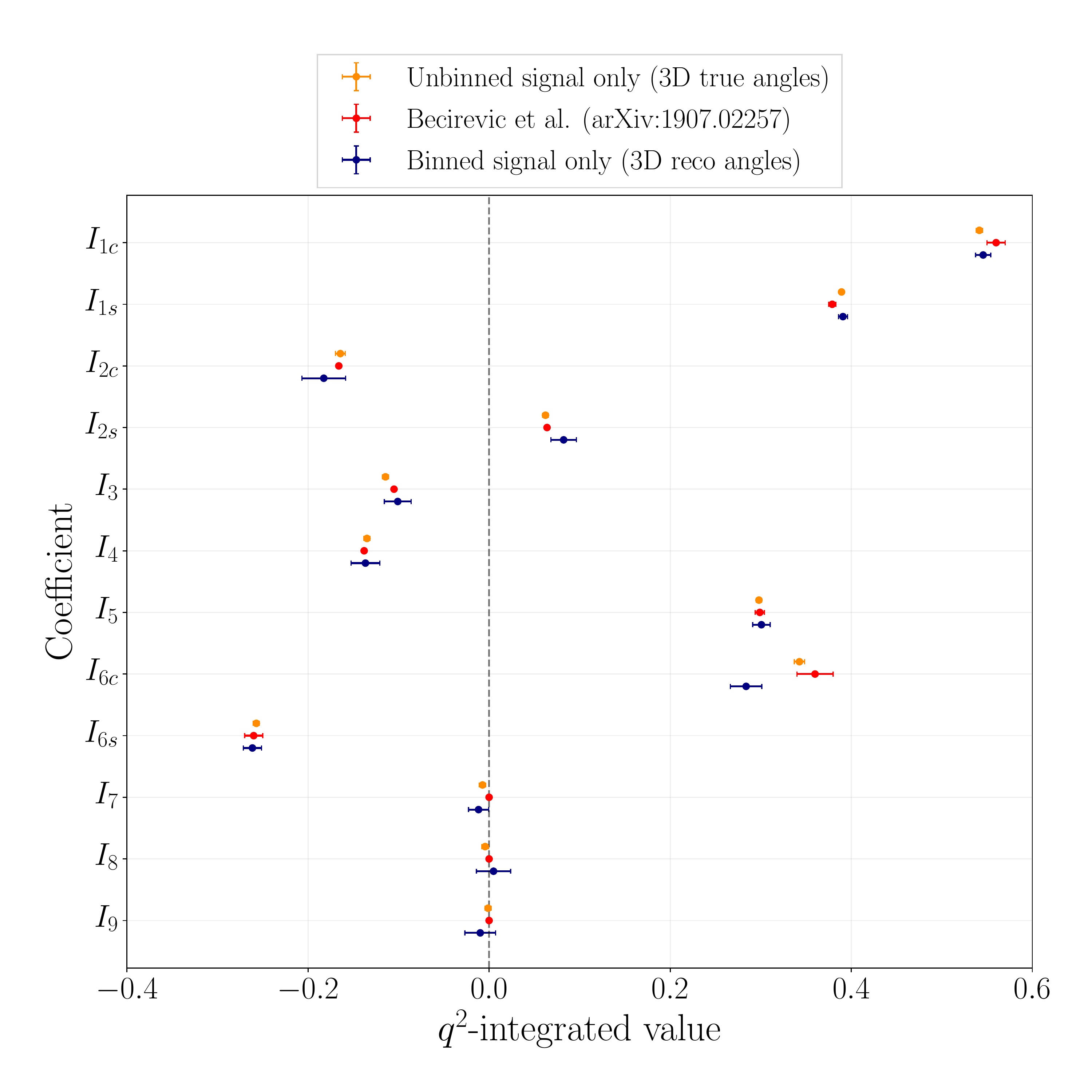}
\caption{Comparison of the signal-only truth-level unbinned parametric fit (orange points), the binned template fit (navy points), and the $q^2$-integrated SM $I_X$ values calculated in Ref.~\cite{Becirevic:2019tpx} (red points). The fitted MC samples are re-weighted to follow the CLN form factor scheme, which has also been used in the calculation of the cited SM values.}
\label{fig:Fedele_comparison}
\end{figure}

It is desirable to show that the template method can correctly recover SM $I_X$ values under an alternative form factor scheme. To demonstrate this, the signal-only template fit is repeated with per-event weights applied to the generated signal sample in order to align it with the CLN form factor scheme~\cite{Caprini:1997mu}. The reweighting is performed using the HAMMER package~\cite{Duell:2016maj,Ligeti:2016npd}. Both the parametric unbinned fit and the binned template fit are rerun; the $h_{I_{X}}$ templates are unaltered as they do not contain any form factor dependence. The $I_X$ coefficients resulting from the fits are displayed in Fig.~\ref{fig:Fedele_comparison}, and are seen to agree well with the SM values calculated using the CLN scheme in Ref.~\cite{Becirevic:2019tpx}. Furthermore, the longitudinal $D^{*-}$ polarisation fraction calculated from this signal-only template fit,
\begin{equation}
  F_L^{D^*} = \frac{3I_{1c}-I_{2c}}{3I_{1c}-I_{2c}+6I_{1s}-2I_{2s}}= 0.446 \pm 0.004\,,
\end{equation}
aligns satisfactorily with the SM expectation noted in Sec.~\ref{sec:intro}.  

\subsection{On binning in ${\boldmath q^2}$}
The weighting procedure can be applied inclusively or in a number of discrete $q^2_{\rm True}$ bins. Assuming simulation accurately models all \mbox{$q^2_{\rm True}\to q^2_{\rm Reco}$} sculpting and bin migration, the model independence of the $I_X$ measurements is preserved. 
Specifically, with $N$ bins in $q^2_{\rm True}$, the event weights in the template definition become $w_{I_X}=R_{I_X}(i,j)$, where $i$ again identifies the true angular bin and $j$ tags the true $q^2_{\rm True}$ bin. Because an event from the $j^{\rm th}$ $q^2_{\rm True}$ bin may migrate to any $k^{\rm th}$ $q^2_{\rm Reco}$ bin, $12N^2$ templates are required. These $h_{j,I_X}^k$ templates are built in the reconstructed angular variables space $(\cos\theta_D,\cos\theta_L,\chi)_{\text{Reco}}$ for each $q^2_{\rm Reco}$ bin by applying the weights.\footnote{The $q^2_{\rm Reco}$ bins are not required to align with the $q^2_{\rm True}$ bins. Since the reconstruction may measure $q^2_{\rm Reco}$ values outside the possible $q^2_{\rm True}$ range, different ranges may be necessary. Also, the number of $q^2_{\rm True}$ bins and $q^2_{\rm Reco}$ bins may differ, though they are both taken to be $N$ in this explanation.} Although there are $12N^2$ templates, there are still only $11N$ freely varying angular coefficients and $N$ signal fractions, $A_j$. This is because the apportioning across the $q^2_{\rm Reco}$ bins is fixed from simulation via the weighting procedure. 
In a generalisation of Eq.~\ref{eq:decay_rate_binned}, the PDF is defined summed over $q^2_{\rm Reco}$ bins and $q^2_{\rm True}$ bins,
\begin{equation}
\begin{split}
P\left((\cos\theta_D,\cos\theta_L,\chi)_{\text{Reco}};q^2_{\rm Reco}\right) &=\sum_j A_j\sum_k\bigg \{ \frac{1}{3}\left(4 - 6y_{j,1s} + y_{j,2c} + 2y_{j,2s}\right)h^k_{j,1_{1c}} \\ 
& + y_{j,1s}h^k_{j,I_{1s}} + y_{j,2c}h^k_{j,I_{2c}} + y_{j,2s}h^k_{j,I_{2s}} \\
&+ y_{j,6c}h^k_{j,I_{6c}} + y_{,j6s}h^k_{j,I_{6s}} + y_{j,3}h^k_{j,I_{3}} + y_{j,9}h^k_{j,I_{9}} \\
&+ y_{j,4}h^k_{j,I_{4}} + y_{j,8}h^k_{j,I_{8}} + y_{j,7}h^k_{j,I_{7}} + y_{j,5}h^k_{j,I_{5}}\bigg \}\,.
\end{split}
\label{eq:decay_rate_binned_qsquare}
\end{equation}
The coefficients are $y^k_{j,X}=f^k_{j,X} I_{j,X}$, where the fractions $f^k_{j,X}$ with $\sum_k f^k_{j,X} = 1$ quantify the bin migration probabilities are derived from simulation through the weighting procedure. 
\begin{sloppypar}
It is stressed that because the $q^2_{\rm True}$ binning is included in the event weights \mbox{$R_{I_X}(i,j)={D_{j,I_{X}}}/{M_j}$}, the dependence on the model used in the simulation is removed in a multi-$q^2$-bin implementation in the same way that it is removed in the inclusive implementation documented in this work. However, as $12N^2$ templates are required just for the signal, large simulated samples would been needed in such an analysis as well as large data sets.
\end{sloppypar}

\section{Dealing with backgrounds}
\label{sec:backgrounds}

The template fit is studied in a more realistic manner by considering backgrounds to the $B^0 \to D^{*-} \tau^+ \nu_\tau$ signal. 
A large number of backgrounds are generated using RapidSim, matching the list of backgrounds considered in recent experimental work~\cite{PhysRevD.97.072013}. There are three categories to consider: (1) prompt $B \to D^{*-}\pi^+\pi^+\pi^-(X)$ backgrounds that are reduced by requiring a displaced three-prong vertex, (2) the dominant double-charm $B \to D^{*-} D_{(s)}^+(X)$ background, and (3) $B \to D^{*-}(X)\,\tau^+\nu_\tau$ feed-down.

Several high branching fraction $B$ decays produce three charged pions at the $B$ decay vertex, with one or more additional particles ($X$) missed in the reconstruction. The prompt backgrounds considered for this study are recorded in Tab.~\ref{tab:prompt_modes} (App.~\ref{sec:bkg_modes}).
These prompt $B \to D^{*-}\pi^+\pi^+\pi^-(X)$ backgrounds are reduced by applying a flight requirement to the $3\pi$ system. In Ref.~\cite{PhysRevD.97.072013}, $\tau$ candidate vertices are required to be displaced from the $B$ vertex along the $z$-axis with a separation of $4\sigma_z$, where the vertex uncertainties of the $B$ and $\tau$ candidates determine the standard deviation $\sigma_z$. 
Vertex fits are not performed in RapidSim so the effect of the flight requirement must be approximated.
Assuming a combined $B$ and $\tau$ vertex uncertainty of 1~mm, the flight requirement is emulated by requiring that all $3\pi$ vertices are displaced 4~mm from their corresponding $B$ decay vertex. This is applied to all RapidSim samples, including the signal sample; the efficiencies on signal decays and prompt $B \to D^{*-}\pi^+\pi^+\pi^-(X)$ decays are similar to those reported in Ref.~\cite{PhysRevD.97.072013}. 

\begin{sloppypar}
The largest source of background remaining after the $\tau$ flight requirement arises from double-charm decays of the type $B \to D^{*-} D_{s}^+(X)$. The $D_s^+$ meson flies before decaying to final states that include three charged pions, mimicking the \mbox{$\tau^+ \to \pi^+\pi^+\pi^- \bar{\nu}_{\tau}$} signature.  
The $B\to D^{*-}D_s^+ (X)$ modes generated for this study are recorded in Tab.~\ref{tab:DstDsX_bkg_modes} (App.~\ref{sec:bkg_modes}) and the $D_s^+$ modes are listed in Tab.~\ref{tab:Ds_bkg_modes} (App.~\ref{sec:bkg_modes}).  
A realistic sample is created for each double-charm $B$ mode by weighting contributions according to measured branching fractions~\cite{PhysRevD.98.030001}. To create a single \mbox{$B \to D^{*-} D_{s}^+(X)$} sample, the subsamples are summed according to the proportions measured in Ref.~\cite{PhysRevD.97.072013}. 
Similarly, \mbox{$B \to D^{*-} D^+(X)$} and \mbox{$B \to D^{*-} D^0(X)$} decays are also present in the background. The $D^0$ and $D^+$ mesons also fly, and often produce three or more charged particles in their decay. The $B\to D^{*-}D^{(0,+)} (X)$ modes generated for this study are recorded in Tab.~\ref{tab:DstDX_bkg_modes} (App.~\ref{sec:bkg_modes}), and the $D^{(0,+)}$ modes are listed in Tab.~\ref{tab:D_bkg_modes} (App.~\ref{sec:bkg_modes}). 
\end{sloppypar}

The $B \to D^{**}\tau \nu_{\tau}$ decay is identical to signal aside from the fact that the charm meson is produced in a higher state of angular momentum. Given this similarity, the feed-down background must be included as a small fixed fraction relative to signal, and treated as a systematic pollution in any phenomenological interpretation. Two contributions are included, namely $B^+ \to D_{1}(2420)^0 \tau^+ \nu_\tau$ and $B^+ \to D_2^*(2460)^0 \tau^+ \nu_\tau$ decays, where the excited charm mesons decay to $D^{*-}\pi^+$ and the additional pion is not reconstructed. The EvtGen models used are given in Tab.~\ref{tab:feed_down_modes} (App.~\ref{sec:bkg_modes}). No $B \to D^{**}\tau\nu_{\tau}$ modes have been observed, so the $D_{1}(2420)^0$ and $D_2^*(2460)^0$ samples are summed in equal proportion.

\subsection{Multivariate classifier}
\label{sec:BDT}

\begin{figure}[h]
\centering
\includegraphics[width = 0.32\textwidth]{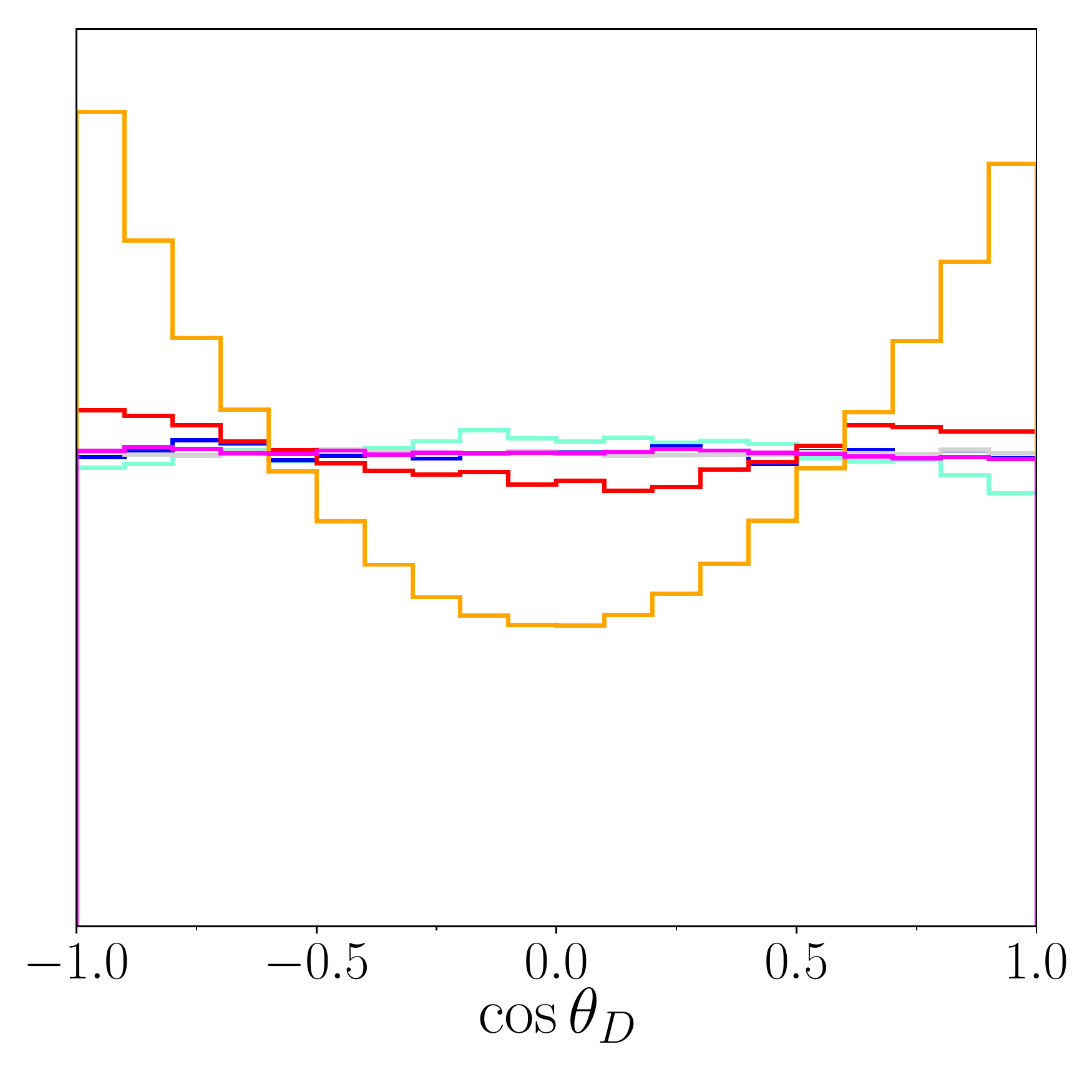}
\includegraphics[width = 0.32\textwidth]{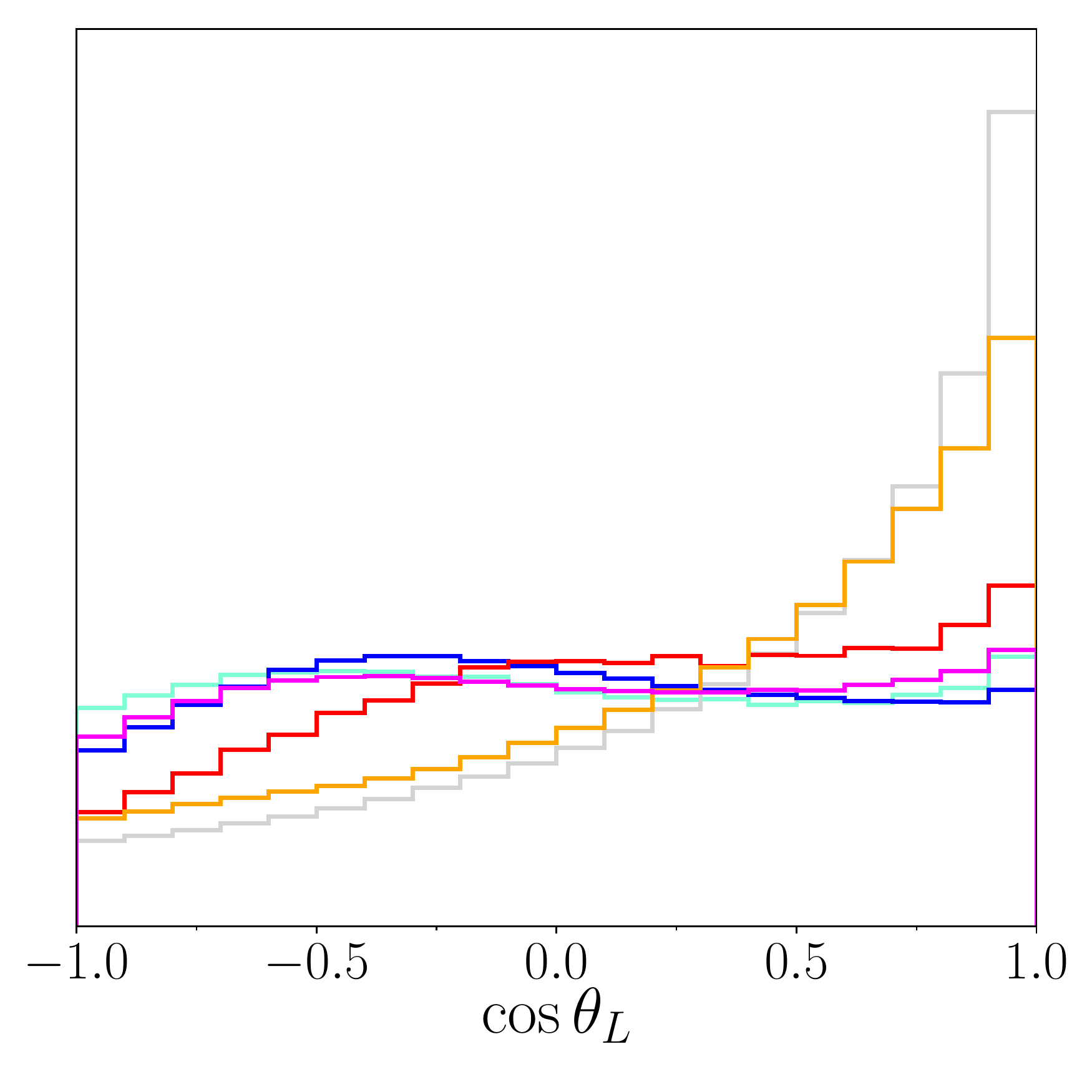}
\includegraphics[width = 0.32\textwidth]{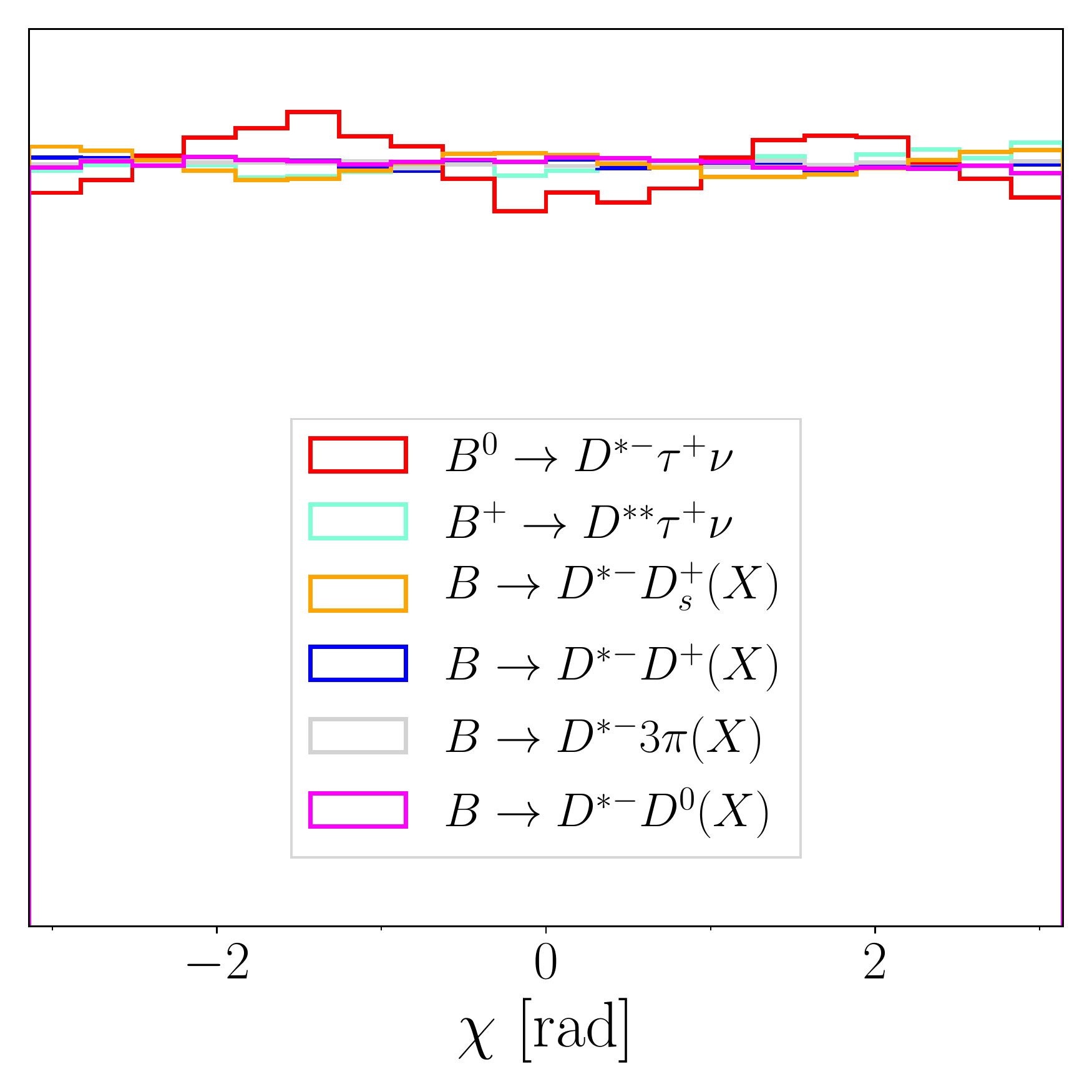}
\caption{Reconstructed angular distributions of the signal and several categories of background. All distributions are normalised to have equal integral.}
\label{fig:bkg_angles}
\end{figure}

The reconstructed angular distributions for each background category are shown in Fig.~\ref{fig:bkg_angles} along with the total signal template. All events are required to fall within the LHCb acceptance and pass the $\tau$ flight requirement. On their own, the reconstructed decay angles do not provide enough signal and background separation to reliably measure the twelve parameters that describe the $B^0 \to D^{*-}\tau^+\nu_{\tau}$ signal. As such, a multivariate classifier is included as a fourth dimension to provide sufficient separation.

To minimise dependence on any underlying model, the classifier is designed to avoid any input variables that relate directly to the $B$ decay kinematics; variables describing the $3\pi$ system are thus preferred. The most appropriate variables include the reconstructed proper lifetime and the invariant mass of the $3\pi$ system, and the invariant mass of the $\pi^+\pi^+$ and $\pi^+\pi^-$ combinations. These variables provide discrimination between $\tau$ candidates and charm mesons, due to their different lifetimes and decay properties. 

\begin{figure}[h]
\centering
\includegraphics[width = 0.32\textwidth]{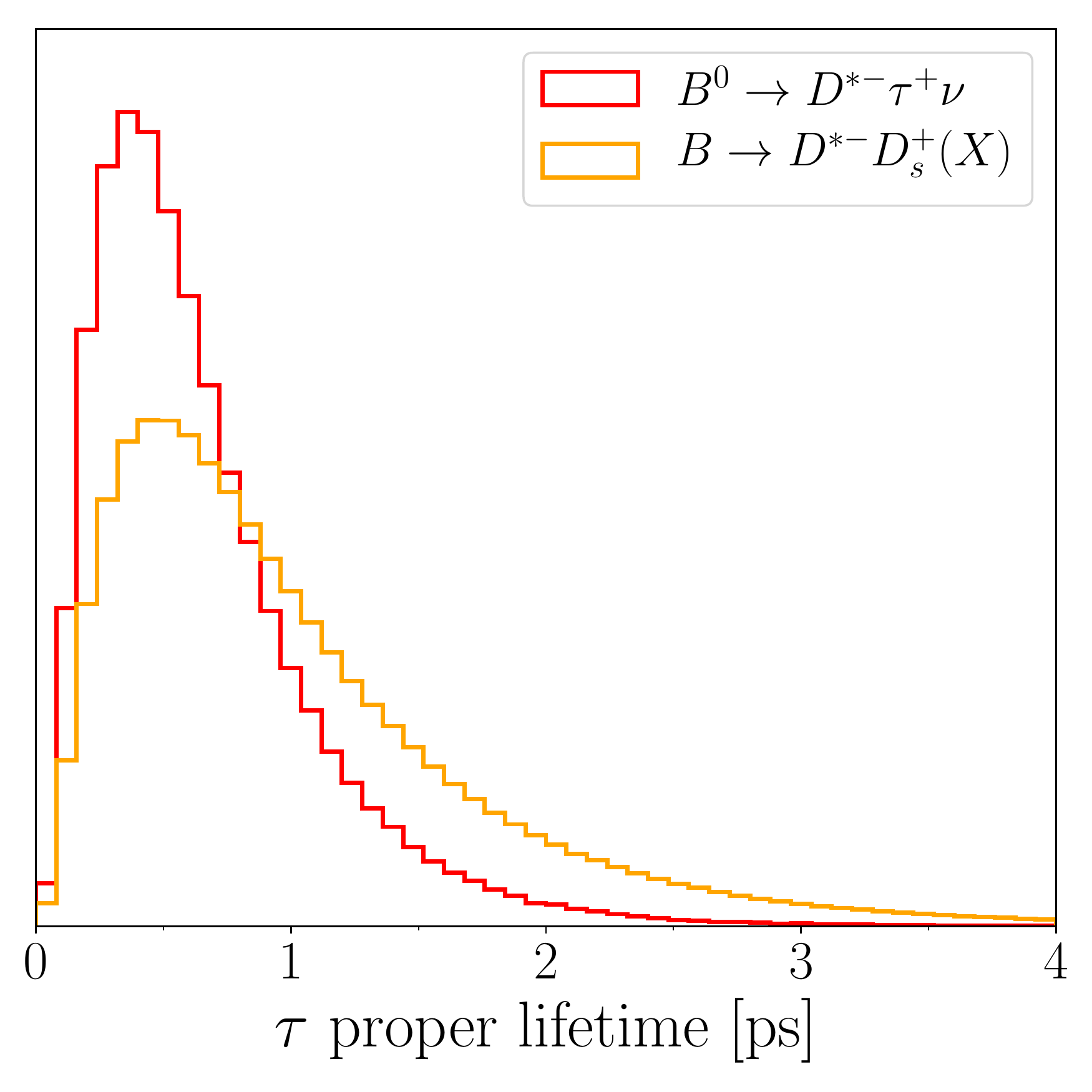}
\includegraphics[width = 0.32\textwidth]{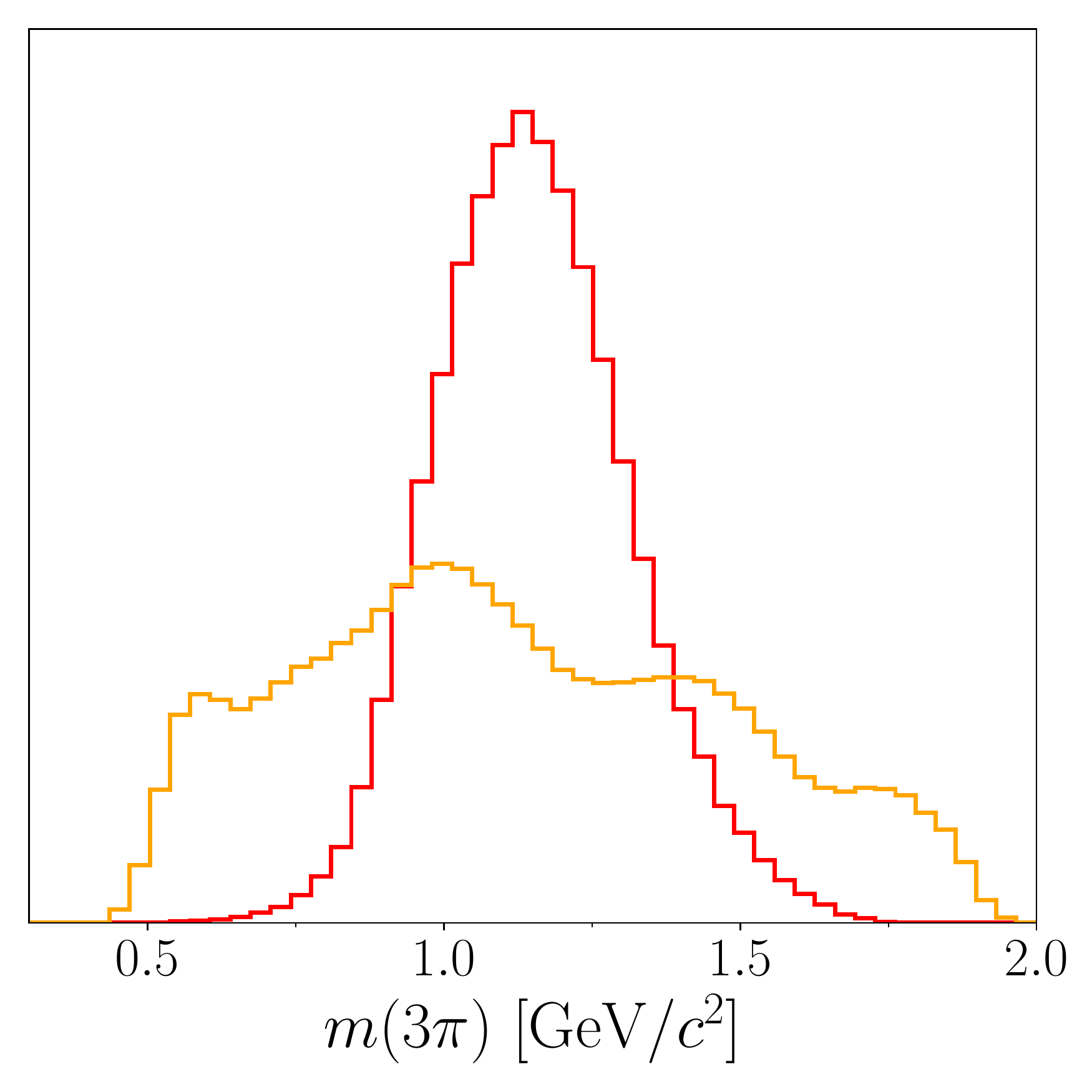}\\
\includegraphics[width = 0.32\textwidth]{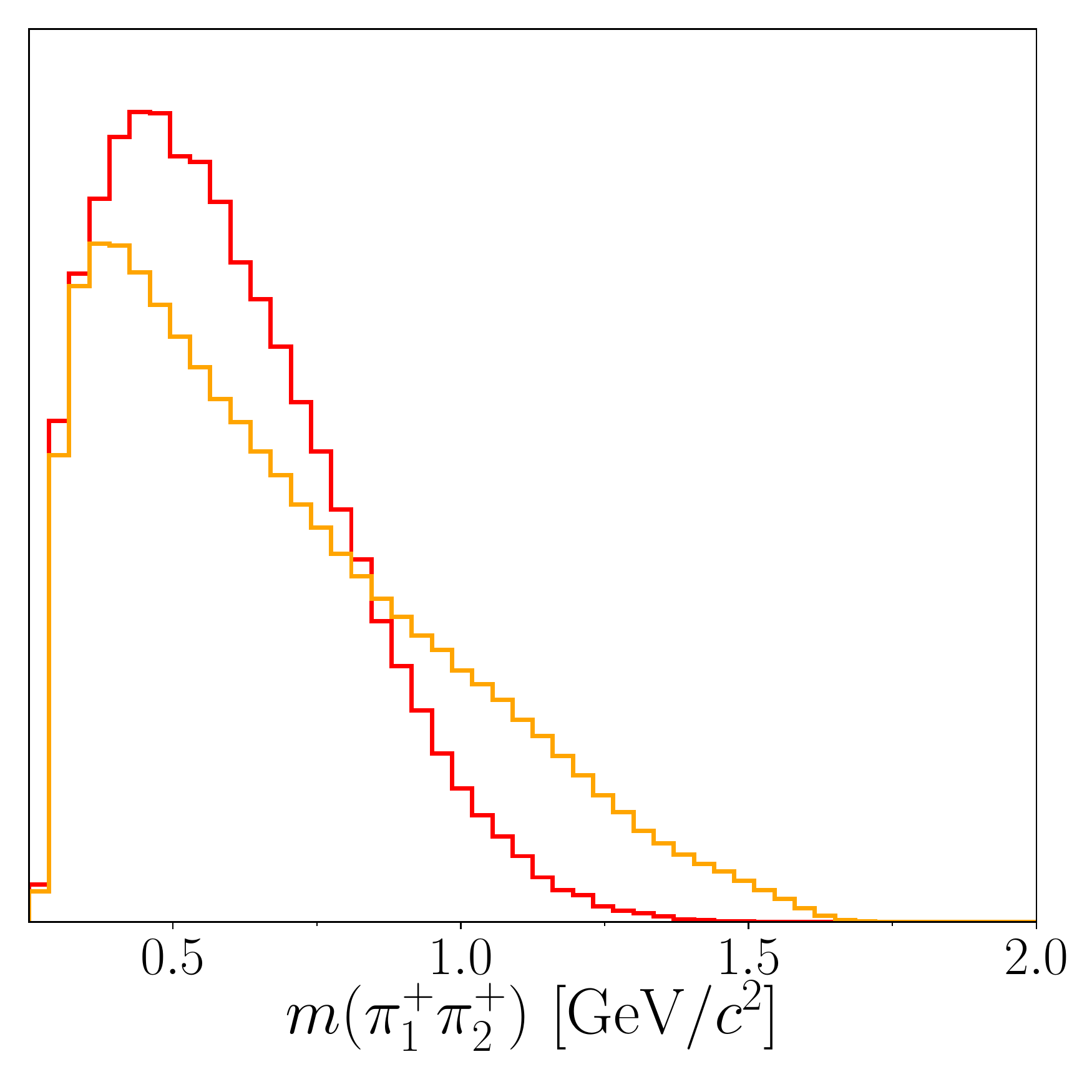}
\includegraphics[width = 0.32\textwidth]{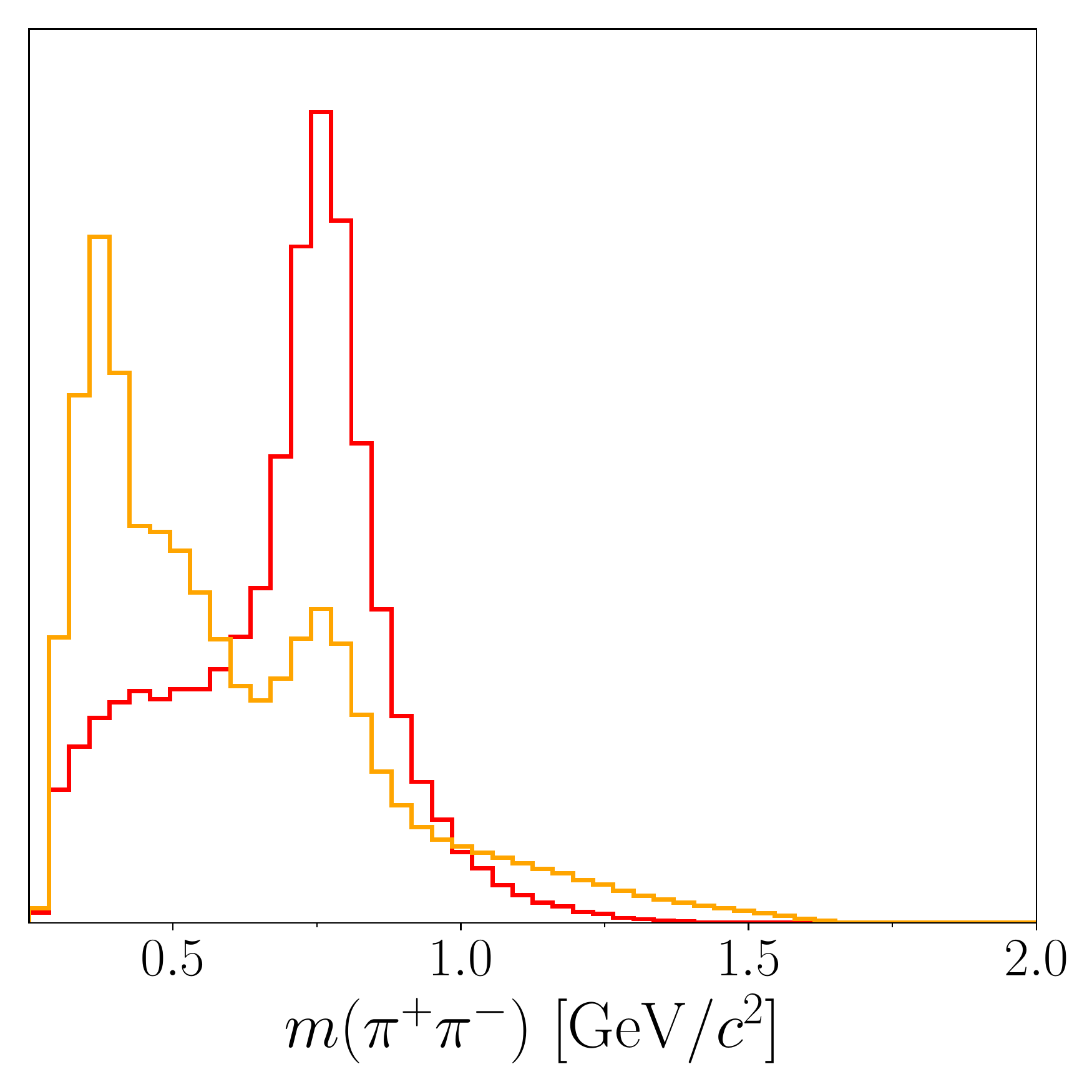}
\caption{Input variables used in the multivariate classifier.}
\label{fig:bdt_inputs}
\end{figure}

\begin{figure}[h]
\centering
\includegraphics[width = 0.34\textwidth]{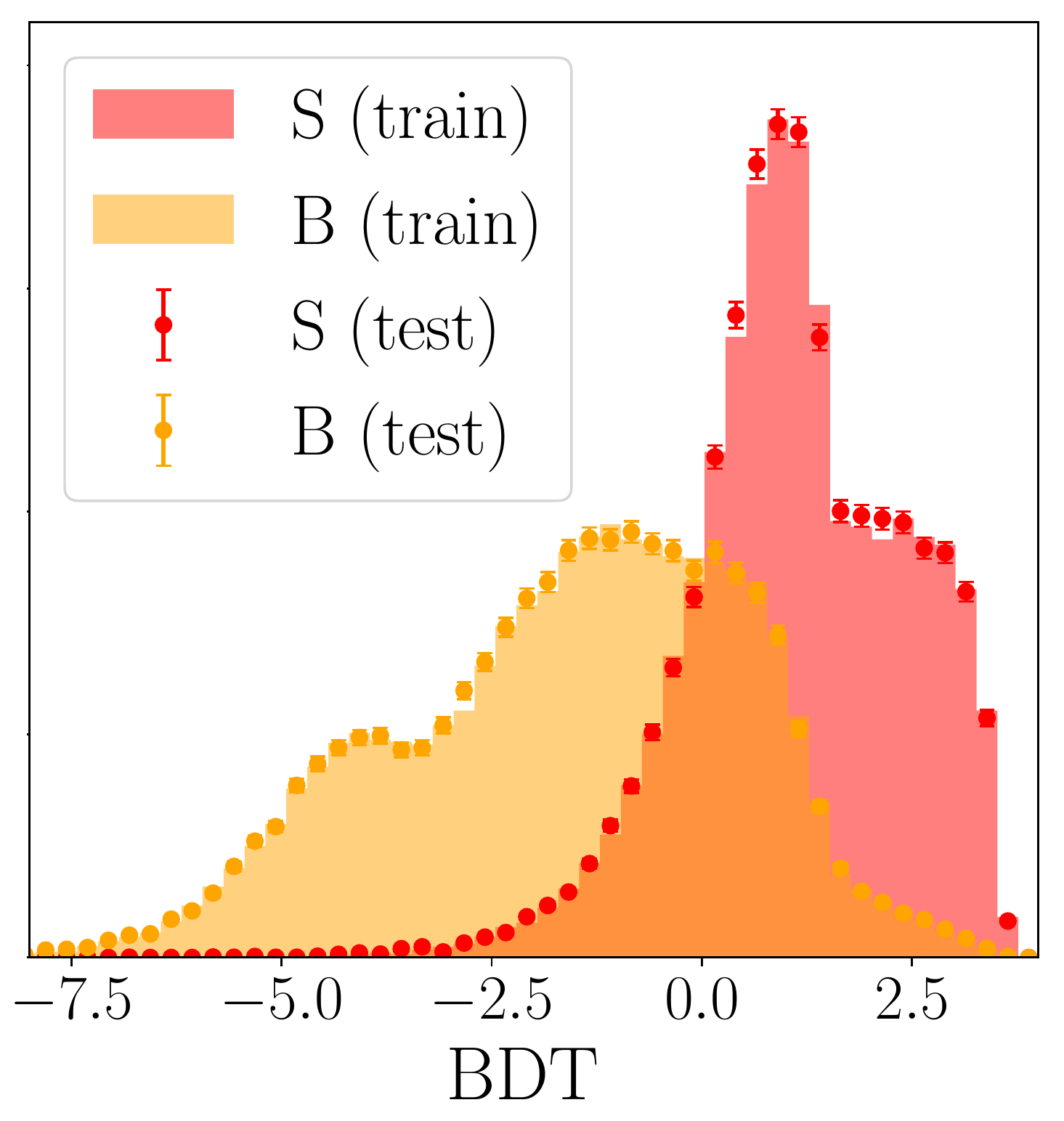}\put(-25,120){(a)}
\includegraphics[width = 0.36\textwidth]{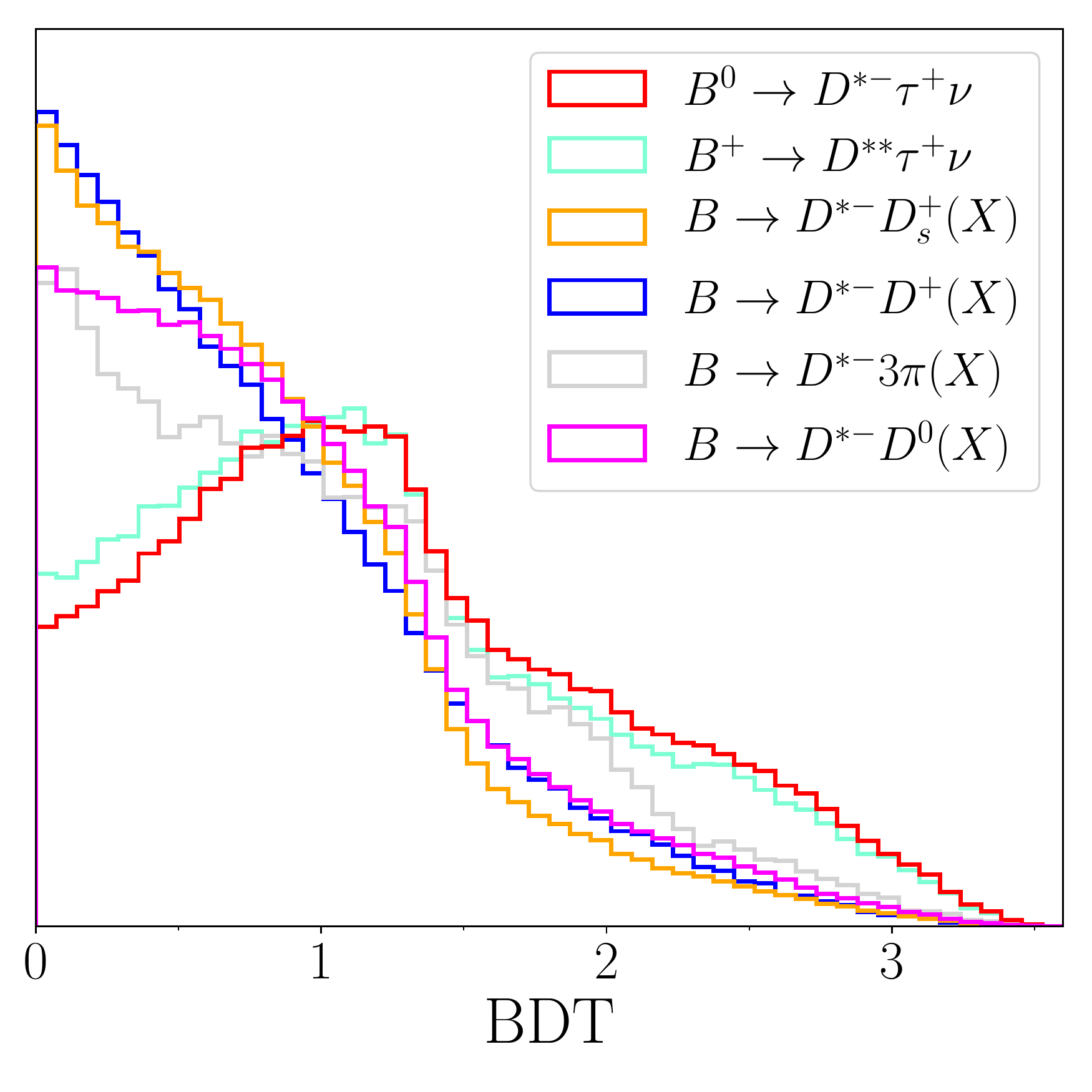}\put(-40,55){(b)}
\caption{(a) BDT distributions for $B^0 \to D^{*-}\tau^+ \nu_\tau$ signal (S) and $B \to D^{*-}D_s^+(X)$ background (B) in the test and training samples. (b) BDT distributions for signal and all categories of background for BDT$>0$.}
\label{fig:bdt}
\end{figure}

A gradient boosted decision tree (BDT) classifier is trained using the scikit-learn package~\cite{scikit-learn}. The total $B \to D^{*-}D_s^+(X)$ background sample described above is used to train the BDT to favour $B^0 \to D^{*-}\tau^+\nu_{\tau}$ decays. The $D_s^+$ sample is used as it is the largest category of background remaining after the $\tau$ flight requirement. The distributions for each input variable in signal and background are shown in Fig.~\ref{fig:bdt_inputs}.
The area under the classifier Receiver Operating Characteristic (ROC) curve is 0.84, and the performance is illustrated in Fig.~\ref{fig:bdt} (a) where the BDT distributions are shown for both in signal and background. The classifier is applied to all generated signal and background samples, and those events with classifier decision values above zero are retained for use in the fit and shown in Fig.~\ref{fig:bdt}. This selection requirement is 80\% efficient on signal while rejecting 70\% of background. 
The feed-down and signal BDT distributions are confirmed to be almost identical.

\subsection{$\tau^+ \to \pi^+\pi^+\pi^-\pi^0\bar{\nu}_{\tau}$ component of the signal}
\label{sec:3pipi0}

The $B^0 \to D^{*-}(\tau^+ \to \pi^+\pi^+\pi^-\pi^0\bar{\nu}_{\tau})\nu_\tau$ decay, where the neutral pion is not reconstructed, is not a background but rather contributes to the total signal. This decay mode differs from the three-prong signal only in the $\tau$ decay and thus has the same $I_X$ coefficients. To benefit from the presence of this mode in the signal, a dedicated sample is generated and processed in a manner identical to the principal $\tau^+ \to \pi^+\pi^+\pi^-\bar{\nu}_{\tau}$ signal.
Small differences in the reconstructed decay angles are observed, due to the additional missing momentum in the $\tau^+ \to \pi^+\pi^+\pi^-\pi^0\bar{\nu}_{\tau}$ decay. The BDT distributions are also not equivalent for the two cases, since the invariant mass variables used in the BDT differ. Following Ref.~\cite{PhysRevD.97.072013}, a total signal sample is created from both the \mbox{$\tau^+ \to \pi^+\pi^+\pi^-\bar{\nu}_{\tau}$} and \mbox{$\tau^+ \to \pi^+\pi^+\pi^-\pi^0\bar{\nu}_{\tau}$} samples, using $f_{3\pi}=78\%$ as a relative fraction. 


\subsection{Final signal plus background sample}
\label{sec:lhc_fit}

To create a realistic total dataset that includes both signal and background events, the generated signal and background samples are summed using the relative fractions listed in Tab.~\ref{tab:sig_bkg_fractions}; the values used are based upon those measured in LHCb data~\cite{PhysRevD.97.072013}. The signal comprises $f_{\text{sig}} = 11.8\%$ of the total sample, while the total feed-down contribution is 11\% of the signal fraction (1.3\% in total). The majority of the sample is composed of \mbox{$B \to D^{*-}D_s^+(X)$} background decays (62.1\%) with smaller fractions assigned to the \mbox{$B \to D^{*-}D^+(X)$} and  \mbox{$B \to D^{*-}D^0(X)$} backgrounds.

\begin{table}[h]
\centering
\scriptsize
\begin{tabular}{c | c | c}
Mode & Fraction & Value \\ \hline
$B^0 \to D^{*-}(\tau^+ \to \pi^+\pi^+\pi^-\bar{\nu}_{\tau})\nu_\tau$ & $f_{3\pi} \times f_{\text{sig}}$ & 0.78 (fixed) $\times$ 0.118 \\
$B^0 \to D^{*-}(\tau^+ \to \pi^+\pi^+\pi^-\pi^0\bar{\nu}_{\tau})\nu_\tau$ & $(1 - f_{3\pi}) \times f_{\text{sig}}$ & 0.22 (fixed) $\times$ 0.118 \\
$B \to D^{*-}D_s^+(X)$ & $f_{D_s^+}$ & 0.621 \\
$B \to D^{*-}D^+(X)$ & $f_{D^+}$ & 0.152 \\
$B \to D^{*-}D^0(X)$ & $f_{D^0}$ & 0.057 (fixed) \\
$B^+ \to D^{**}(\tau^+ \to \pi^+\pi^+\pi^-\bar{\nu}_{\tau})\nu_\tau$ & $f_{D^{**}} \times f_{3\pi} \times f_{\text{sig}}$ & 0.11 (fixed) $\times$ 0.78 (fixed) $\times$ 0.118 \\
$B^+ \to D^{**}(\tau^+ \to \pi^+\pi^+\pi^-\pi^0\bar{\nu}_{\tau})\nu_\tau$ & $f_{D^{**}} \times (1 - f_{3\pi}) \times f_{\text{sig}}$ & 0.11 (fixed) $\times$ 0.22 (fixed) $\times$ 0.118 \\
$B \to D^{*-}\pi^+\pi^+\pi^-(X)$ & $(1 - f_{\text{sig}} - f_{D_s^+} - f_{D^+} - f_{D^0} - f_{D^{**}})$ & 0.039 (constrained by $\sum f = 1$)
\end{tabular}
\caption{Fractions used to construct the total data sample. The fractions that are fixed in the fit are labelled so.}
\label{tab:sig_bkg_fractions}
\end{table}

\subsection{Four-dimensional fit}

To reliably measure the signal fraction $f_{\text{sig}}$ and $I_X$ coefficients, a four-dimensional binned maximum-likelihood fit to the total dataset is performed using the PDF
\begin{equation}
\begin{split}
P((\cos\theta_D,\cos\theta_L,\chi)_{\text{Reco}},\text{BDT}) &= f_{\text{sig}} h_{\text{sig}}\\
&+ f_{D_s^+}h_{D_s^+} + f_{D^+}h_{D^+} + f_{D^0}h_{D^0} + f_{D^{**}}h_{D^{**}} \\
&+ (1 - f_{\text{sig}} - f_{D_s^+} - f_{D^+} - f_{D^0} - f_{D^{**}})h_{\text{prompt}}\,,
\end{split}
\label{eq:decay_rate_binned_total}
\end{equation}
where $h_{\text{sig}}$ represents the signal PDF, defined in \eqref{eq:decay_rate_binned}. In the fit, $f_{\text{sig}}$, $f_{D_s^+}$, and $f_{D^+}$ freely vary, as do the eleven $I_X$ coefficients within $h_{\text{sig}}$. The fractions $f_{D^0}$ and $f_{D^{**}}$ are fixed, matching the procedure in Ref.~\cite{PhysRevD.97.072013}. The prompt fraction is constrained such that the fractions sum to unity.

\begin{sloppypar}
The twelve signal template histograms $h_{I_{X}}$ are created following a procedure that is essentially identical to that described in Sec.~\ref{sec:templates}, but where the additional BDT dimension is included and a signal sample containing both the \mbox{$\tau^+ \to \pi^+\pi^+\pi^-\bar{\nu}_{\tau}$} and \mbox{$\tau^+ \to \pi^+\pi^+\pi^-\pi^0\bar{\nu}_{\tau}$} modes is used. For each background mode $\beta$, a template $h_{\beta}$ is created by filling a four-dimensional histogram with the ($(\cos\theta_D,\cos\theta_L,\chi)_{\text{Reco}}$,BDT) distribution of the total $\beta$ sample. Large generated samples exceeding one million events are used to create all of the templates, in order to avoid statistical uncertainties on the template bin contents. The number of bins $n_{\text{bins}}$ in each fit dimension is chosen to be equal and is determined using
$n_{\text{bins}} = \text{Int}\left[\left(N_{\text{sig}}\Big/{25}\right)^{\frac{1}{4}}+\frac{1}{2}\right]
\label{eq:4d_bins}
$, where $N_{\text{sig}}$ is the anticipated number of signal events in the sample. The bin boundaries are not uniform, but are chosen such that each bin is populated with approximately 25 signal events. An alternative binning based on 50 signal events per bin gives consistent results.
\end{sloppypar}

\section{Expected precision}

\subsection{Hadron collider scenario}
\label{sec:4d_fit_results}

The four-dimensional fit is applied to datasets corresponding to three scenarios: \mbox{$N_{\text{sig}} = 8000$}, \mbox{$N_{\text{sig}} = 40,000$} and \mbox{$N_{\text{sig}} = 100,000$}. These are calculated by extrapolating the yield measured in Ref.~\cite{PhysRevD.97.072013} to 9 fb$^{-1}$ (Runs $1 + 2$), 23 fb$^{-1}$ (Runs 1--3), and 50 fb$^{-1}$ (Runs 1--4), as anticipated by LHCb~\cite{TheLHCbCollaboration:2320509}. The 23 and 50 fb$^{-1}$ scenario yield expectations account for additional improvements in the upgrade LHCb detector performance relative to LHCb in Run 1 and 2.  
The signal and background fractions listed in Tab.~\ref{tab:sig_bkg_fractions} are used to create the data sample for each of these cases, maintaining the same signal purity throughout.

\paragraph*{9 fb$\mathbf{^{-1}\ (N_{\text{sig}} = 8000)}$:}
\label{sec:9fb}

\begin{figure}
\centering
\includegraphics[width = 0.35\textwidth]{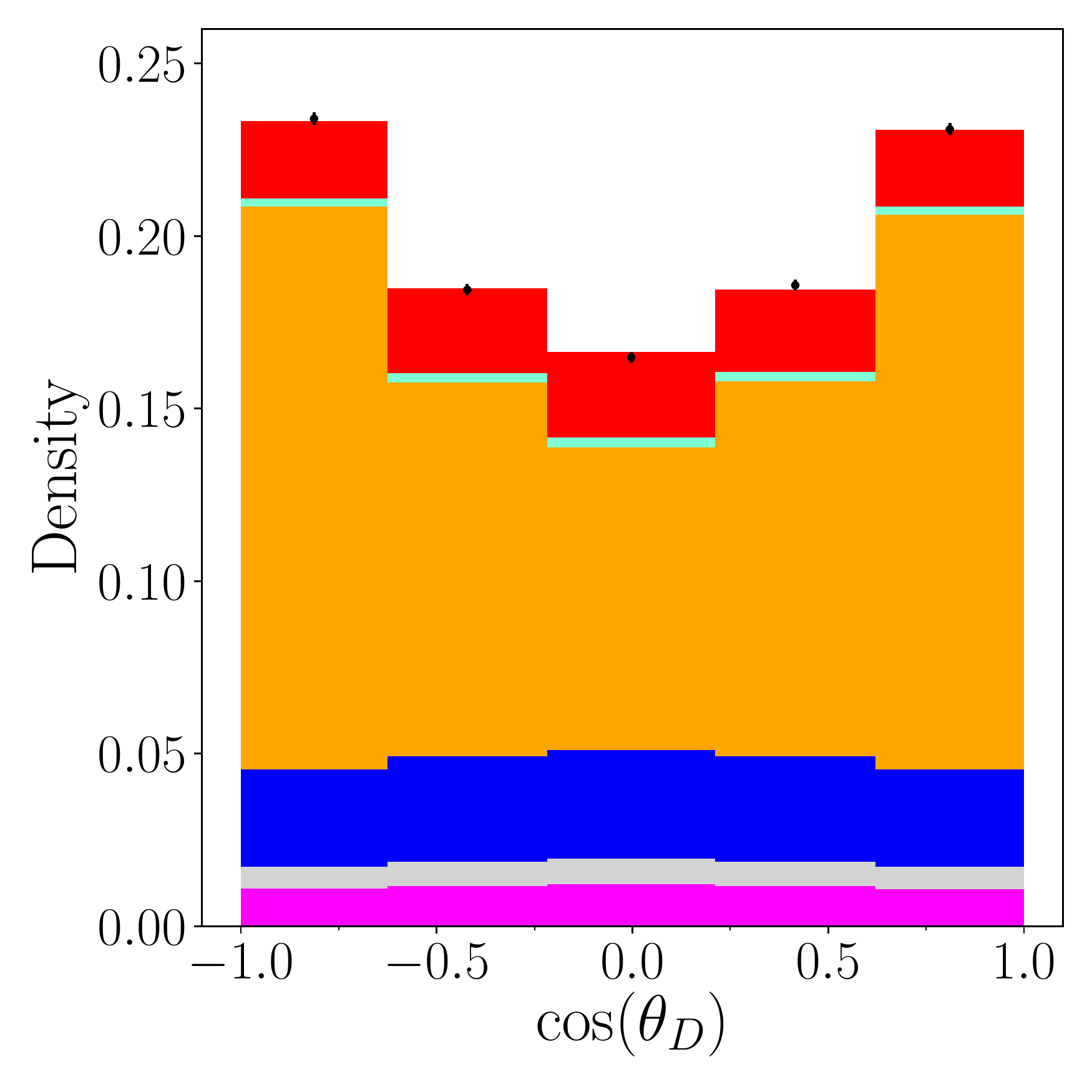}
\includegraphics[width = 0.35\textwidth]{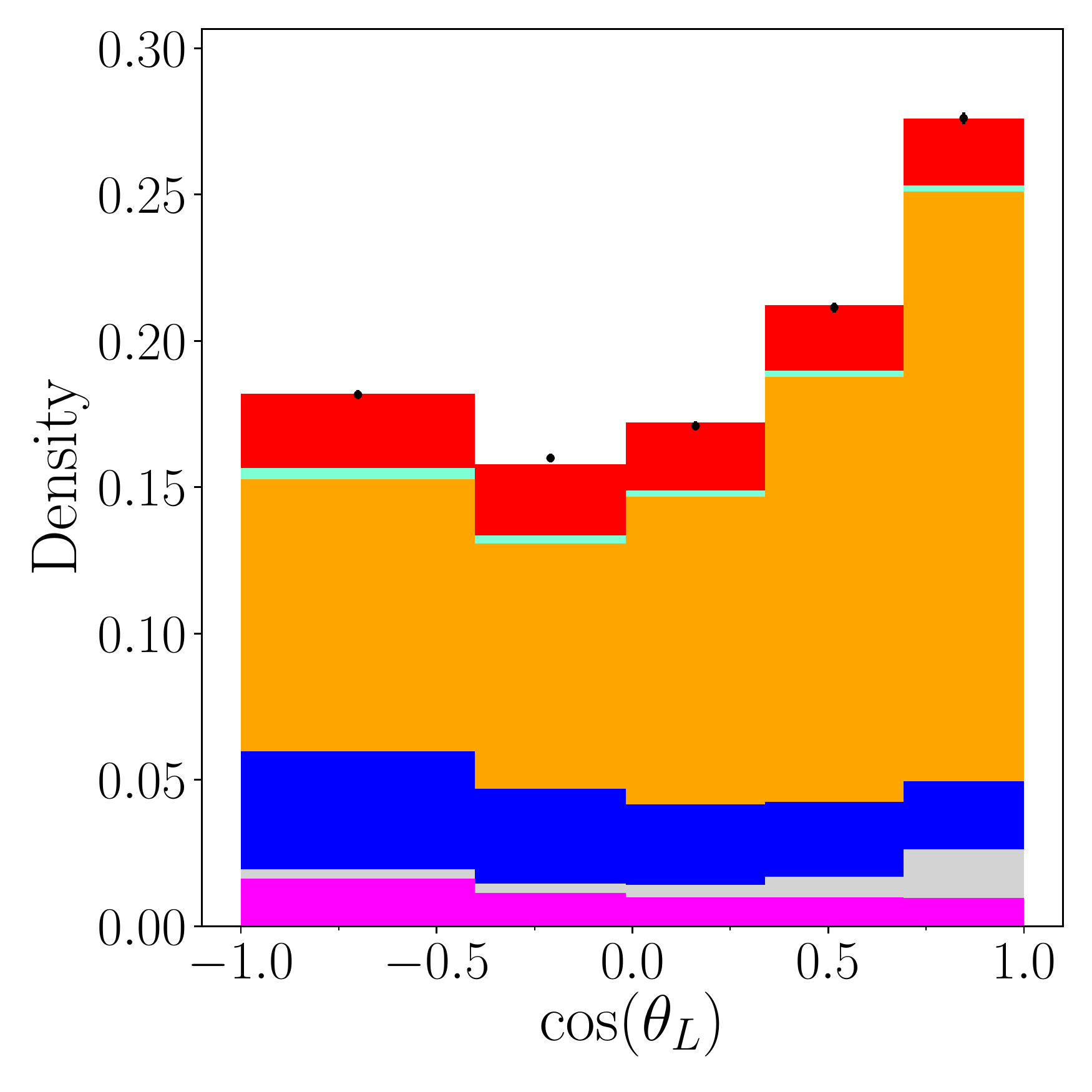}\\
\includegraphics[width = 0.35\textwidth]{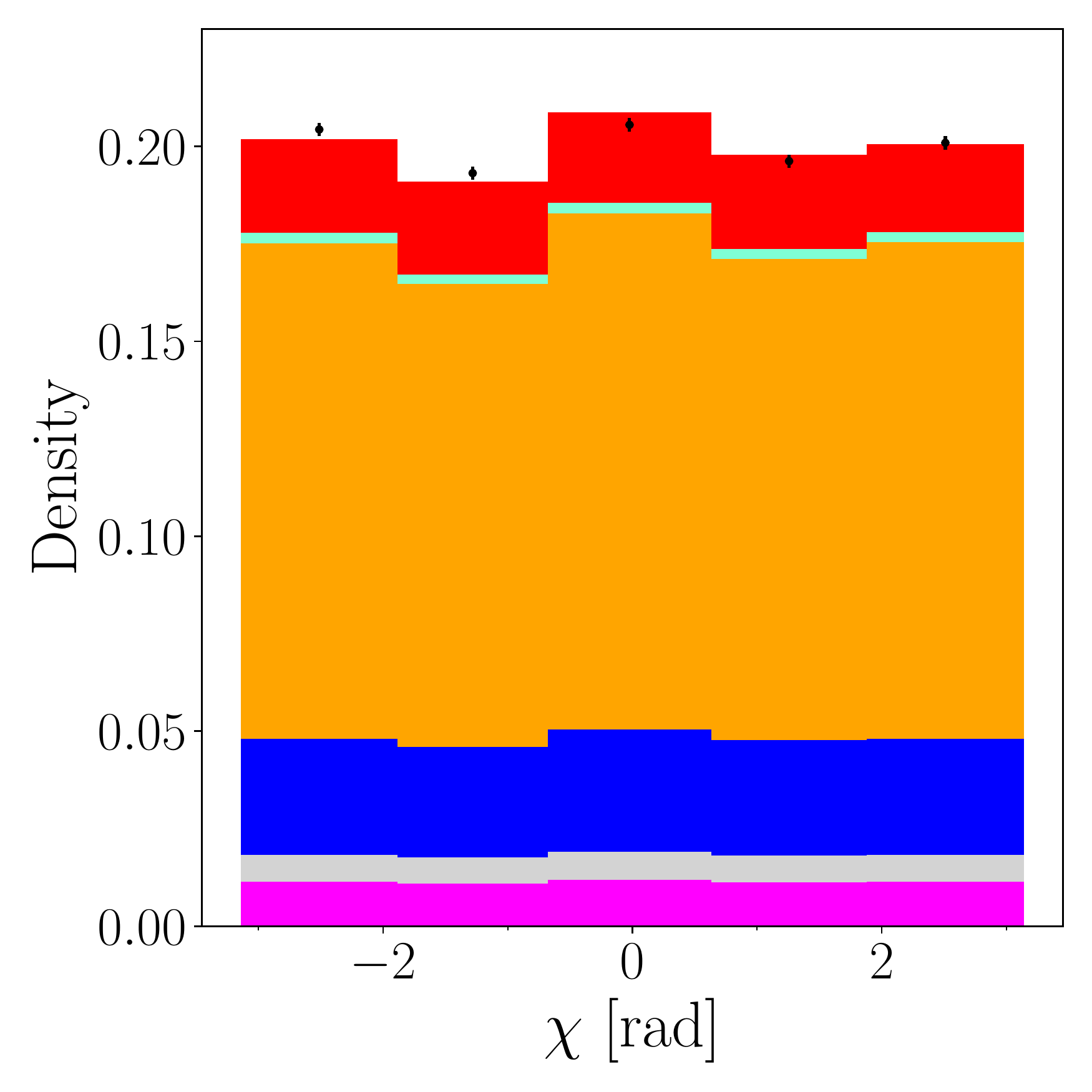}
\includegraphics[width = 0.35\textwidth]{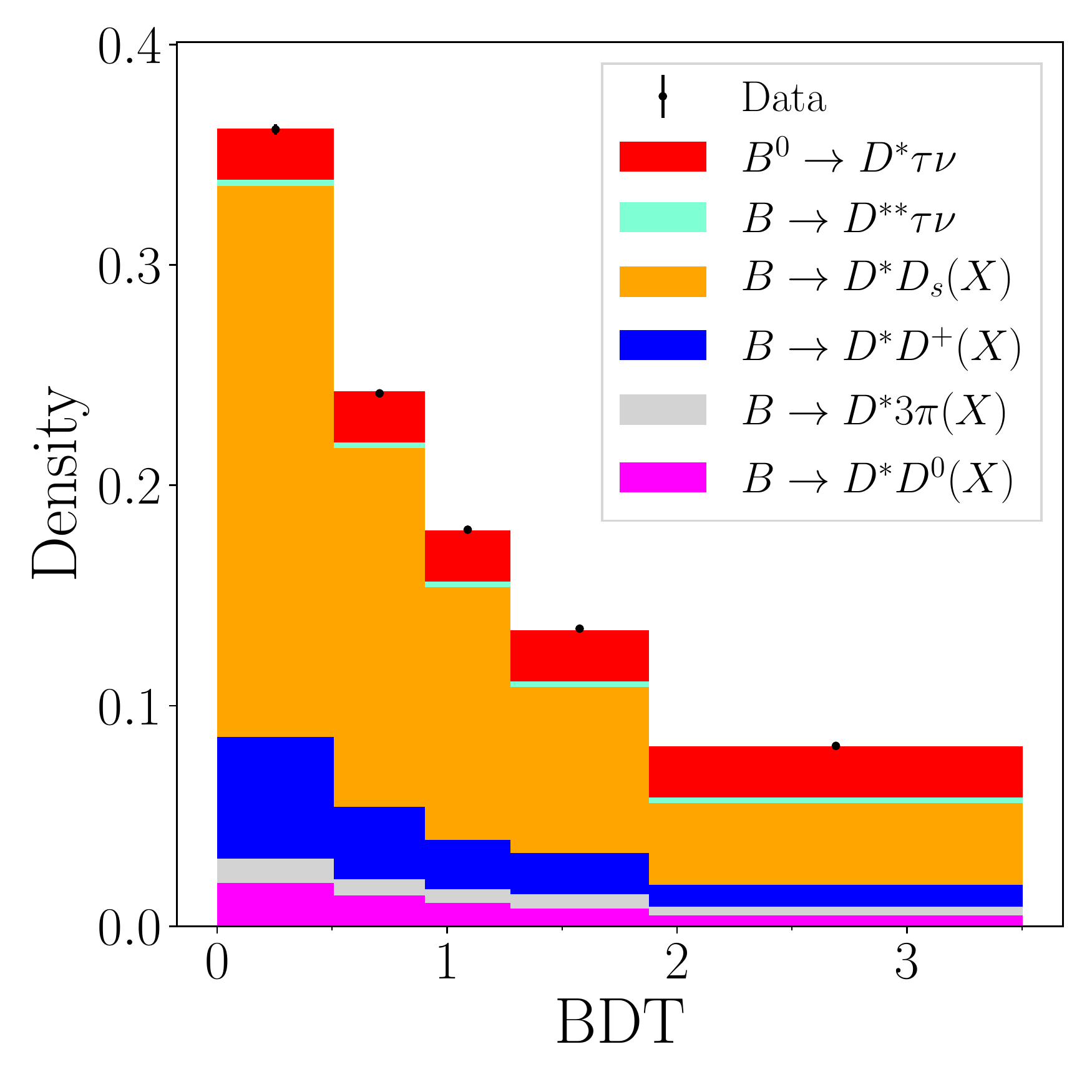}
\caption{One-dimensional projections of the \mbox{$N_{\text{sig}}=8,000$} binned fit, where the solid points represent the data and the filled histograms represent each fit component. The total $B^0 \to D^{*-}\tau^+\nu_\tau$ signal, given by the sum of all twelve angular terms, is shown in red.}
\label{fig:8k_projections}
\end{figure}

\begin{figure}[t!]
\centering
\includegraphics[width = 0.7\textwidth] {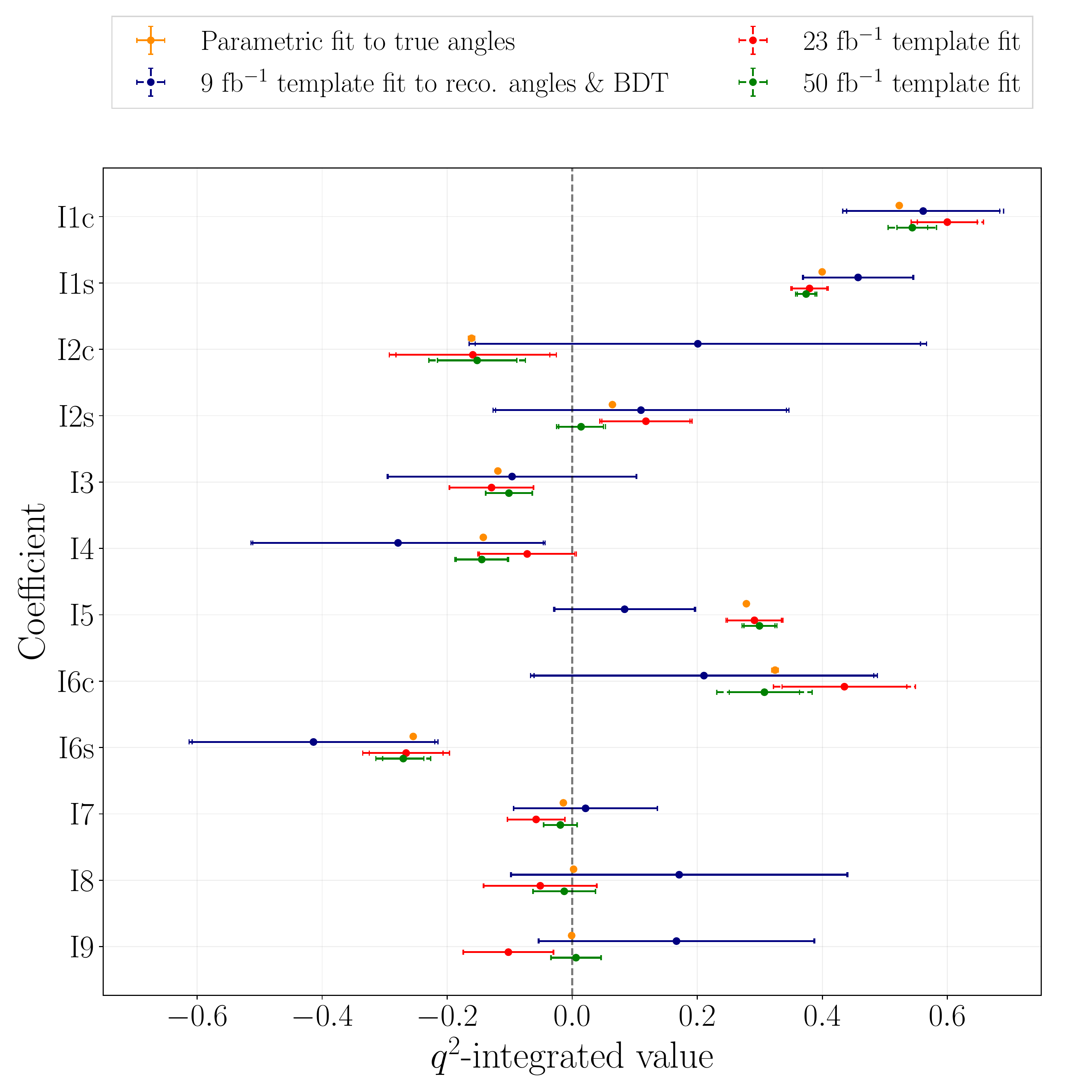}
\caption{Comparison of the signal-only unbinned parametric fit to the truth-level angles (orange) and the LHCb scenario template fits. The fit uncertainties dominate the results, but systematic uncertainties are included in the error bars.
\label{fig:all_results_systs}}
\end{figure}

The 4D template fit functions stably in this lowest-statistics case, and the resulting fit projections are shown in Fig.~\ref{fig:8k_projections}. The signal fraction is measured to be $f_{\text{sig}} = 0.116 \pm 0.010$ (8.6\% relative uncertainty), which agrees with the input value $f_{\text{sig}} = 0.118$. The $I_X$ values are measured with large uncertainties but remain compatible with the true values, as shown in Fig.~\ref{fig:all_results_systs}. Using the $I_X$ results for this sample, the derived value of $F_L(D^*)$ is $0.368 \pm 0.047$, where the uncertainty quoted is statistical only. It will thus be possible to make a competitive measurement of $F_L(D^*)$ using the 9 fb$^{-1}$ data.

\paragraph*{23 fb$\mathbf{^{-1}\ (N_{\text{sig}} = 40,000)}$:}
\label{sec:23fb}
\begin{sloppypar}
Increasing the signal yield by a factor five gives a strong improvement in the $I_X$ measurements, as shown in Fig.~\ref{fig:all_results_systs}; the derived value of $F_L(D^*)$ is $0.489 \pm 0.018$. The signal fraction is \mbox{$f_{\text{sig}} = 0.117 \pm 0.003$} (2.6\% relative uncertainty), which has reduced more than a naive $\sqrt{N_{\text{sig}}}$ scaling due to the larger number of bins in the 23 fb$^{-1}$ fit relative to the 9 fb$^{-1}$ fit. Both fits operate with binning schemes that require approximately 25 signal events per bin, and the increased number of bins in the 23 fb$^{-1}$ fit provides greater differentiation between the $I_X$ signal components and the backgrounds.
\end{sloppypar}

\paragraph*{50 fb$\mathbf{^{-1}\ (N_{\text{sig}} = 100,000)}$:}
\label{sec:50fb}

\begin{sloppypar}
With the largest dataset, the $I_X$ values are measured with absolute statistical uncertainties in the range 0.01--0.06, as shown in Fig.~\ref{fig:all_results_systs}. The statistical correlation matrix for the fit is provided in App.~\ref{sec:corr_matrix}; the other fit scenarios show a consistent pattern of parameter correlations.
The signal fraction is measured to be \mbox{$f_{\text{sig}} = 0.116 \pm 0.002$} (1.5\% relative uncertainty). Given the considerable reduction in $I_X$ uncertainty between the 23 and 50 fb$^{-1}$ scenarios, it is well motivated to continue performing measurements of this type during Run 4 of the LHC. This is highlighted by the derived value of $F_L(D^*)$ value, which is found to be $0.446 \pm 0.010$.
\end{sloppypar}

\subsection{Fit stability validation}
\label{sec:toys}

\begin{sloppypar}
To demonstrate the stability and accuracy of the three fit scenarios, many pseudo-experiments (``toys'') based on the fits are run. Using the template PDFs and the yields from the 9, 23, and 50 fb$^{-1}$ fits, toy datasets are generated where the number of events is independently determined in each bin according to Poisson variations of the bin content. The template fit is applied to each toy dataset, and pull distributions are created for all freely varying fit parameters. All pull distributions have mean values close to zero and widths close to unity, as expected for an unbiased fit returning the appropriate uncertainties. 
\end{sloppypar}



\subsection{$B$-factory scenario}
\label{sec:bfac_fit}

Complementary to LHCb, the Belle II experiment~\cite{SHWARTZ2015233,Kou:2018nap} can use an anticipated 50 ab$^{-1}$ dataset to measure the angular coefficients in $B \to D^{*}\tau\nu_{\tau}$ decays. 
The $b\bar{b}$ production cross section is much lower in $e^+e^-$ collisions compared to $pp$, but the well-defined initial state and the absence of other tracks in $B\Bbar$ events give considerable advantages when reconstructing final states with missing particles. By reconstructing the second $B$ meson, the momentum of a signal $B^0$ meson can be well defined in spite of the missing neutrino. 

To estimate the performance of the template fit at an $e^+e^-$ experiment, the signal and background samples generated to emulate the LHCb resolution and acceptance are reprocessed using the true $B$ meson four-vector in the decay angle calculations rather than relying on the estimation in Eq.~\eqref{eq:b_max}. Vertex and track momentum resolutions are assumed to be similar, so an overall 10-20\% advantage in angular resolution is determined. 

\begin{sloppypar}
The three-prong $\tau$ decay mode has not yet been used in a $B$-factory semitauonic analysis, so the signal yields at Belle II must be approximated. 
Belle have used the \mbox{$\tau^+ \to \pi^+ \nu_\tau$} and $\tau^+ \to \rho^+ \nu_\tau$ modes with hadronic tagging to measure the $\tau$ polarisation~\cite{Hirose:2016wfn}. Signal yields of $N(B^0 \to D^{*-}\tau^+\nu_{\tau}) = 88 \pm 11$ and \mbox{$N(B^+ \to D^{*0}\tau^+\nu_{\tau}) = 210 \pm 27$} are reported in 0.77 ab$^{-1}$ of data. Assuming near-perfect track finding efficiency at \mbox{Belle II}, such that the three-prong modes are reconstructed with a similar efficiency as the one-prong, a total tagged sample of \mbox{$N_{\text{sig}}(B^0 + B^+) \approx 7000$} three-prong events is estimated in 50 ab$^{-1}$ of Belle II data.
\end{sloppypar}

In Ref.~\cite{Hirose:2016wfn}, the Belle Collaboration reported a signal purity of 18.6\%. Although the combinatorial background at \mbox{Belle II} and LHCb differ, the $B$ backgrounds generated in Sec.~\ref{sec:backgrounds} are still the most important. Thus, a data sample is created containing 7000 signal events with 18.6\% purity, where the relative background fractions remain the same as those used in Sec.~\ref{tab:sig_bkg_fractions}.

\paragraph*{Results for 50 ab$\mathbf{^{-1}}$ of $\mathbf{e^+e^-}$ data $\mathbf{(N_{\text{sig}} = 7000)}$:}

\begin{figure}[h]
\centering
\includegraphics[width = 0.6\textwidth]{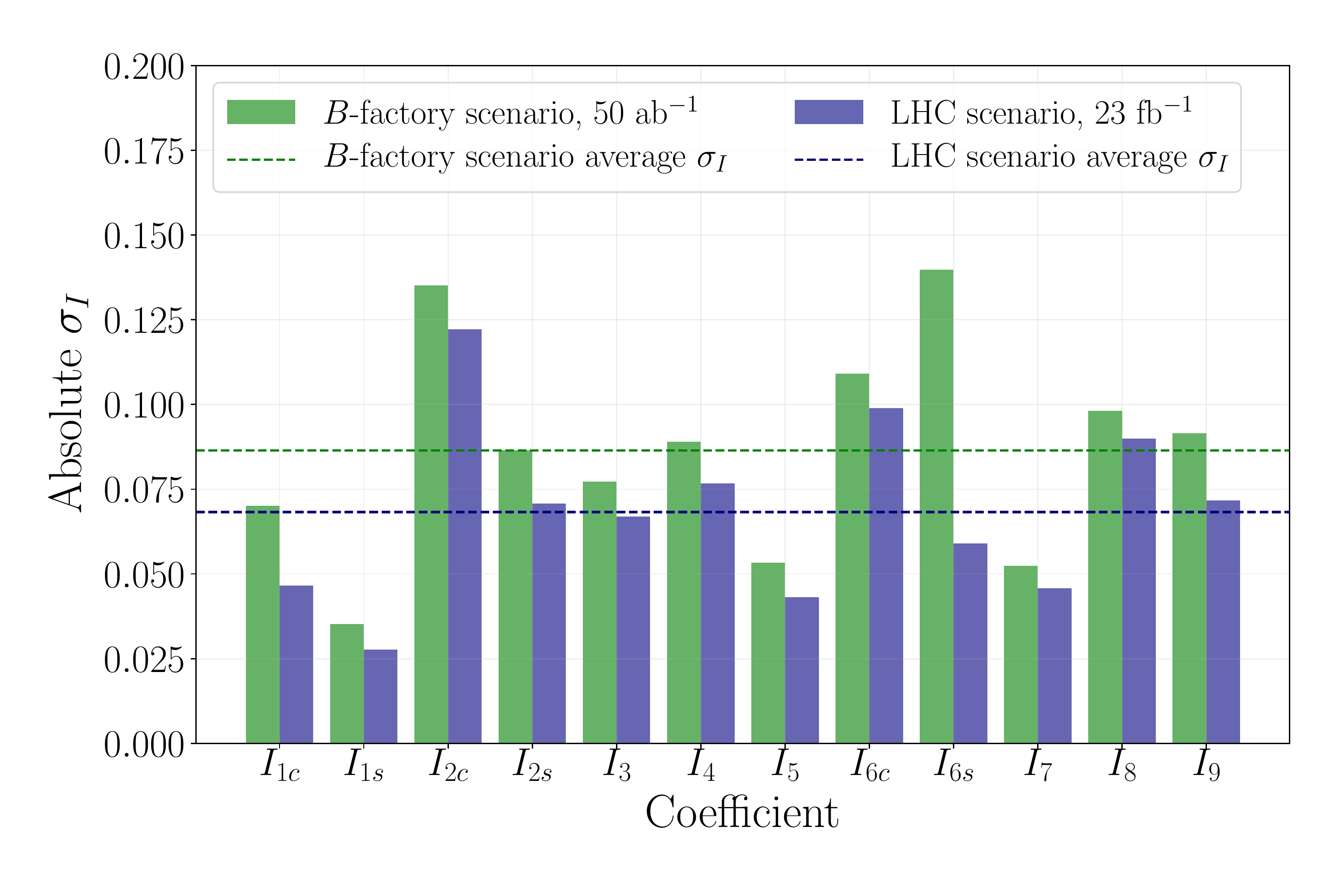}
\caption{Comparison of absolute $I_X$ coefficient statistical uncertainties in the \mbox{$N_{\text{sig}}=40,000$} hadron collider template fit (navy) and the \mbox{$N_{\text{sig}}=7,000$} $B$-factory fit (green). The average uncertainties over all $I_X$ coefficients are indicated by the dotted lines.}
\label{fig:reco_I_BFac}
\end{figure}

\begin{sloppypar}
The four-dimensional template fit to the $B$-factory sample is performed in $((\cos\theta_D,\cos\theta_L,\chi)_{\text{Reco}},\text{BDT})$ variable space, where the decay angles are calculated using the true $B$ meson four-vector to mimic the benefit of the hadronic tagging. The number of bins in each dimension is chosen in the same manner as the LHCb scenario fits. The signal fraction is measured to be $f_{\text{sig}} = 0.195 \pm 0.014$ (7.0\% relative uncertainty) and is consistent with the input value. The uncertainties on the $I_X$ measurements are compared to the 23 fb$^{-1}$ LHCb scenario in Fig.~\ref{fig:reco_I_BFac}. 
Even though the $B$-factory signal yield is lower, the overall $I_X$ precision is competitive due to the higher purity and constraint on the initial state from the tagging of the other $B$ decay.
\end{sloppypar}

\subsection{Systematic uncertainties}
\label{sec:systematics}

The dominant systematic uncertainty comes from the assumed accuracy of the  templates used to model the background. Measured branching fractions are used to define the contribution from each background decay, so these are varied within their uncertainties to determine the appropriate uncertainty. Similarly, fixed fractions are used to define the feed-down contribution, which has not yet been confirmed experimentally and thus a 40\% variation around $f_{D^{**}} = 0.11$ is used.
Smaller variations in the angular coefficient measurements are seen when the number of bins in the weighting procedure is varied from the default $30^3$ binning.
The total systematic uncertainties are found to be small relative to the statistical uncertainties, even in the highest yield case. The systematic uncertainties are shown to modestly increase the error bars in Fig.~\ref{fig:all_results_systs}.

Further systematic effects not quantified here include those due to  limited MC statistics and imperfect simulation of experimental data. The Beeston-Barlow~\cite{Barlow:1993dm} method can account for the first effect, while the second must be addressed through the development of control channels such as $B^0 \to D^{*-}D_s^+$ decays at LHCb or an inclusive study of $D_{(s)}\to 3\pi X$ decays at BESIII. 

\subsection{Determination of $R(D^*)$}
It is possible to convert the measured signal fraction $f_{\text{sig}}$ to a value of $\mathcal{B}(B^0 \to D^{*-} \tau^+ \nu_\tau)$ and thus $R(D^*)$ if $\epsilon_{\text{sig}}$, the total signal efficiency, is known. Typically, $\epsilon_{\text{sig}}$ is calculated from simulated samples that accurately model detector and selection inefficiencies. 
However, the simulated sample used to estimate $\epsilon_{\text{sig}}$ must be generated using a specific model, and so $\epsilon_{\text{sig}}$ is unavoidably model-dependent. This leads to an additional systematic uncertainty that must be considered in the extrapolation to $R(D^*)$, particularly if measurements of the $I_X$ coefficients diverge from their SM expectations.

It should be noted that the model-independent strategy cannot be competitive with the accuracy with which model-dependent fits can measure $R(D^*)$. For example, the statistical uncertainty on $f_{\text{sig}}$ in the 9 fb$^{-1}$ LHCb scenario is 8.6\%. The statistical uncertainty of a model-dependent fit to the same dataset may be under 3\%, based on an extrapolation of the statistical uncertainty in Ref.~\cite{PhysRevD.97.072013}. The inferior precision is due to the fact that (1) the model-independent fit has twelve parameters to describe the signal rather than one overall yield, and (2) the angular variables are less discriminating between signal and background than the model-dependent variables currently used such as $q^2$. To confirm these assertions, a test is performed using the generated 9 fb$^{-1}$ LHCb dataset, where all of the $I_X$ coefficients are fixed and the signal-background separation is artificially improved by 20\%. In this test, the fit uncertainty on $f_{\text{sig}}$ indeed reduces from 8.6\% to 3.3\%.

\section{Conclusion}
\label{sec:summary}

A model-independent method for measuring the angular coefficients of $B^0 \to D^{*-}\tau^+\nu_{\tau}$ decays is demonstrated. Reconstruction bias and resolution effects caused by the missing neutrino are handled using a binned template fit to the decay angles, with a BDT classifier included to improve signal-background separation. 
A realistic background mixture is introduced, and the template fit is found to be statistically unbiased and model-independent even with the current LHCb statistics (\mbox{Run 1 + 2}). Due to the cleaner $e^+e^-$ environment and a better-constrained $B$ meson reconstruction, \mbox{Belle II} should perform competitively despite lower signal yields.

The template fit is directly applicable to the isospin partner decay $B^+ \to \Dbar^{*0} \tau^+ \nu_\tau$, if the neutral vector meson can be efficiently reconstructed. 
By extension, template angular analysis of the pseudoscalar semitauonic decays $B^+ \to \Dzb \tau^+ \nu_\tau$ and $B^0 \to D^- \tau^+ \nu_\tau$ is motivated for experimental reasons. Here, the decay rate depends only on $\cos\theta_L$,
\begin{equation}
\frac{d^2\Gamma}{dq^2 d(\cos\theta_L)} = a + b\cos\theta_L + c\cos^2\theta_L\,,
\end{equation}
where $a$, $b$, and $c$ are the $q^2$-dependent angular coefficients~\cite{Becirevic:2019tpx}. 
However, a full angular analysis would enable the measurement of $I_X$ coefficients in the $D^{*-}$ feed-down in addition to the $(a,b,c)$ coefficients of the pseudoscalar modes.

The template procedure should be applied as a null test to $B \to D^{(*)}l\nu$ ($l \in \{e,\mu\}$) decays, since they are also governed by Eq.~\eqref{eq:decay_rate}. 
And in all cases, $\CP$ conservation, which could be violated if additional NP processes interfere with the single SM amplitude~\cite{London:2019wpw}, can be verified by splitting according to the $\tau$ lepton charge.
Finally, the template method is ideal for angular analysis that searches for right-handed currents in $B_s^0 \to K^{*-} \mu^+ \nu_\mu$ and $B^+ \to \rho^0 \mu^+ \nu_\mu$ decays, which suffer similar complications from neutrinos in the final state.

\FloatBarrier

\clearpage
\appendix

\section{Decay angle definitions}
\label{sec:angles}

\begin{figure}[h!]
\centering
\includegraphics[width = 0.8\textwidth]{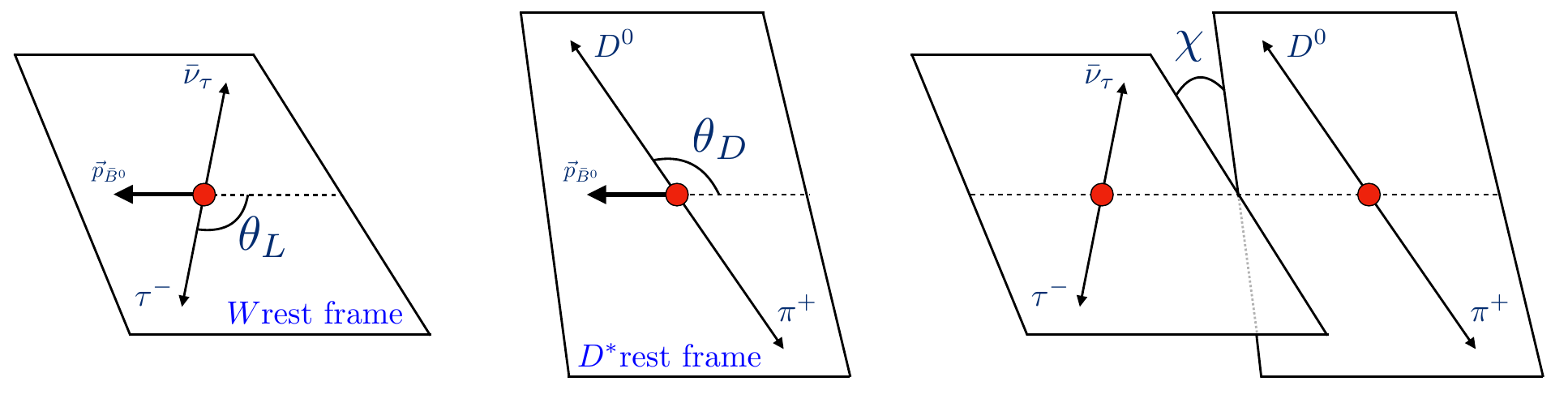}
\caption{Diagram of the three $B^0 \to D^{*-} \tau^+ \nu$ decay angles.}
\label{fig:decay_angle_diagram}
\end{figure}

In this work, $\theta_D$ is defined as the angle between the direction of the $\Dzb$ meson and the direction opposite that of the $B^0$ meson in in the $D^{*-}$ meson rest frame. The angle $\theta_L$ is defined as the angle between the direction of the $\tau^+$ lepton and the direction opposite that of the $B^0$ meson in in the mediator ($W^+$) rest frame. The angle $\chi$ is the angle between the plane containing the $\tau^+$ and $\nu_{\tau}$ and the plane containing the $\Dzb$ and pion from the $D^{*-}$ in the $B^0$ rest frame. The three decay angles are displayed graphically in Fig.~\ref{fig:decay_angle_diagram}. Explicitly, the decay angles are defined following the definitions in Ref.~\cite{Aaij2013}
\begin{align}
\cos\theta_D &= \Big(\hat{p}_{D^0}^{(D^{*-})} \Big) \cdot \Big(\hat{p}_{D^{*-}}^{(B^0)} \Big) = \Big(\hat{p}_{D^0}^{(D^{*-})} \Big) \cdot \Big(-\hat{p}_{B^0}^{(D^{*-})} \Big)\,,\\
\cos\theta_L &= \Big(\hat{p}_{\tau^+}^{(W^+)} \Big) \cdot \Big(\hat{p}_{W^+}^{(B^0)} \Big) = \Big(\hat{p}_{\tau^+}^{(W^+)} \Big) \cdot \Big(-\hat{p}_{B^0}^{(W^+)} \Big)\,,\\
\cos\chi &= \Big(\hat{p}_{\tau^+}^{(B^0)} \times \hat{p}_{\nu_{\tau}}^{(B^0)} \Big) \cdot \Big(\hat{p}_{D^0}^{(B^0)} \times \hat{p}_{\pi^-}^{(B^0)} \Big)
\end{align}
where the $\hat{p}_{X}^{(Y)}$ are unit vectors describing the direction of a particle $X$ in the rest frame of the system $Y$.  In every case the particle momenta are first boosted to the $B^0$ rest frame. In this basis, the angular definition for the $\bar{B}^0$ decay is a $CP$ transformation of that for the $B^0$ decay.

\FloatBarrier
\clearpage
\section{Background models}
\label{sec:bkg_modes}

{
\renewcommand{\arraystretch}{1.25}
\begin{table}[h!]
\centering
\scriptsize
\begin{tabular}{l | l}
Mode & EvtGen models \\
\hline
$B^0 \to D^{*-}\pi^+\pi^+\pi^-\pi^0$ & $B^0$ (PHSP)\\\hline
$B^0 \to D^{*-}(\omega \to \pi^+\pi^-\pi^0)\pi^+$ & $B^0$ (PHSP),\\
& $\omega$ (OMEGA\_DALITZ)\\\hline
$B^0 \to D^{*-}\pi^+\pi^+\pi^-\pi^+\pi^-$ & $B^0$ (PHSP)\\\hline
$B^+ \to D^{*-}\pi^+\pi^+\pi^-\pi^+$ & $B^+$ (PHSP)
\end{tabular}
\caption{$B^+ \to D^{*-}\pi^+\pi^+\pi^-(X)$ modes generated in order to construct the prompt background sample.}
\label{tab:prompt_modes}
\end{table}
}

{
\renewcommand{\arraystretch}{1.25}
\begin{table}[h!]
\centering
\scriptsize
\begin{tabular}{l | l}
Mode & EvtGen models \\
\hline
$B^0 \to D^{*-} D_s^+$ & $B^0$ (SVS)\\\hline
$B^0 \to D^{*-} (D_s^{*+} \to D_s^+ \gamma)$ & $B^0$ (SVV\_HELAMP 0.4904 0.0 0.7204 0.0 0.4904 0.0), \\
& $D_s^{*+}$ (VSP\_PWAVE)\\
$B^0 \to D^{*-} (D_s^{*+} \to D_s^+ \pi^0)$ & $B^0$ (""),\\ & $D_s^{*+}$ (VSS)\\\hline
$B^0 \to D^{*-} (D_{s1}(2460)^+ \to D_s^+\gamma)$ & $B^0$ (SVV\_HELAMP 1 0 1 0 1 0), \\
& $D_{s1}(2460)^+$ (VSP\_PWAVE) \\
$B^0 \to D^{*-} (D_{s1}(2460)^+ \to (D_s^{*+} \to D_s^+\gamma) \pi^0)$ & $B^0$ (""),\\
& $D_{s1}(2460)^+$ (PHSP), \\
& $D_s^{*+}$ (VSP\_PWAVE) \\\hline
$B^+ \to (D_{1}(2420)^0 \to D^{*-} \pi^+) D_s^+$ & $B^+$ (SVS), \\
& $D_{1}(2420)^0$ (VVS\_PWAVE  0 0 0 0 1 0)\\\hline
$B^+ \to (D_{1}(2420)^0 \to D^{*-} \pi^+) (D_s^{*+} \to D_s^+\gamma)$ & $B^+$ (SVV\_HELAMP 0.48 0.0 0.734 0.0 0.48 0.0),\\
& $D_{1}(2420)^0$ (""), \\
& $D_s^{*+}$ (VSP\_PWAVE)\\
$B^+ \to (D_{1}(2420)^0 \to D^{*-} \pi^+) (D_s^{*+} \to D_s^+\pi^0)$ & $B^+$ (""),\\
& $D_{1}(2420)^0$ (""), \\
& $D_s^{*+}$ (VSS)\\\hline
$B^+ \to (D_1(2420)^0 \to D^{*-}\pi^+)(D_{s1}(2460)^+ \to D_s^+\gamma)$ & $B^+$ (SVV\_HELAMP 1.0 0.0 1.0 0.0 1.0 0.0),\\ & $D_1(2420)^0$ (VVS\_PWAVE  0 0 0 0 1 0), \\
& $D_{s1}(2460)^+$ (VSP\_PWAVE) \\
$B^+ \to (D_1(2420)^0 \to D^{*-}\pi^+) (D_{s1}(2460)^+ \to (D_s^{*+} \to D_s^+\gamma) \pi^0)$ & $B^+$ (""),\\
& $D_1(2420)^0$ (""), \\
& $D_{s1}(2460)^+$ (PHSP), \\
& $D_s^{*+}$ (VSP\_PWAVE) \\

\end{tabular}
\caption{Background modes generated to construct the $B \to D^{*-}D_s^+(X)$ sample. In all cases, $D^{*-} \to \Dzb \pi^-$ is generated according to the VSS model, and $\Dzb \to K^+\pi$ is generated according to phase space. Where indicated, "" means that the same model was used as in the previous row.}
\label{tab:DstDsX_bkg_modes}
\end{table}
}

{
\renewcommand{\arraystretch}{1.25}
\begin{table}
\centering
\scriptsize
\begin{tabular}{l | l}
Mode & EvtGen models \\
\hline
$D_s^+ \to (\eta \to \pi^+ \pi^- \pi^0)\pi^+$ & $D_s^+$ (PHSP),\\
& $\eta$ (ETA\_DALITZ) \\\hline
$D_s^+ \to (\omega \to \pi^+ \pi^- \pi^0)\pi^+$ & $D_s^+$ (SVS), \\
& $\omega$ (OMEGA\_DALITZ)\\\hline
$D_s^+ \to (\eta \to \pi^+ \pi^- \pi^0)(\rho^+ \pi^+ \pi^0)$ & $D_s^+$ (SVS),\\
& $\eta$ (ETA\_DALITZ), \\
& $\rho^+$ (VSS) \\\hline
$D_s^+ \to (\omega \to \pi^+ \pi^- \pi^0)(\rho^+ \pi^+ \pi^0)$ & $D_s^+$ (PHSP),\\
& $\omega$ (OMEGA\_DALITZ), \\
& $\rho^+$ (VSS) \\\hline
$D_s^+ \to (\rho^0 \to \pi^+ \pi-)(\rho^0 \to \pi^+ \pi-)\pi^+$ & $D_s^+$ (PHSP),\\
& $\rho^0$ (VSS)\\\hline
$D_s^+ \to (\omega \to \pi^+ \pi^- \pi^0)\pi^+\pi^+\pi^-$ & $D_s^+$ (PHSP), \\
& $\omega$ (OMEGA\_DALITZ)\\\hline
$D_s^+ \to (\eta' \to (\eta \to \pi^+\pi^-\pi^0) \pi^+\pi^-) \pi^+$ & $D_s^+$ (PHSP),\\
& $\eta'$ (PHSP), \\
& $\eta$ (ETA\_DALITZ) \\
$D_s^+ \to (\eta' \to (\rho^0 \to \pi^+\pi^-)\gamma)\pi^+$ & $D_s^+$ (""), \\
& $\eta'$ (SVP\_HELAMP 1.0 0.0 1.0 0.0), \\
& $\rho^0$ (VSS)\\\hline
$D_s^+ \to (\eta' \to (\eta \to \pi^+\pi^-\pi^0) \pi^+\pi^-) (\rho^+ \to \pi^+ \pi^0)$ & $D_s^+$ (SVS),\\
& $\eta'$ (PHSP), \\
& $\eta$ (ETA\_DALITZ), \\
& $\rho^+$ (VSS)\\
$D_s^+ \to (\eta' \to (\rho^0 \to \pi^+\pi^-)\gamma)(\rho^+ \to \pi^+ \pi^0)$ & $D_s^+$ (""),\\
& $\eta'$ (SVP\_HELAMP 1.0 0.0 1.0 0.0), \\
& $\rho^{0,+}$ (VSS)\\

\end{tabular}
\caption{$D_s^+$ decays generated for constructing the $B \to D^{*-}D_s^+(X)$ sample. Where indicated, "" means that the same model was used as in the previous row.}
\label{tab:Ds_bkg_modes}
\end{table}
}

{
\renewcommand{\arraystretch}{1.25}
\begin{table}
\centering
\scriptsize
\begin{tabular}{l | l}
Mode & EvtGen models \\
\hline
$B^- \to D^{*-}D^0\bar{K}^0$ & $B^-$ (PHSP)\\ \hline
$B^- \to D^{*-}(D^{*0} \to D^0 \pi^0)\bar{K}^0$ & $B^-$ (PHSP), \\
& $D^{*0}$ (VSS)\\
$B^- \to D^{*-}(D^{*0} \to D^0 \gamma)\bar{K}^0$ & $B^-$ (PHSP), \\
& $D^{*0}$ (VSP\_PWAVE)\\\hline
$B^0 \to D^{*-} D^0 K^+$ & $B^0$ (PHSP)\\\hline
$B^0 \to D^{*-}(D^{*0} \to D^0 \pi^0)K^+$ & $B^0$ (PHSP),\\ 
& $D^{*0}$ (VSS)\\
$B^0 \to D^{*-}(D^{*0} \to D^0 \gamma)K^+$ & $B^0$ (PHSP),\\ 
& $D^{*0}$ (VSP\_PWAVE)\\\hline
$B^0 \to D^{*-} (D^{*+} \to D^0 \pi^+)\bar{K}^0$ & $B^0$ (PHSP), \\
& $D^{*+}$ (VSS) \\ \hline
$B^0 \to D^{*-} D^+ \bar{K}^0$ & $B^0$ (PHSP) \\ \hline
$B^- \to D^{*-} D^+ K^-$ & $B^0$ (PHSP) \\
\end{tabular}
\caption{Background modes generated to construct the $B \to D^{*-}D^0(X)$ and $B \to D^{*-}D^+(X)$ samples. In all cases, $D^{*-} \to \Dzb \pi^-$ is generated according to the VSS model, and $\Dzb \to K^+\pi$ is generated according to phase space.}
\label{tab:DstDX_bkg_modes}
\end{table}
}

{
\renewcommand{\arraystretch}{1.25}
\begin{table}
\centering
\scriptsize
\begin{tabular}{l | l}
Mode & EvtGen models \\
\hline
$D^0 \to K^- \pi^+\pi^+\pi^-$ & $D^0$ (PHSP)\\\hline
$D^0 \to K^- \pi^+\pi^+\pi^-\pi^0$ & $D^0$ (PHSP)\\\hline
$D^+ \to K_s^0\pi^+\pi^+\pi^-$ & $D^+$ (PHSP)\\\hline
$D^+ \to \pi^+\pi^+\pi^-\pi^0$ & $D^+$ (PHSP)\\\hline
\end{tabular}
\caption{$D^{0,+}$ decays generated for constructing the $B \to D^{*-}D^{0,+}(X)$ sample.}
\label{tab:D_bkg_modes}
\end{table}
}

{
\renewcommand{\arraystretch}{1.25}
\begin{table}
\centering
\scriptsize
\begin{tabular}{l | l}
Mode & EvtGen models \\
\hline
$B^+ \to (D_1(2420)^0 \to D^{*-}\pi^+) \tau^+\nu_{\tau}$ & $B^+$ (ISGW2),\\
& $D_1(2420)^0$ (VVS\_PWAVE 0.0 0.0 0.0 0.0 1.0 0.0)\\\hline
$B^+ \to (D_2^*(2460)^0 \to D^{*-}\pi^+) \tau^+\nu_{\tau}$ & $B^+$ (ISGW2),\\
& $D_2^*(2460)^0$ (TVS\_PWAVE 0.0 0.0 1.0 0.0 0.0 0.0)\\
\end{tabular}
\caption{$B^+ \to D^{**0}\tau^+\nu_{\tau}$ modes generated for constructing the feed down background sample. As these modes have not been observed, they are added together in equal proportion to create the total sample.}
\label{tab:feed_down_modes}
\end{table}
}

\FloatBarrier
\section{Correlation matrix}
\label{sec:corr_matrix}

\begin{table}[h!]
\centering
\scriptsize
\begin{tabular}{c|c c c c c c c c c c c c c c}
 & $I_{1s}$ & $I_{2c}$ & $I_{2s}$ & $I_{3}$ & $I_{4}$ & $I_{5}$ & $I_{6c}$ & $I_{6s}$ & $I_{7}$ & $I_{8}$ & $I_{9}$ & $f_{\text{sig}}$ & $f_{D_s^+}$ & $f_{D^+}$ \\ \hline
$I_{1s}$ & \phantom{-}1.00 & -0.26 & \phantom{-}0.86 & \phantom{-}0.21 & -0.17 & \phantom{-}0.00 & \phantom{-}0.03 & \phantom{-}0.23 & \phantom{-}0.00 & -0.01 & -0.00 & \phantom{-}0.16 & \phantom{-}0.45 & -0.26 \\
$I_{2c}$ & -0.26 & \phantom{-}1.00 & -0.35 & -0.26 & -0.07 & -0.07 & -0.04 & \phantom{-}0.01 & \phantom{-}0.00 & \phantom{-}0.01 & \phantom{-}0.02 & \phantom{-}0.13 & -0.08 & \phantom{-}0.14 \\
$I_{2s}$ & \phantom{-}0.86 & -0.35 & \phantom{-}1.00 & \phantom{-}0.02 & -0.05 & \phantom{-}0.11 & \phantom{-}0.06 & \phantom{-}0.10 & \phantom{-}0.00 & -0.01 & -0.02 & \phantom{-}0.08 & \phantom{-}0.20 & -0.04 \\
$I_{3}$ & \phantom{-}0.21 & -0.26 & \phantom{-}0.02 & \phantom{-}1.00 & -0.19 & -0.03 & -0.21 & -0.22 & -0.01 & \phantom{-}0.03 & -0.00 & -0.14 & \phantom{-}0.39 & -0.44 \\
$I_{4}$ & -0.17 & -0.07 & -0.05 & -0.19 & \phantom{-}1.00 & \phantom{-}0.05 & -0.01 & \phantom{-}0.04 & -0.00 & \phantom{-}0.01 & \phantom{-}0.01 & -0.45 & -0.01 & -0.17 \\
$I_{5}$ & \phantom{-}0.00 & -0.07 & \phantom{-}0.11 & -0.03 & \phantom{-}0.05 & \phantom{-}1.00 & -0.09 & -0.36 & -0.01 & -0.00 & \phantom{-}0.01 & -0.08 & \phantom{-}0.01 & \phantom{-}0.04 \\
$I_{6c}$ & \phantom{-}0.03 & -0.04 & \phantom{-}0.06 & -0.21 & -0.01 & -0.09 & \phantom{-}1.00 & -0.13 & \phantom{-}0.01 & -0.03 & -0.00 & \phantom{-}0.08 & \phantom{-}0.03 & \phantom{-}0.03 \\
$I_{6s}$ & \phantom{-}0.23 & \phantom{-}0.01 & \phantom{-}0.10 & -0.22 & \phantom{-}0.04 & -0.36 & -0.13 & \phantom{-}1.00 & -0.01 & -0.00 & \phantom{-}0.00 & \phantom{-}0.20 & -0.05 & -0.00 \\
$I_{7}$ & \phantom{-}0.00 & \phantom{-}0.00 & \phantom{-}0.00 & -0.01 & -0.00 & -0.01 & \phantom{-}0.01 & -0.01 & \phantom{-}1.00 & -0.07 & -0.35 & -0.00 & \phantom{-}0.01 & -0.00 \\
$I_{8}$ & -0.01 & \phantom{-}0.01 & -0.01 & \phantom{-}0.03 & \phantom{-}0.01 & -0.00 & -0.03 & -0.00 & -0.07 & \phantom{-}1.00 & -0.05 & -0.01 & -0.01 & \phantom{-}0.00 \\
$I_{9}$ & -0.00 & \phantom{-}0.02 & -0.02 & -0.00 & \phantom{-}0.01 & \phantom{-}0.01 & -0.00 & \phantom{-}0.00 & -0.35 & -0.05 & \phantom{-}1.00 & \phantom{-}0.01 & \phantom{-}0.00 & -0.01 \\
$f_{\text{sig}}$ & \phantom{-}0.16 & \phantom{-}0.13 & \phantom{-}0.08 & -0.14 & -0.45 & -0.08 & \phantom{-}0.08 & \phantom{-}0.20 & -0.00 & -0.01 & \phantom{-}0.01 & \phantom{-}1.00 & \phantom{-}0.08 & -0.24 \\
$f_{D_s^+}$ & \phantom{-}0.45 & -0.08 & \phantom{-}0.20 & \phantom{-}0.39 & -0.01 & \phantom{-}0.01 & \phantom{-}0.03 & -0.05 & \phantom{-}0.01 & -0.01 & \phantom{-}0.00 & \phantom{-}0.08 & \phantom{-}1.00 & -0.79 \\
$f_{D^+}$ & -0.26 & \phantom{-}0.14 & -0.04 & -0.44 & -0.17 & \phantom{-}0.04 & \phantom{-}0.03 & -0.00 & -0.00 & \phantom{-}0.00 & -0.01 & -0.24 & -0.79 & \phantom{-}1.00 \\
\end{tabular}
\caption{Statistical correlation matrix for the $N=100,000$ hadron collider binned fit. All other fits show a consistent pattern of correlations.}
\label{tab:stat_corr_100k}
\end{table}

\clearpage
\bibliography{refs} 
\bibliographystyle{JHEP}

\section*{Acknowledgements}

We wish to thank Daniel Craik, Dean Robinson, Michele Papucci, and Tom Blake for their advice and assistance. We are particularly indebted to Marco Fedele for several fruitful exchanges. The work is supported by the Science and Technology Facilities Council (STFC, United Kingdom) and the Centre National de la Recherche Scientifique (CNRS, France).

\end{document}